\documentclass[a4paper,one column]{IEEEtran}

\usepackage{fixltx2e,fix-cm}

\usepackage{amssymb}
\usepackage{amsmath}
\usepackage{makeidx}
\usepackage{multicol}
\usepackage[abs]{overpic}
\usepackage{tabularx}
\usepackage[numbers,sort&compress]{natbib}
\usepackage{amsfonts}
\usepackage{color}
\usepackage{array,color}
\usepackage{tabularx}
\usepackage{multirow}
\usepackage{graphics}
\usepackage{epsfig}
\usepackage{graphicx}
\usepackage{makeidx}
\usepackage{multicol}
\usepackage{amsthm}
\usepackage{bm}
\usepackage{booktabs}
\usepackage{ulem}
\usepackage{enumerate}
\usepackage{subfigure}
\usepackage{algorithm}
\usepackage{algorithmic}
\usepackage{amssymb}
\usepackage{amsmath}
\usepackage{graphicx}
\usepackage{subfigure}
\usepackage{makeidx}
\usepackage{multicol}

\newtheorem{theorem}{Theorem}[section]
\newtheorem{example}[theorem]{Example}

\newtheorem{definition}[theorem]{Definition}
\newtheorem{corollary}[theorem]{Corollary}

\newcommand {\Data} [1]{\mbox{${#1}$}}  

\newcommand {\Vector} [1]{\Data {\mathbf {#1}}}
\newcommand {\Tdata} [1]{\Data {\tilde {#1}}} 
\newcommand {\Fdata} [1]{\Data {\mathbb {#1}}} 

\newcommand {\Mul} [2]{\Data{ {#1} \times {#2}}}  
\newcommand {\Muls} [2]{\Data{ {#1} \! \times \!{#2}}}  

\newcommand {\Index} [2]{\Data{ {#1}_{\TI {#2}}}}  

\newcommand {\Equs} [2]{\Data{ {#1}\! =\! {#2}}}  

\newcommand {\SI}[1] {\small{#1}}
\newcommand {\TI}[1] {\tiny {#1}}
\newcommand {\Text}[1] {\text {#1}}

\newcommand {\VLES}[1]{\Index {\tau} {\SI{\Index {}{ \Text{#1}}}}}

\newcommand {\VPbus}[1]{\Index {P}{\Text{Node-}{#1}}}

\newcommand {\Vgam}[1]{\Index {\gamma}{#1}}

\newcommand {\Cur}[2] {\mbox {\Data {#1}-\Data {#2}}}

\newcommand {\STE}[1] {\Fdata {E}{\Data{({#1})}}}

\newcommand {\ROMAN}[1] {\uppercase\expandafter{\romannumeral#1}}

\def \FuncC #1#2{
\begin{equation}
{#2}
\label {#1}
\end{equation}
}

\definecolor{Orange}{RGB}{249,106,027}
\definecolor{sOrange}{RGB}{251,166,118}
\definecolor{ssOrange}{RGB}{254,213,190}

\definecolor{Blue}{RGB}{008,161,217}
\definecolor{sBlue}{RGB}{090,206,249}
\definecolor{ssBlue}{RGB}{200,239,253}

\begin{document}
%
\title{Spatio-Temporal Big Data Analysis for Smart Grids Based on Random Matrix Theory: A Comprehensive Study}


%

\author{{\bf{Robert Qiu}}$^{1,2,3}$  , {\bf{Lei Chu}}$^{2,3}$, {\bf{Xing He}}$^{2,3}$, {\bf{Zenan Ling}}$^{2,3}$, {\bf{Haichun Liu}}$^{2,3}$
       \\
       $^1$Tennessee Technological University, Cookeville, TN 38505 USA.
       \\
       $^2$Department of Electrical Engineering, Shanghai Jiaotong University, Shanghai 200240, China.
       \\
       $^3$Research Center for Big Data Engineering and Technology, State Energy Smart Grid Resarch and Development Center.
       \\
       \thanks{Dr.  Qiu's  work  is  supported  by  N.S.F. of  China  No.61571296 and N.S.F.  of  US  Grant No.  CNS-1247778, No.  CNS-1619250 }
       }



\IEEEtitleabstractindextext{%
\begin{abstract}
A cornerstone of the smart grid is the advanced monitorability on its assets and operations. Increasingly pervasive installation of the phasor measurement units (PMUs) allows the so-called synchrophasor measurements to be taken roughly 100 times faster than the legacy supervisory control and data acquisition (SCADA) measurements, time-stamped using the global positioning system (GPS) signals to capture the grid dynamics. On the other hand, the availability of low-latency two-way communication networks will pave the way to high-precision real-time grid state estimation and detection, remedial actions upon network instability, and accurate risk analysis and post-event assessment for failure prevention.

In this chapter, we firstly modelling spatio-temporal PMU data in large scale grids as random matrix sequences. Secondly, some basic principles of random matrix theory (RMT), such as asymptotic spectrum laws, transforms, convergence rate and free probability, are introduced briefly in order to the better understanding and application of RMT technologies. Lastly, the case studies based on synthetic data and real data are developed to evaluate the performance of the RMT-based schemes in different application scenarios (i.e., state evaluation and situation awareness).
\end{abstract}

}

\maketitle

\IEEEpeerreviewmaketitle
\IEEEdisplaynontitleabstractindextext

\section{Introduction}
\label{Intro}

\subsection{Perspective on Smart Grids}

The modern power grid is one of the most complex engineering systems in existence; the North American power grid is recognized as the supreme engineering achievement in the 20th century \cite{doe2003grid}. The complexity of the future's electrical grid is ever increasing: 1) the evolution of the grid network, especially the expansion in size;  2) the penetration of renewable/distributed resources, flexible/controllable electronic units, or even prosumers with dual load-generator behavior \cite{grijalva2011prosumer}; and 3) the revolution of the operation mechanism, e.g. demand-side management. Also, the financial, the environmental and the regulatory constraints are pushing the electrical grid towards its stability limit.

\begin{figure*}[htbp]
 \centering
 \subfigure[G1, Small-scale isolated grid (1900--1950)]
 {\label{fig:G3sa}
 \includegraphics[width=0.31\textwidth]{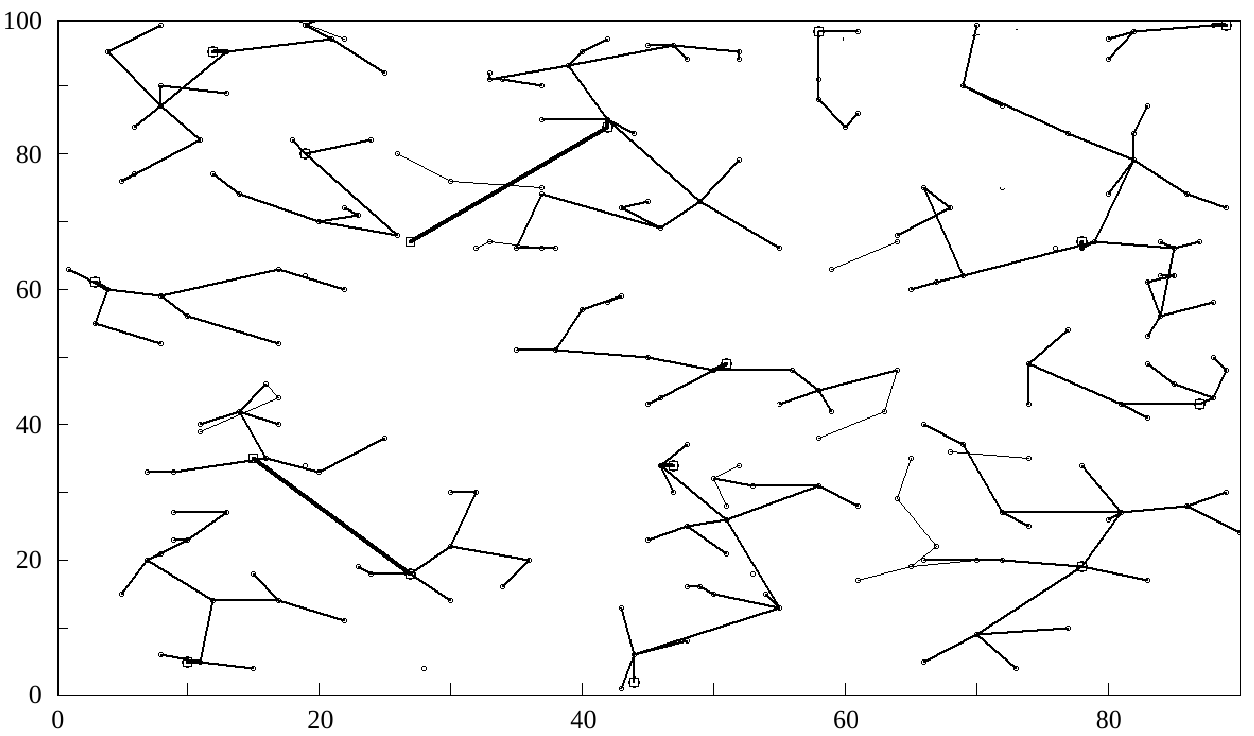}
 }
 \subfigure[G2, Large-scale interconnected grid (1960--2000)]
 {\label{fig:G3sb}
 \includegraphics[width=0.31\textwidth]{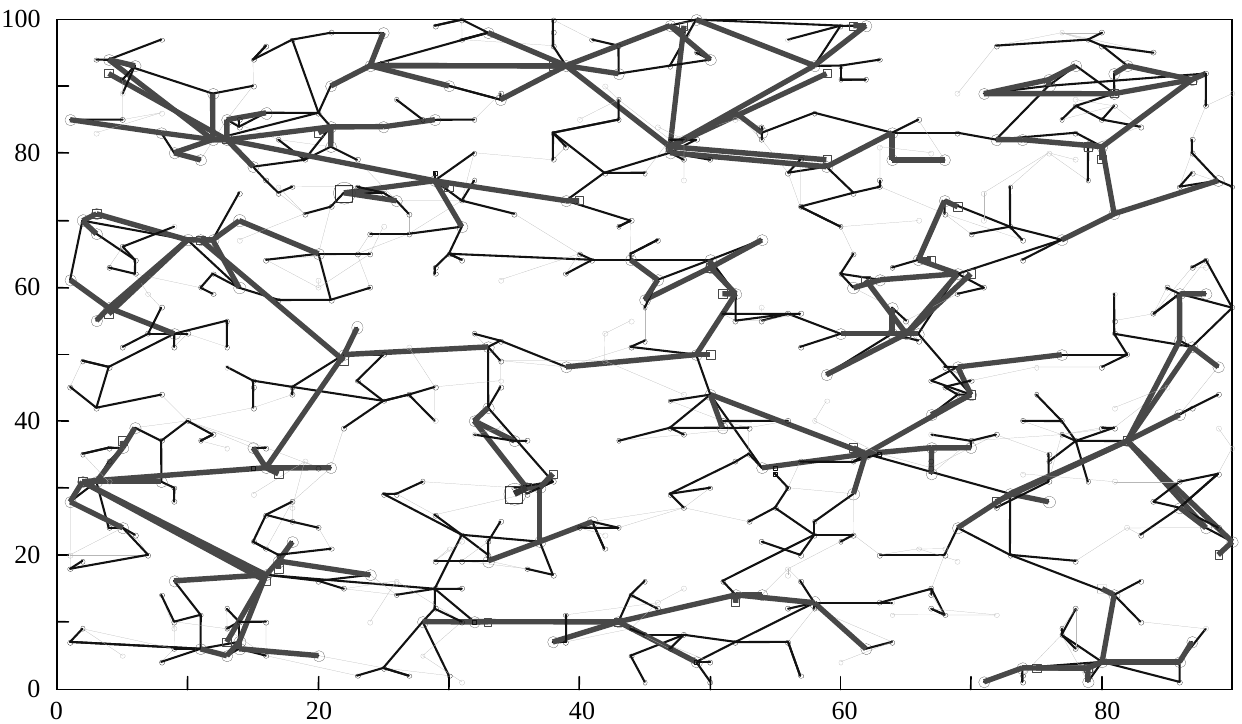}
 }
 \subfigure[G3, Smart grid (2000--2050)]
 {\label{fig:G3sc}
 \includegraphics[width=0.31\textwidth]{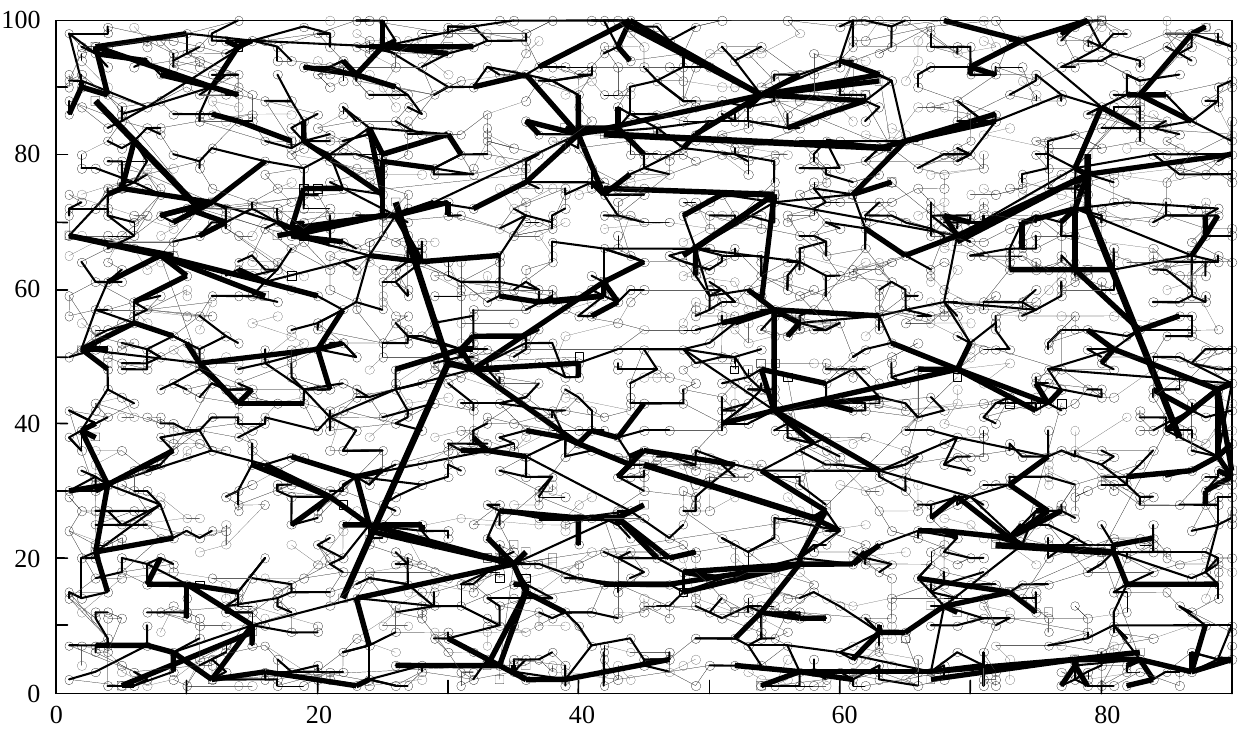}
 }
 \caption{Topologies of Grid Network.}
 \label{fig:G3s}
 \end{figure*}

\begin{figure}[htbp]
\centering
\includegraphics[width=0.6\textwidth]{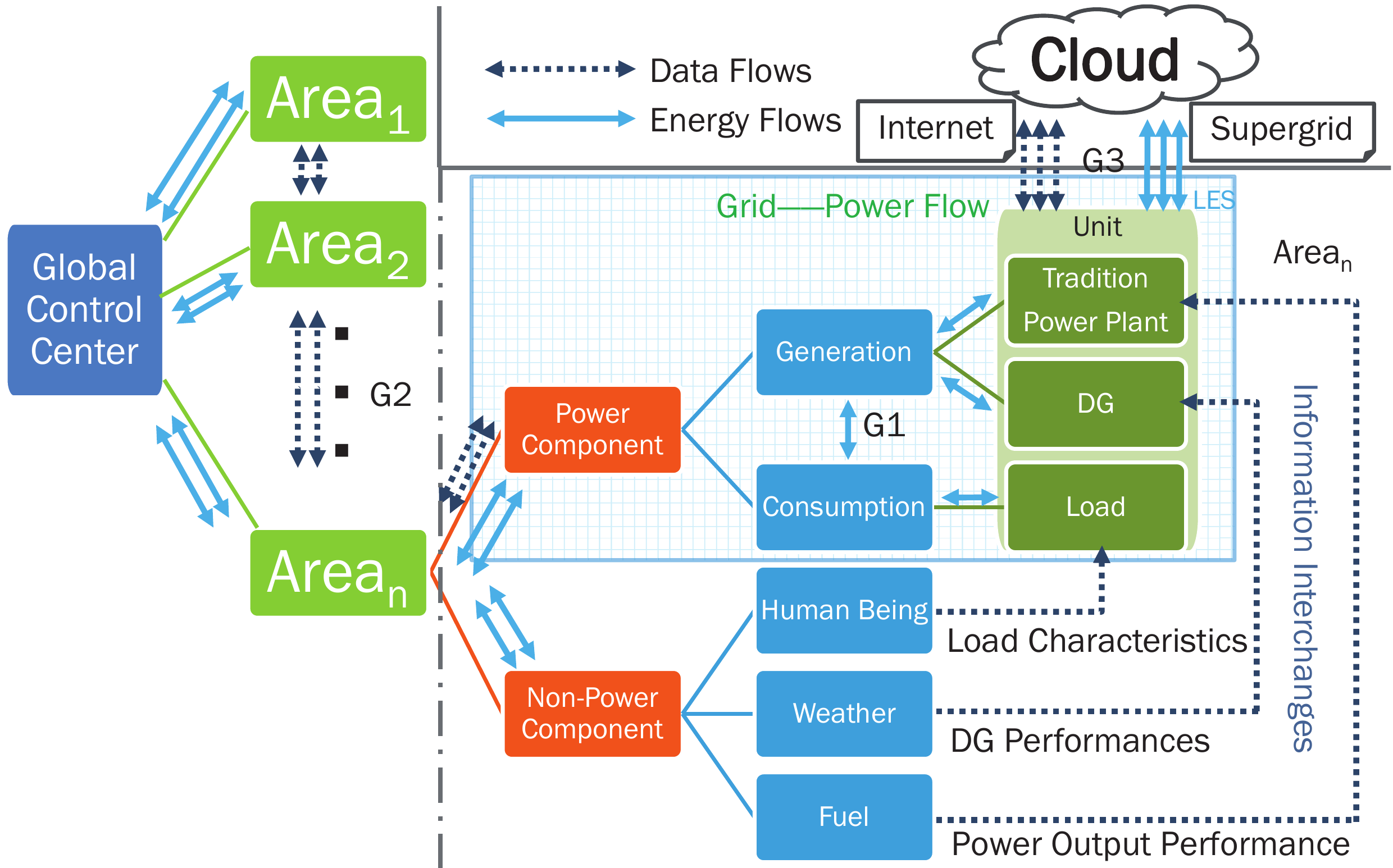}

\caption{Data flows and energy flows for three generations of power systems. The single lines, double lines, and triple lines indicate the flows of G1, G2, and G3, respectively.}
\label{fig:flow}
\end{figure}

\begin{figure}[htbp]
\centering
\includegraphics[width=0.95\textwidth]{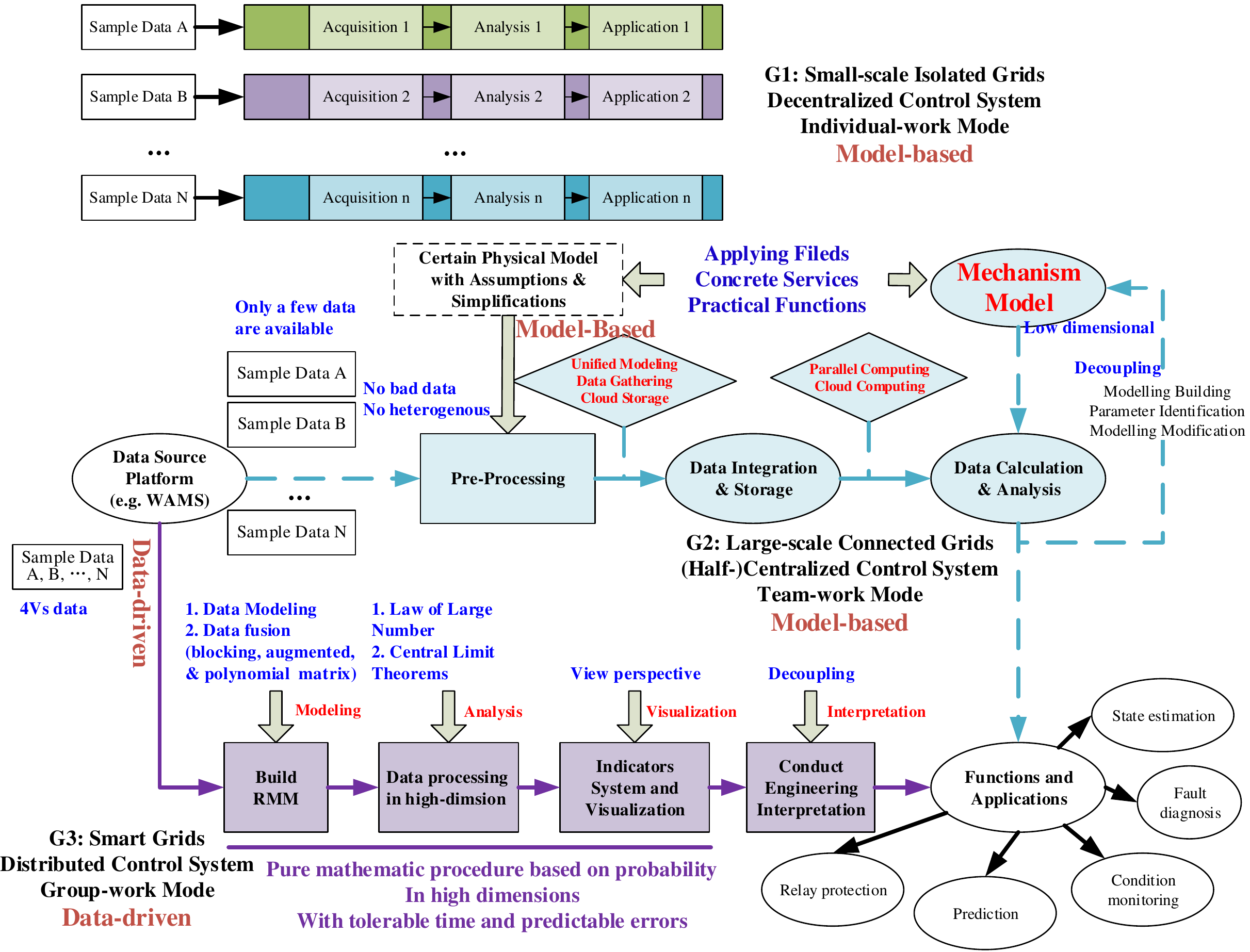}

\caption{Data management systems and work modes for three ages of power systems. The above, middle, and below parts indicate the data management systems and the work modes of G1, G2, and G3, respectively. For G1, each grid works independently. For G2, global and local control centers are operating under the team-work mode. For G3, the group-work mode breaks through the regional limitation for energy.}
\label{fig:procedure}
\end{figure}

Generally, power grids have experienced three ages---G1, G2, and G3 \cite{zhou2013review}. The network structures are depicted in Fig. \ref{fig:G3s} \cite{mei2014theevolution}. Their data flows and energy flows, as well as corresponding data management systems and work modes, are quite different \cite{he2014power}, which are shown in Fig. \ref{fig:flow} and Fig. \ref{fig:procedure}, respectively.

$\bf{G1: \ Small-scale \ isolated \ grids}$

G1 was developed from power system around 1900 to 1950, featured by small-scale isolated grids. For G1, units interchange energy and data within the isolated grid to keep generation-consumption balance. The units are most controlled by themselves, i.e., operating under individual-work mode. As shown in Fig. \ref{fig:G3sa}, each apparatus collects designated data, and makes corresponding decisions only with its own application. The individual-work mode works with an easy logic and little information communication. However, it means few advanced functions and inefficient utilization of resources. It is only suitable for small grids or isolated islands.

$\bf{G2: \ Large-scale \ interconnected \ grids}$

G2 was developed from power grids about 1960 to 2000, featured by zone-dividing large-scale interconnected grids. For G2, units interchange energy and data with the adjacent ones. The units are dispatched by a control center, i.e. operating under team-work mode. The regional team leaders, such as local dispatching centers, substations, and microgrid control centers, aggregate their own team-members (i.e. units in the region) into a standard black-box model. These standard models will be further aggregated by the global control center  for control or prediction purposes. The two aggregations above are achieved by four steps: data monitoring, data pre-processing, data storage, and data processing. The description above can be summarized by by dotted blue lines in Fig. \ref{fig:procedure}. In general, the team-work mode conducts model-based analysis, and mainly concerns system stability rather than individual benefit; it does not work well for smart grids with 4Vs data.

$\bf{G3: \ Smart \ grids}$

The development of G3 was launched at the beginning of the 21st century; and for China, it is expected to be completed around 2050 \cite{zhou2013review}. Fig. \ref{fig:G3sc} shows that the clear-cut partitioning is no longer suitable for G3, as well as the team-work mode which is based on the regional leader. For G3, the individual units, rather than the regional center (if still exists), play a dominant role. They are well self-control with high intelligence, resulting in much more flexible flows for both energy exchange and data communication \cite{zhang2014economic}. Accordingly, the group-work mode is proposed. Under this mode, the individuals freely operate under the supervision of the global control centers \cite{he2014power}. VPPs \cite{ai2011multi-agent}, MMGs \cite{he2012research}, for instance, are typically G3 utilities. These group-work mode utilities provide a relaxed environment to benefit both individuals and the grids: the former (i.e. individuals), driven by their own interests and characteristics, are able to create or join a relatively free group to benefit mutually from sharing their own superior resources; meanwhile, these utilities are often big and controllable enough to be good customers or managers to the grids.

\subsection{The Role of Data in Future Power Grid}
\label{RDFPG}

\begin{figure}[htbp]
\centering
\includegraphics[width=0.4\textheight]{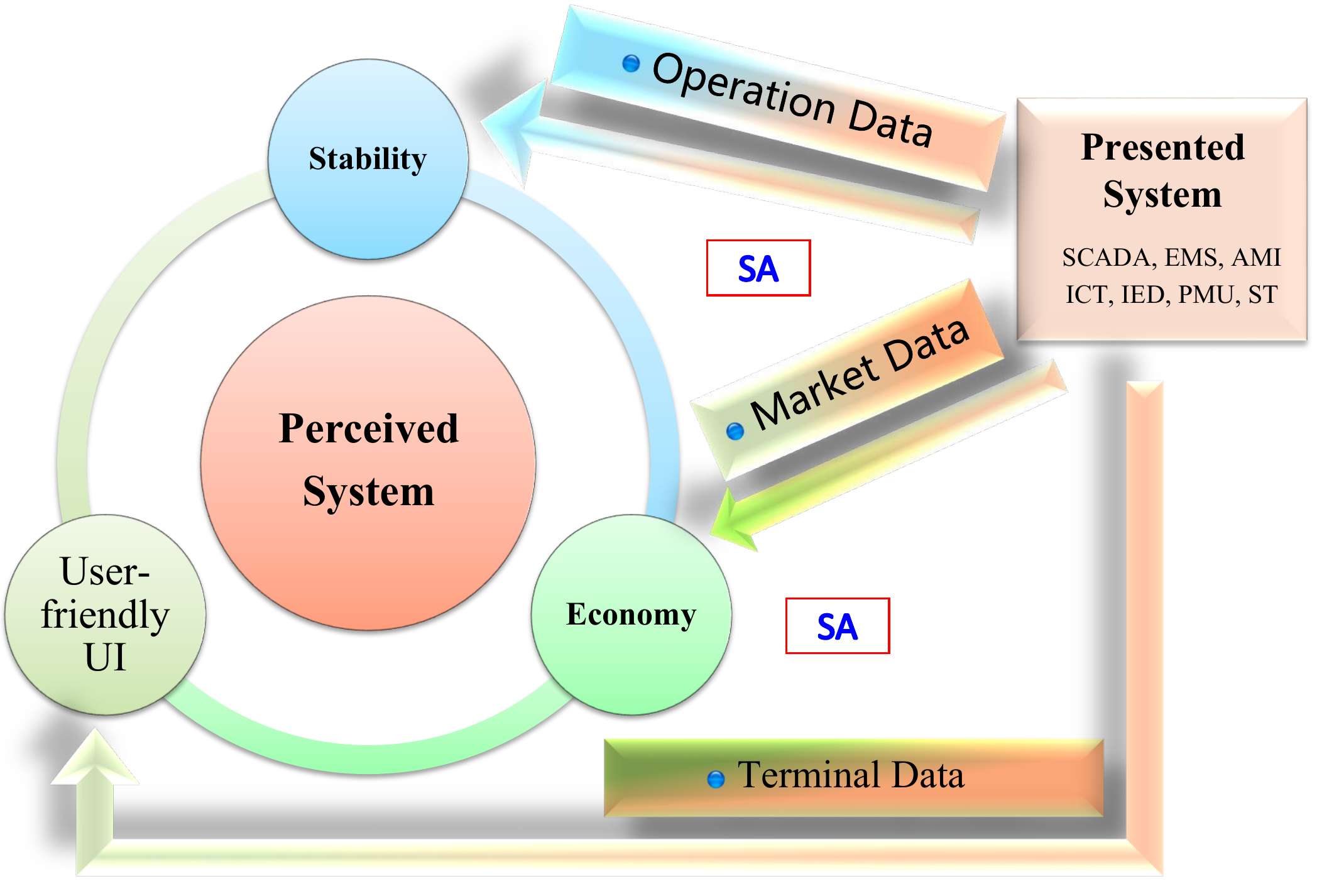}
\caption{Smart grid with 4Vs data and its SA.}
\label{fig:bdandsg}
\end{figure}

Data are more and more easily accessible in smart grids.  Fig. \ref{fig:bdandsg} shows numerous data sources: Information Communication Technology (ICT), Advanced Metering Infrastructure (AMI), Supervisory Control and Data Acquisition (SCADA), Sensor Technology (ST), Phasor Measurement Units (PMUs), and Intelligent Electronic Devices (IEDs) \cite{xu2013power}. Hence, data with features of volume, velocity, variety, and veracity (i.e. 4Vs data) \cite{IBM2014fourv} are inevitably generated and daily aggregated.
Particularly, the "4Vs" are elaborated as follows:
\begin{itemize}
\item \textit{Volume.} There are massive data in power grids. The so-called curse of dimensionality \cite{moulin2004support} occurs inevitably. The world wide small-scale roof-top photovoltaics (PVs) installation reached 23 GW at the end of 2013, and the growth is predicted to be 20 GW per year until 2018. The up-take of electric vehicles (EVs) also continues to grow. At least 665,000 electric-drive light-duty vehicles, 46,000 electric buses and 235 million electric two-wheelers were in the worldwide market in early 2015 \cite{parag2016electricity}.
\item \textit{Velocity.}   The resource costs (time, hardware, human, etc.) for big data analytics should be tolerable. To sever on-line decision-makings, massive data must be processed within a fraction of second.
\item \textit{Variety.} The data in various formats are often derived from diverse departments. In the view of data management, sampling frequency of source data, processing speed and service objects are not completely accord.
\item \textit{Veracity.} For a massive data source, there often exist realistic bad data, e.g. incomplete, inaccurate, asynchronous, and unavailable. For system operations, decisions such as protection, should be highly reliable.
\end{itemize}

As mentioned above, smart grids are always huge in size and complex in topology; big data analytics and data-driven approach become natural solutions for the future grid \cite{Lynch2008Big,Staff2011Dealing,Khan2015Cognitive,Chu2017Massive}.
Driven by data analysis in high-dimension, big data technology works out data correlations (indicated by statistical parameters) to gain insight to the inherent mechanisms.
Actually, big data technology has already been successfully applied as a powerful data-driven tool for numerous phenomena, such as quantum systems \cite{brody1981random}, financial systems \cite{laloux2000random,chen2012business}, biological systems \cite{howe2008big}, as well as wireless communication networks \mbox{\cite{qiu2013bookcogsen, qiu2014Intial70N, qiu2014MIMO}}.
For smart grids, data-driven approach and data utilization are current stressing topics, as evidenced in the special issue of  "Big Data Analytics for Grid Modernization" \cite{bda2016tsg}. This special issue is most relevant to our book in spirit. Several SA topics are discussed as well. We highlight anomaly detection and classification \cite{7460963, 7452675},  social media such as Twitter in \cite{7444207}, the estimation of active ingredients such as PV installations \cite{7456317, 7347457} and finally the real-time data for online transient stability evaluation \cite{7445227}.
In addition, we point out researchs about the improvement in wide-area monitoring, protection and control (WAMPAC) and the utilization of PMU data \cite{phadke2008wide,terzija2011wide,xie2012distributed, Khan2017Requirements}, together with the fault detection and location \cite{jiang2012pmu, venugopal2013novel, al2014fully}. Xie et al., based on Principal component analysis (PCA), proposes an online application for early event detection by introducing a reduced dimensionality \cite {xie2014dimensionality}. Lim et al. studies the quasi-steady-state operational problems relevant to the voltage instability phenomena \cite {lim2016svd}. These works provide primary exploration of the big data analysis in smart grid. Furthermore, a brief account for random matrix theory (RMT) which can be seen as basic analysis tools for spatial-temporal grid data processing, is elaborated in the following subsection.

\subsection{A Brief Account for RMT}

The last two decades have seen the rapid growth of RMT in many science fields. The brilliant mathematical works in RMT shed light on the challenges from classical statistics. In this subsection, we present a brief introduction to the main development of the RMT. The application-related account, with particular attention paid to recently rising RMT-based technology that are relevant for smart grid, is elaborated in Section \ref{ASG}.

The research of random matrices began with the work of Wishart in 1928 which focused on the distribution of the sample covariance matrices. The first asymptotic results on the limit spectrum of large random matrices (energy levels of nuclei) were obtained by Wigner in 1950s in a series of works \cite{wigner1955lower, Wigner1958On, wigner1963problem, Wigner1967Random} which ultimately lead to the well-known Semi-Circle Law \cite{Wigner1993Characteristic}. Another breakthrough was presented in \cite{Marchenko1967Distribution} that studied the distribution of eigenvalues for empirical covariance matrices. Based on these excellent works, RMT became a vibrant research direction of its own. Plenty of brilliant works that branched off the early physical and statistical applications were put forward in the last decades. For the sake of brevity, here we only show two remarkable results that turned out to be related to a large number of research hotspots in economics, communications and smart grid. One of the most striking progress is the discovery of the Tracy Widom distribution of extreme eigenvalues and another one is the single ring law which described the limit spectrum of eigenvalues of non-normal square matrices \cite{Guionnet2009The}. Interested readers are referred to monographs \cite{Anderson2009An, Tao2012Topics, qiu2015smart} for more details.

We will end this section by providing the structure of the remainder of this chapter.

Firstly, Section \ref{RMT} gives a tutorial account of existing mathematical works that are relevant to the statistical analysis of random matrices arising in smart grids. Specially, Section \ref{LDRM} introduces data collected from the widely applied phasor measurement unit and data modelling using linear and nonlinear combination of random matrices. Section \ref{ASL} focuses on asymptotic spectrum laws of the major types of random matrices. Section \ref{TF} presents some three dominant transforms which play key roles in describing the limit spectra of random matrices. Recent results on the convergence rate to the asymptotic limits are contained in the Section \ref{CR}.  Section \ref{FP} is dedicated to the free probability theory which is demonstrated as a practical tool for smart grids.

Secondly, we begin with some representative problems arisen from widely deployment of synchronous phasor measurement units that capture various features of interest in smart grids. We then show how random matrix theories have been used to characterize the data collected from synchronous phasor measurement and tackle the problems in the era of "Big Data". In particular, Section \ref{HTSG} provides some basis hypothesis tests that remain fundamental to research into the behaviour of the data in smart grids. Section \ref{DDMSA} concerns stability assessment from some recently developed data driven methods that based on RMT. Section \ref{SALES} focuses on situation awareness for smart grids from linear eigenvalue statistics. Early event detection problem is studied in details using free probability in Section \ref{EEDFP}.

\section{RMT: A Practical and Powerful Big Data Analysis Tool}
\label{RMT}

In this section, we provide a comprehensive existing mathematical results that are associated with the analysis of statistics of random matrices arising in smart grid. We also describe some new results on random matrices and other data-driven methods which were inspired by problems of engineering interest.

\subsection{Modelling Grid Data using Large Dimensional Random Matrices}
\label{LDRM}

Before the comprehensive utilization of RMT framework, we try to build a model for spatio-temporal PMU data using large dimensional random matrices.

It is well accepted that the transient behavior of a large electric power system can be illustrated by a set of differential and algebraic equations (DAEs) as follows \cite{bollen2000understanding}:
\begin{eqnarray}
\label{eqC1}
{{{\bf{\dot x}}}^{\left( t \right)}} &=& f\left( {{{\bf{x}}^{\left( t \right)}},{{\bf{u}}^{\left( t \right)}},{{\bf{h}}^{\left( t \right)}},w} \right)\\
\label{eqC2}
0 &=& g\left( {{{\bf{x}}^{\left( t \right)}},{{\bf{u}}^{\left( t \right)}},{{\bf{h}}^{\left( t \right)}},w} \right)
\end{eqnarray}
where ${{\bf{x}}^{\left( t \right)}} \in {\mathcal{C}}^{m \times p}$ are the power state variables, e.g., rotor speeds and the dynamic states of loads, ${{\bf{u}}^{\left( t \right)}}$ represent the system input parameters, ${{\bf{h}}^{\left( t \right)}}$ define algebraic variables, e.g., bus voltage magnitudes, $w$ denote the time-invariant system parameters. $t \in \mathcal{R}$, $m$ and $p$ are the sample time, number of system variables and bus, respectively. The model-based stability estimators \cite{lee2016voltage, ghanavati2016identifying} focus on linearization of nonlinear DAEs in \eqref{eqC1} and \eqref{eqC2} which gives
\begin{equation}
\label{eqC3}
\left[ {\begin{array}{*{20}{c}}
{\Delta {\bf{\underline{\dot x}}}}\\
{\Delta {\bf{\underline{\dot u}}}}
\end{array}} \right]{\rm{ = }}\left[ {\begin{array}{*{20}{c}}
{\bf{A}}&{ - {{\bf{f}}_{\bf{u}}}{\bf{g}}_{\bf{u}}^{ - 1}{{\bf{g}}_{\bf{h}}}}\\
{\bf{0}}&{ - {\bf{E}}}
\end{array}} \right]\left[ {\begin{array}{*{20}{c}}
{\Delta {\bf{\underline{x}}}}\\
{\Delta {\bf{\underline{u}}}}
\end{array}} \right]{\rm{ + }}\left[ {\begin{array}{*{20}{c}}
\bf{0}\\
{\bf{C}}
\end{array}} \right]{\bf{\xi }},
\end{equation}

where ${{\bf{f}}_{\bf{x}}}$, ${{\bf{f}}_{\bf{u}}}$ are the Jacobian matrices of ${{\bf{f}}}$ with respect to ${\bf{\underline{x}}}, {\bf{\underline{u}}}$ and ${\bf{A}} = {{\bf{f}}_{\bf{x}}} - {{\bf{f}}_{\bf{u}}}{\bf{g}}_{\bf{u}}^{ - 1}{{\bf{g}}_{\bf{x}}}$. $\bf{E}$ is a diagonal matrix whose diagonal entries equal $t_{cor}^{-1}$ and $t_{cor}$ is the correction time of the load fluctuations. $\bf{C}$ denotes a diagonal matrix whose diagonal entries are nominal values of the corresponding active $(P)$ or reactive $(Q)$ of loads; $\bf{\xi}$ is assumed to be a vector of independent Gaussian random variables.

It is noted that estimating the system stability by solving the equation \eqref{eqC3} is becoming increasingly more challenging \cite{qiu2015smart} as a consequence of the steady growth of the parameters, say, $t$, $p$ and $m$. Besides, the assumption that $\bf{\xi}$ follows Gaussian distribution would restrict the practical application.

On the other hand, as a novel alternative, the lately advanced data driven estimators \cite{Xu2015A, xie2014dimensionality, lim2016svd, ghanavati2016identifying} can assess stability without knowledge of the power network parameters or topology. However, these estimators are based on the analysis of individual window-truncated PMU data. In this chapter, we seek to provide a method with ability of continuous learning of power system from spatio-temporal PMU data.

\begin{figure}[!htp]
\centering
{
\includegraphics[width=0.7\textwidth]{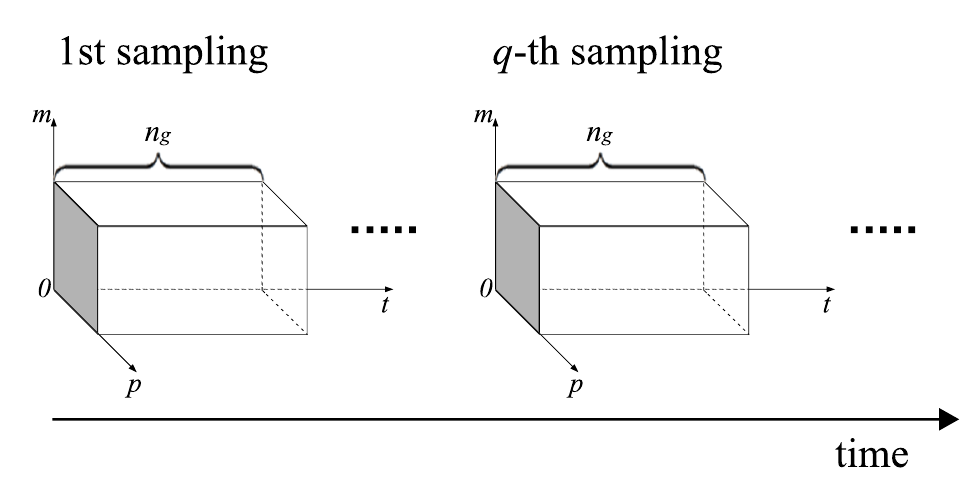}
}
\caption{{\label{Cfig1}} Conceptual representation of the structure of the spatio-temporal PMU data.}
\end{figure}

Firstly, we provide a novel method for modelling the spatio-temporal PMU data. Fig. \ref{Cfig1} illustrates the conceptual representation of the structure of the spatio-temporal PMU data. More specifically, let $p$ denote the number of the available PMUs across the whole power network, each providing $c$ measurements. At $i$th time sample, a total of $\kappa = p \times c$ measurements, say ${\bf{z}}_i$, are collected. With respect to each PMU, the $c$ measurements could contain many categories of variables, such as voltage magnitude, power flow and frequency, etc. In this chapter, we develop PMU data analysis assuming each type of measurements is independent. That is, we assume that at each round of analysis, $\kappa :=p$. Given $q$ time periods of $T$ seconds with $K$ Hz sampling frequency in $k$th data collection. Let $n_g=T \times K$ and  ${{\bf{Z}}_{ig}} = \left\{ {{{\bf{z}}_{i1}}, \cdots ,{{\bf{z}}_{{in_g}}}} \right\}, i = 1,2, \cdots, n$, a sequence of large random matrix
\begin{equation}
\label{eqC4}
\left\{ {\underbrace {{{\bf{Z}}_{11}},{{\bf{Z}}_{12}}, \cdots ,{{\bf{Z}}_{1q}}}_{q \ {\rm{ window - truncated \ data}}}, \cdots ,\underbrace {{{\bf{Z}}_{n1}},{{\bf{Z}}_{n2}}, \cdots ,{{\bf{Z}}_{nq}}}_{q \ {\rm{ window - truncated \ data}}}} \right\}
\end{equation}
is obtained to represent the collected voltage PMU measurements.

As illustrated in Fig. \ref{Cfig1}, ${{\bf{Z}}_i}$ is a large random matrix with independent identical distributed entries. Here we also include other forms of basic random matrices that are relevant to the applications in smart grids as follows.

{\bf{Gaussian Unitary Ensemble (GUE)}}: Let $\bf{Y}$ be $N \times N$ Wigner matrix or so called Gaussian unitary ensemble \rm{GUE}, and
 ${\bf{Y}} = {\left\{ {{w_{ij}}} \right\}_{1 \le i,j \le N}}$. $\bf{Y}$ satisfies the following
conditions:
\begin{enumerate}
	\item The entries of $\bf{Y}$ are $\rm{i.i.d}$ Gaussian variables.
	\item For $1 \le i \le j \le N$, ${\mathop{\rm Re}\nolimits} \left( {{Y_{ij}}} \right)$ and
${\mathop{\rm Re}\nolimits} \left( {{Y_{ij}}} \right),$ are $\rm{i.i.d.}$ with distribution
$ N\left( {0,\frac{1}{2}{\sigma ^2}} \right)$.
	\item For any $i,j$ in $\left\{ {1,2, \cdots ,N} \right\}$, ${Y_{ij}} = {\bar Y_{ji}}$.
	\item The diagonal entries of $\bf{Y}$ are real random variable with distribution
$ N\left( {0,{\sigma ^2}} \right)$.
\end{enumerate}

For the convenience of analysis, we can denote GUE as
${\bf{Y}} = \frac{1}{2}\left( {{\bf{X}} + {{\bf{X}}^H}} \right)$. Besides, the joint $\rm{p.d.f.}$
 of ordered eigenvalues of GUE
$\left( {{\lambda _1} \ge {\lambda _2} \ge  \cdots  \ge {\lambda _N}} \right)$ is
\cite{wegmann1976asymptotic,qiu2012cognitive}
\begin{equation}
\label{GaussianMatrices:eq2}
{2^{ - N{{} \mathord{\left/
{\vphantom {{} {}}} \right.\kern-\nulldelimiterspace} {}}2}}{\pi ^{ - {N^2}{{} \mathord{\left/
{\vphantom {{} {}}} \right.
\kern-\nulldelimiterspace} {}}2}}\exp \left[ { - \frac{{{\mathop{\rm Tr}\nolimits} {{\bf{Y}}^2}}}{2}}
\right]
\end{equation}

{\bf{Laguerre Unitary Ensemble (LUE)}}: Let ${\left\{ {{X_{ij}}} \right\}_{1 \le i \le M,1 \le j \le N}}$
be $\rm{i.i.d.}$ Gaussian random variables with
$\mathbb{E}\left( {{X_{ij}}} \right) = 0$ and $
\mathbb{E}X_{ij}^2 = \frac{1}{2}\left( {1 + {\delta _{ij}}} \right)$. The so called Wishart matrix or
Laguerre unitary ensemble \rm{LUE} can be expressed as
${\bf{W}} = \frac{1}{N}{\bf{X}}{{\bf{X}}^{\mathop{\rm H}\nolimits} }$.
The $\rm{p.d.f.}$ of ${\bf{W}}$ for $N\geq M$ is \cite{Anderson2009An, qiu2012cognitive}
\begin{equation}
\label{GaussianMatrices:eq3}
\frac{{{\pi ^{ - M\left( {M - 1} \right)/2}}}}{{\det \sum {\prod\nolimits_{i = 1}^M
{\left( {N - i} \right)!} } }}\exp \left[ { - {\mathop{\rm Tr}\nolimits}
\left\{ {\bf{W}} \right\}} \right]\det {{\bf{W}}^{N - M}}.
\end{equation}

{\bf{Large random matrix polynomials}:}
\[{\bf{M}} = f\left( {\bf{A}} \right) + g\left( {\bf{B}} \right),\] where $f,g$ are analytical functions, ${\bf{A}}$ is GUE, and ${\bf{B}}$ is LUE. See more details in Section \ref{FP}.

\subsection{Asymptotic Spectrum Laws}
\label{ASL}

In this subsection, we provide a brief introduction to the asymptotic spectrum laws of the large basic random matrices as shown in the Section \ref{LDRM}. There are remarkable results describing the asymptotic spectrum laws. Here special attention is paid to the limit behavior of marginal eigenvalues as the data dimensions tend to infinity.

We start with the GUE matrix ${\bf{X}} \in \mathbb{R}^N$ whose entries are independent identical distributed zero-mean (real or complex) Gaussian ensembles. As shown in \cite{Anderson2009An}, as $N \rightarrow \infty $, the empirical distribution of eigenvalues of $\frac{1}{{\sqrt N }}{\bf{X}}$ converges to the well-known semi-circle law whose density can be represented as

\begin{equation}
\label{ASL:eq1}
\rho \left( \lambda  \right) = \left\{ {\begin{array}{*{20}{c}}
{{\textstyle{1 \over {2\pi }}}\sqrt {4 - {\lambda ^2}} }&{{\rm{if}}}&{\left| x \right| \le 2}\\
0&{{\rm{if}}}&{\left| x \right| > 2}
\end{array}} \right.
\end{equation}

Also shown in \cite{wigner1955lower}, the same result could be obtained for a symmetric  ${\bf{X}}$ whose diagonal entries are 0 and whose lower-triangle entries are independent and take the values $\pm 1$ with equal probability.

If no attempt is made to symmetrize the square matrix ${\bf{X}}$, then the eigenvalues of $\frac{1}{{\sqrt N }}{\bf{X}}$ are asymptotically uniformly distributed on the unit circle of the complex plane. This is referred as the well-known Griko circle law which is elaborated in the following Theorem.

\begin{theorem}[Circular Law \cite{Tao2012Topics}] Let $\xi$ be a complex random variable with mean zero and unit variance. For each $N\ge 1.$ Let ${\bf X}_N$ be an iid random matrix of size $N$ with atom variable $\xi.$ Then, for any bounded and continuous function $f:\mathbb{C} \to \mathbb{C},$\[\int_\mathbb{C} {f\left( z \right)} d{\mu _{\frac{1}{{\sqrt N }}{{\mathbf{X}}_N}}}\left( z \right) \to \frac{1}{\pi }\int_\mathbb{U} {f\left( z \right)} {d^2}z\]
almost surely as $N\to \infty$ where $\mathbb U$ is the unit disk in the complex plane $\left| z \right| \leqslant 1$ and ${d^2}z = dxdy,$ with $z = x + iy.$
\end{theorem}

Semi-circle law and Circular Law explain the asymptotic property of large random matrices with independent entries. However, as illustrated in Section \ref{RDFPG}, the key issues in smart grid involve the singular values of rectangle large random matrix ${\bf{W}} \in \mathbb{R}^{N\times T}$. The LUE matrices ${\bf{W}} = \frac{1}{T}{{\bf{Y}}^H}{\bf{Y}}$ have dependent eigenvalues of interest even if ${\bf{Y}}$ has independent entries. Let the matrix aspect ratio $c=T/N$, the asymptotic theory of singular values of ${\bf{W}}$ was presented by the landmark work \cite{Marchenko1967Distribution} as follows.

As $N,T \rightarrow \infty$ and $c \leq 1$, the limit distribution of the eigenvalues of ${\bf{W}}$ converges to the so-called Marcenko-Pastur law whose density function is

\begin{equation}
\label{ASL:eq2}
\rho \left( \lambda  \right) = {\left( {1 - \frac{1}{c}} \right)^ + }\delta \left( \lambda  \right) + \frac{{\sqrt {{{\left( {\lambda  - a} \right)}^ + }{{\left( {b - \lambda } \right)}^ + }} }}{{2\pi c\lambda }},
\end{equation}
where ${\left( x \right)^ + } = \max \left( {0,x} \right)$ and \[a = {\left( {1 - \sqrt c } \right)^2},b = {\left( {1 + \sqrt c } \right)^2}\].

Analogously, when $c \geq 1$, the limit distribution of the eigenvalues of ${\bf{W}}$ converges to

\begin{equation}
\label{ASL:eq3}
\tilde \rho \left( \lambda  \right) = {\left( {1 - c} \right)^ + }\delta \left( \lambda  \right) + \frac{{\sqrt {{{\left( {\lambda  - a} \right)}^ + }{{\left( {b - \lambda } \right)}^ + }} }}{{2\pi \lambda }}.
\end{equation}

In addition to above Wigner's semicircle law and Marchenko-Pastur law, we are also interested in the Single Ring Law developed by Guionnet, Krishnapur and Zeitouni (2011) \cite{Guionnet2009The}. It describes the empirical distribution of the eigenvalues of a large generic matrix with prescribed singular values, i.e. an $N \times N$ matrix of the form ${\mathbf{A}} = {\mathbf{UTV}},$ with $\bf U;V$ some independent Haar-distributed unitary matrices and $\bf T$ a deterministic matrix whose singular values are the ones prescribed. More precisely, under some technical hypotheses, as the dimension $N$ tends to infinity, if the empirical distribution of the singular values of $\bf A$ converges to a compactly supported limit measure $\Theta $ on the real line, then the empirical eigenvalues distribution of $\bf A$ converges to a limit measure $\mu$ on the complex plane which depends only on  $\Theta. $ The limit measure $\mu$ is rotationally invariant in $\mathbb C$ and its support is the annulus $S: = \left\{ {z \in \mathbb{C};a \leqslant \left| z \right| \leqslant b} \right\}$ with $a, b \ge 0$ such that
\begin{equation}
\label{ASL:eq4}
	{a^{ - 2}} = \int {{x^{ - 2}}} d\Theta \left( x \right){\text{   and   }}{{\text{b}}^2} = \int {{x^2}} d\Theta \left( x \right).
\end{equation}

\subsection{Transforms}
\label{TF}

The transforms of large random matrices are specially useful to study the limit spectral properties and to tackle the problems of polynomial calculation of random matrices. In this subsection, we will review the useful transforms including Stieltjes transform, R transform and S transform suggested by problems of interest in power grid \cite{cao2015random, qiu2015smart}.

We begin with the Stieltjes transform of $\bf{X}$ that is defined as follows.

\begin{definition}
\label{Transforms:def1}
Let $\bf{X}$ be a random matrix with distribution $F\left(  \cdot  \right)$. Its Stieltjes transform is defined as
\begin{equation}
\label{Transforms:eq1}
G\left( z \right) = \left\{ {\frac{1}{N}{\rm{Tr}}\left[ {{{\left( {z{\bf{I}} - {\bf{X}}} \right)}^{ - 1}}} \right]} \right\} = \int_{\mathbb{R}} {\frac{1}{{x - z}}dF\left( x \right)},
\end{equation}
where $z \in {\mathbb{C}}^- $and  ${\mathbb{C}}^- = \left\{ {z \in C:{\mathop{\rm Im}\nolimits} \left( z \right) < 0} \right\}$ and $\bf{I}$ represents the identity matrix of dimension $N$.
\end{definition}

An important application of Stieltjes transform is that its $close$ relationship with the limit spectrum density of $\bf{X}$.
\begin{theorem}
\label{Transforms:thm1}
Let $\bf{X}$ be a $N\times N$ random hermitian matrix and its Stieltjes transform is
$G\left( z \right)$, the corresponding eigenvalue density $\rho \left( x \right)$ can be
expressed as:
\begin{equation}
\label{Transforms:eq3}
\rho \left( x \right) =  - \frac{1}{\pi }\mathop {\lim }\limits_{{\mathop{\rm Im}
\nolimits} z \to 0} {\mathop{\rm Im}\nolimits} \left\{ {G\left( z \right)} \right\}.
\end{equation}
\end{theorem}

It is noted that the signs of the ${{\mathop{\rm Im}\nolimits} \left( z \right)}$ and $G\left( z \right)$ coincide. This property should be emphasized in the following examples where the sign of the square root should be chosen.

For GUE and LUE matrices, the corresponding Stieltjes transforms are shown in the following examples.

\begin{example}
\label{Transforms:ex1}
Let $\bf{X}$ be an GUE matrix and its limit spectral density is defined in \eqref{ASL:eq1}, the Stieltjes transform of $\bf{X}$  is
\[
G\left( z \right) = \frac{1}{{2\pi }}\int_{ - 2}^2 {\frac{{\sqrt {4 - {x^2}} }}{{x - z}}dx
 = \frac{1}{2}} \left[ {z - \sqrt {{z^2} - 4} } \right].
\]
\end{example}

\begin{example}
\label{Transforms:ex2}

Let $\bf{W}$ be an LUE matrix and its limit spectral density is defined in \eqref{ASL:eq2} and \eqref{ASL:eq3}.
The corresponding Stieltjes transform can be represented as

\[G\left( z \right) = \frac{1}{{2\pi }}\int_a^b {\frac{{\rho \left( x \right)}}{{\left( {x - z} \right)}}dx = \frac{{ - \beta  - z + 1 \pm \sqrt {{z^2} - 2\left( {\beta  + 1} \right)z + {{\left( {\beta  - 1} \right)}^2}} }}{{2\beta z}}} \]

and

\[G\left( z \right) = \frac{1}{{2\pi }}\int_a^b {\frac{{{\rho ^{'}}\left( x \right)}}{{\left( {x - z} \right)}}dx = \frac{{\beta  - z - 1 \pm \sqrt {{z^2} - 2\left( {\beta  + 1} \right)z + {{\left( {\beta  - 1} \right)}^2}} }}{{2z}}} ,\]
respectively.
\end{example}

Another two important transforms which we elaborate in the following are R transform and S transform. The
key point of these two transforms is that R/S transform enable the characterization of the limiting
spectrum of a sum/product of random matrices from their individual limiting spectra. These properties would turn out to be extremely useful in the following subsection. We start with blue function, that is, the functional inverse of the Stieltjes transform $G\left( z \right)$ which is defined as
\[B\left( {G\left( z \right)} \right) = z\]
and then the R transform is simply defined by

\[R\left( \omega  \right) = B\left( \omega  \right) - \frac{1}{w}\].

Two important prosperities of R transform are shown in the following.

$\bf{Additivity \ law}:$
let ${R_{\bf{A}}}\left( z \right)$, ${R_{\bf{B}}}\left( z \right)$ and ${R_{\bf{A}+\bf{B}}}\left( z \right)$ be the R transforms of matrices $\bf{A}$, $\bf{B}$ and $\bf{A}+\bf{B}$, respectively. We have
\begin{equation}
\label{Transforms:eq6}
{R_{{\bf{A} + \bf{B}}}}\left( z \right) = {R_{\bf{A}}}\left( z \right) + {R_{\bf{A}}}\left( z \right).
\end{equation}

$\bf{Scaling \ property}:$
For any $\alpha>0$,
\begin{equation}
\label{Transforms:eq7}
{R_{\alpha {\bf{X}}}}\left( z \right) = \alpha {R_{\bf{X}}}\left( {\alpha z} \right).
\end{equation}

Additivity law can be easily understood in terms of Feynman diagrams, we refer interested readers to references \cite{qiu2015smart} for details. The above properties of R transform enable us to do the linear calculation of the asymptotic spectrum of random matrices.

Another important transform of engineering significance in RMT is the S transform. S transform is related to the R transform which is defined by

\begin{equation}
\label{Transforms:eq8}
S\left( z \right) = \frac{1}{{R\left( {zS\left( z \right)} \right)}}.
\end{equation}

An interesting property of S transform is that the S transform of the product of two independent random matrices equals the product of the S transforms:

\begin{equation}
\label{Transforms:eq10}
{S_{{\bf{AB}}}}\left( z \right) = {S_{\bf{A}}}\left( z \right){S_{\bf{B}}}\left( z \right).
\end{equation}

Note that \eqref{Transforms:eq10} is known as multiplication law of S transform. For the sake of brevity, see Section \ref{FP} for more details.

\subsection{Convergence Rate}
\label{CR}

In this section, we investigate the spectral asymptotics for GUE and LUE matrices. We are motivated by the practical problems introduced in \cite{qiu2015smart}. Let $F(x)$ be the empirical spectral distribution function of GUE or LUE matrices and $G(x)$ be the distribution function of the limit law (semicircle law for GUE matrices and Marchenko-Pastur law for LUE matrices). Here, we study the convergence rate of expected empirical distribution function $\mathbb{E}F(x)$ to $G(x)$. Specially, the bound
\begin{equation}
\label{eq1:Convergence for Spectra}
\Delta  = \left| \mathbb{E}{F\left( x \right) - G\left( x \right)} \right|,
\end{equation}
is mainly concerned in the following.

The rate of convergence for the expected spectral distribution of GUE matrices has attracted numerous attention due to its increasingly appreciated importance in applied mathematics and statistical physics.
Wigner initially looked into the convergence of the spectral distribution of GUE matrices \cite{mehta2004random}. Bai \cite{bai1993convergenceI} conjectured that the optimal bound for $\Delta$ in GUE case should be of order $n^{-1}$.  Bai and coauthors in \cite{bai1997note} proved that $\Delta  = O\left( {{N^{{{ - 1} \mathord{\left/
 {\vphantom {{ - 1} 3}} \right.
 \kern-\nulldelimiterspace} 3}}}} \right)$. Gotze and Tikhomirov in \cite{gotze2003rate} improved the result in \cite{bai1997note} and proved that $\Delta  = O\left( {{N^{{{ - 1} \mathord{\left/
 {\vphantom {{ - 1} 2}} \right.
 \kern-\nulldelimiterspace} 2}}}} \right)$. Bai et al. in \cite{bai2003convergence} also showed that $\Delta  = O\left( {{N^{{{ - 1} \mathord{\left/
 {\vphantom {{ - 1} 2}} \right.
 \kern-\nulldelimiterspace} 2}}}} \right)$ on the condition that the 8th moment of $\bf{X}$ satisfies $\sup E{\left| {{X_{ij}}} \right|^8} < \infty$. Girko in \cite{girko2002extended} stated as well that $\Delta  = O\left( {{N^{{{ - 1} \mathord{\left/
 {\vphantom {{ - 1} 2}} \right.
 \kern-\nulldelimiterspace} 2}}}} \right)$ assuming uniform bounded 4th moment of $\bf{X}$. Recently, Gotze and Tikhomirov proved an optimal bound as follows.
\begin{theorem}
\label{thm1:Convergence for Spectra}
There exists a positive constant $C$ such that, for any $N \geq 1 $,
\begin{equation}
\label{eq2:Convergence for Spectra}
\Delta  \le C{N^{ - 1}}.
\end{equation}
\end{theorem}
The convergence of the density (denoted by $g(x)$) of standard semicircle law to the expected spectral density $p(x)$ is proved by Gotze and Tikhomirov in the following Theorem.
\begin{theorem}
\label{thm2:Convergence for Spectra}
There exists a positive constant $\varepsilon$ and $C$ such that, for any $x \in \left[ { - 2 + {N^{ - \frac{1}{3}}}\varepsilon ,2 - {N^{ - \frac{1}{3}}}\varepsilon } \right]$,
\begin{equation}
\label{eq3:Convergence for Spectra}
\left| {p\left( x \right) - g\left( x \right)} \right| \le \frac{C}{{N\left( {4 - {x^2}} \right)}}.
\end{equation}
\end{theorem}

For LUE matrix $\bf{W}$ with spectral distribution function $F(x)$, let $\beta  = \frac{N}{M}$ as $N,M \to \infty $, it is well known that $\mathbb{E}F(x)$ convergences to the Marchenko-Pastur law $H(x)$ with density
\begin{equation}
\label{eq4:Convergence for Spectra}
h\left( x \right) = \frac{1}{{2\pi \beta x}}\sqrt {\left( {x - a} \right)\left( {b - x} \right)},
\end{equation}
where $a = {\left( {1 - \sqrt \beta  } \right)^2},b = {\left( {1 + \sqrt \beta  } \right)^2}$. The bound
\begin{equation}
\label{eq5:Convergence for Spectra}
\Delta  = \left| \mathbb{E}{F\left( x \right) - H\left( x \right)} \right|,
\end{equation}
for the convergence rate is shown in the following theorems.

\begin{theorem}
\label{thm3:Convergence for Spectra}
For $\beta  = \frac{N}{M}$, there exist some positive constant $\beta _1$ and $\beta _2$ such that $0 < {\beta _1} \le \beta  \le {\beta _2} < 1$, for all $N \geq 1 $. Then there exists a positive constant $C$ depending on $\beta _1$ and $\beta _2$ and  for any $N \geq 1 $
\begin{equation}
\label{eq6:Convergence for Spectra}
\Delta  \le C{N^{ - 1}}.
\end{equation}
\end{theorem}

Considering the case $\beta < 1$, a similar result is shown in Theorem \ref{thm5:Convergence for Spectra}.

\begin{theorem}
\label{thm5:Convergence for Spectra}
For $\beta  = \frac{N}{M}$, there exists some positive constant $\beta _1$ and $\beta _2$ such that $0 < {\beta _1} \le \beta  \le {\beta _2} < 1$, for all $N \geq 1 $. Then there exists a positive constant $C$ and $\varepsilon$ depending on $\beta _1$ and $\beta _2$ and  for any $N \geq 1 $ and $x \in \left[ {a + {N^{ - \frac{2}{3}}}\varepsilon ,b - {N^{ - \frac{2}{3}}}\varepsilon } \right]$
\begin{equation}
\label{eq8:Convergence for Spectra}
\left| {p\left( x \right) - h\left( x \right)} \right| \le \frac{C}{{N\left( {x - a} \right)\left( {b - x} \right)}}.
\end{equation}
\end{theorem}

Interested readers are referred to \cite{gotze2005rate} for technical details and Section \ref{SALES} for applications in smart grid.

\subsection{Free Probability }
\label{FP}

Free probability theory, initiated  in 1983 by Voiculescu in \cite{Dan1985Symmetries}, together with the results published in \cite{Dan1991Limit} regarding asymptotic freeness of random matrices, has established a new branch of theories and tools in random matrix theory. Here, we provide some of the basic principles and then examples to enhance the understanding and application of the free probability theory.

Let ${{x}_{1}},\ldots ,{{x}_{n}}$ be selfadjoint elements which are freely independent.
Consider a selfadjoint polynomial $p$ in n non-commuting variables and let $P$ be the
element $P=p({{x}_{1}},\ldots ,{{x}_{n}})$. Now  we introduce the method \cite{Belinschi2013Analytic} \cite{Speicher2015Polynomials} to obtain the distribution of $P$ out of the
distributions of ${{x}_{1}},\ldots ,{{x}_{n}}$.

Let $\mathcal{A}$ be a unital algebra and $\mathcal{B}\subset \mathcal{A}$ be a subalgebra containing the unit. A linear map\[E:\mathcal{A}\to \mathcal{B}\] is a conditional expectation if
\[E[b]=b  \ \quad for\;all \;b\in \mathcal{B}\]
and
\[E[{{b}_{1}}a{{b}_{2}}]={{b}_{1}}E[a]{{b}_{2}}  \ \quad for\;all \;a\in \mathcal{A} \ \;for\;all\; {{b}_{1}},{{b}_{2}}\in \mathcal{B}\]

An operator-valued probability space consists of $\mathcal{B}\subset \mathcal{A}$ and a conditional expectation $E:\mathcal{A}\to \mathcal{B}$. Then, random variables ${{x}_{i}}\in \mathcal{A}(i\in I)$ are free with respect to $E$ (or free with amalgamation over $\mathcal{B}$ ) if $E[{{a}_{1}}\ldots {{a}_{n}}]=0$  whenever ${{a}_{i}}\in \mathcal{B}<{{x}_{j(i)}}>$ are polynomials in some ${{x}_{j(i)}}$ with coefficients from $\mathcal{B}$ and $E[{{a}_{i}}]=0$ for all $i$ and $j(1)\ne j(2)\ne \cdots \ne j(n)$. For a random variable $x\in \mathcal{A}$, we denote the operator-valued Cauchy transform:\[G(b):=E[{{(b-x)}^{-1}}]   (b\in \mathcal{B})\] whenever $(b-x)$ is invertible in $\mathcal{B}$. In order to have some nice analytic behaviour, we assume that both $\mathcal{A}$ and $\mathcal{B}$ are ${{C}^{*}}$-algebras in the following; $\mathcal{B}$ will usually be of the form $\mathcal{B}={{M}_{N}}(\mathbb{C})$, the $N\times N$-matrices. In such a setting and for $x={{x}^{*}}$, this $G$ is well-defined and a nice
analytic map on the operator-valued upper halfplane:
\[{{\mathbb{H}}^{+}}(B):=\{b\in B|(b-b*)/(2i)>0\}\]
and it allows to give a nice description for the sum of two free selfadjoint elements.
In the following we will use the notation\[h(b):=\frac{1}{G(b)}-b\]
 \newtheorem{fp1}{\textbf{Theorem}}[section]
 \begin{theorem}(\cite{Belinschi2013Analytic})
  Let $x$  and $y$ be selfadjoint operator-valued random variables free over $\mathcal{B}$. Then there exists a Frechet analytic map
$\omega :{{\mathbb{H}}^{+}}(\mathcal{B} )\to {{\mathbb{H}}^{+}}(\mathcal{B})$ so that

$\bullet\Im {{\omega }_{j}}(b)\ge \Im b$ for all  $b\in {{\mathbb{H} }^{+}}(\mathcal{B} )$, $j\in \left\{ 1,2 \right\}$

$\bullet{{G}_{x}}({{\omega }_{1}}(b))={{G}_{y}}({{\omega }_{2}}(b))={{G}_{x+y}}(b)$

Moreover, if $b\in {{\mathbb{H}}^{+}}(\mathcal{B} )$ , then ${{\omega }_{1}}(b)$  is the unique fixed point of the map.
${{f}_{b}}:{{\mathbb{H}}^{+}}(\mathcal{B} )\to {{\mathbb{H}}^{+}}(\mathcal{B} ),  {{f}_{b}}(\omega )={{h}_{y}}({{h}_{x}}(\omega )+b)+b, $
 and ${{\omega }_{1}}(b)\text{=}\underset{n\to \infty }{\mathop{\lim }}\,{{f}_{b}}^{on}(\omega )$ for any $\omega \in {{\mathbb{H}}^{+}}(\mathcal{B} )$, where $f_{b}^{on}$ means the n-fold composition of ${{f}_{b}}$ with itself. Same statements hold for ${{\omega }_{\text{2}}}(b)$, replaced by $\omega \to {{h}_{x}}({{h}_{y}}(\omega )+b)+b.$
  \label{fp1}
  \end{theorem}

Let $\mathcal{A}$ be a complex and unital ${{C}^{*}}$-algebra and let selfadjoint elements ${{x}_{1}},\ldots {{x}_{n}}\in \mathcal{A}$. $\mathcal{A}$ is given. Then, for any non-commutative polynomial $p\in \mathbb{C}<{{{\bf X}}_{1}},\ldots ,{{\bf X}_{n}}>$,we get an operator $P=p({{x}_{1}}\ldots {{x}_{n}})\in \mathcal{A}$ by evaluating $p$ at $({{x}_{1}},\ldots ,{{x}_{n}})$ .In this situation, knowing a linearization trick \cite{Anderson2011Convergence} means to have a procedure that leads finally to an operator \[{{L}_{p}}={{b}_{0}}\otimes 1+{{b}_{1}}\otimes {{x}_{1}}+\cdots {{b}_{n}}\otimes {{x}_{n}}\in {{M}_{N}}(\mathbb{C})\otimes \mathcal{A}\]
for some matrices ${{b}_{0}},\ldots ,{{b}_{n}}\in {{M}_{N}}(\mathbb{C})$ of dimension$N$ , such that $z-P$  is invertible in $\mathcal{A}$ if and only if $\Lambda (z)-{{L}_{p}}$ is invertible in ${{M}_{N}}(\mathbb{C})\otimes \mathcal{A}$ . Hereby, we put \[\Lambda (z)=\left[ \begin{matrix}
   z & 0 & \cdots  & 0  \\
   0 & 0 & \cdots  & 0  \\
   \vdots  & \vdots  & \ddots  & \vdots   \\
   0 & 0 & \cdots  & 0  \\
\end{matrix} \right]     \quad    for \;all\; {z}\in {\mathbb{C}}.\]
Let $p\in \mathbb{C}<{{\bf X}_{1}},\ldots ,{{\bf X}_{n}}>$ be given. A matrix
\[{{L}_{p}}:=\left[ \begin{matrix}
   0 & u  \\
   v & {\bf Q}  \\
\end{matrix} \right]\in {{M}_{N}}(\mathbb{C})\otimes \mathbb{C}<{{\bf X}_{1}},\ldots ,{{\bf X}_{n}}>\] ,
\\where

 \begin{itemize}
\item  $ N\in \mathbb{N}$ is an integer,

 \item $ {\bf Q}\in {{M}_{N-1}}(\mathbb{C})\otimes \mathbb{C}<{{\bf X}_{1}},\ldots ,{{\bf X}_{n}}>$ is invertible,

 \item and $u$ is a row vector and $v$ is a column vector, both of size $N-1$ with entries in $\mathbb{C}<{{\bf X}_{1}},\ldots ,{{\bf X}_{n}}>$,
 \end{itemize}

 is called a linearization of $p$, if the following conditions are satisfied:
\begin{enumerate}
\item There are matrices ${{b}_{0}},\ldots ,{{b}_{n}}\in {{M}_{N}}(\mathbb{C})$, such that

${{L}_{p}}={{b}_{0}}\otimes 1+{{b}_{1}}\otimes {{\bf X}_{1}}+\cdots {{b}_{n}}\otimes {{\bf X}_{n}}$
i.e. the polynomial entries in $\bf Q$, $u$ and $v$ all have degree $\le 1$ .
\item It holds true that $p=-u{{\bf Q}^{-1}}v$ .
\end{enumerate}

To introduce the following corollary,which will enable us to shift ${\Lambda _\varepsilon }(z)$ for $z \in \mathbb{C}^{+}$ to a point
\[{\Lambda _\varepsilon }(z): = \left[ {\begin{array}{*{20}{c}}
z&{}&{}&{}\\
{}&{i\varepsilon }&{}&{}\\
{}&{}& \ddots &{}\\
{}&{}&{}&{i\varepsilon }
\end{array}} \right]\]
lying inside the domain $\mathbb{H}^{+}(M_N (\mathbb{C}))$ in order to get access to all analytic tools that are available there.

\begin{corollary}
Let $(\mathcal{A},\phi )$ be a ${{C}^{*}}$-probability space and let elements ${{x}_{1}},\ldots ,{{x}_{n}}\in \mathcal{A}$ be given. For any seladjoint $p\in \mathbb{C}<{{X}_{1}},\ldots ,{{X}_{n}}>$ that has a selfadjoint linearization\[{{L}_{p}}={{b}_{0}}\otimes 1+{{b}_{1}}\otimes {{\bf X}_{1}}+\cdots {{b}_{n}}\otimes {{\bf X}_{n}}\in{{M}_{N}}(\mathbb{C})\otimes \mathbb{C}<{{\bf X}_{1}},\ldots ,{{\bf X}_{N}}>\]
with matrices ${{b}_{0}},\ldots,{{b}_{n}}\in{{M}_{N}}(\mathbb{C}){{}_{\text{sa}}}$, we put $P=p({{x}_{1}},\ldots ,{{x}_{n}})$ and
\[{{L}_{P}}={{b}_{0}}\otimes 1+{{b}_{1}}\otimes {{x}_{1}}+\cdots {{b}_{n}}\otimes {{x}_{n}}\in{{M}_{N}}(\mathbb{C})\otimes \mathcal{A}\]
Then, for each $z\in {{\mathbb{C}}^{+}}$ and all sufficiently small $\varepsilon >0$ , the operators $z-P\in \mathcal{A}$ and ${{\Lambda }_{\varepsilon }}(z)-{{L}_{p}}\in {{M}_{N}}(\mathbb{C})\otimes \mathcal{A}$ are both invertible and we have \[\underset{\varepsilon \to 0}{\mathop{\lim }}\,{{\left[ \mathbb{E}({{({{\Lambda }_{\varepsilon }}(z)-{{L}_{p}})}^{-1}}) \right]}_{1,1}}={{G}_{P}}(z)\]
Hereby, \[\mathbb{E}{{M}_{N}}(\mathbb{C})\otimes \mathcal{A}\to {{M}_{N}}(\mathbb{C})\] denotes the conditional expectation given by
$\mathbb{E}\text{=i}{{\text{d}}_{{{M}_{N}}(\mathbb{C})}}\otimes \phi $.
\label{fp2}
\end{corollary}

Let $(\mathcal{A},\phi )$ be a non-commutative ${{C}^{*}}$-probability space, ${{x}_{1}},\ldots ,{{x}_{n}}\in \mathcal{A}$ selfadjiont elements which are freely independent, and $p\in \mathbb{C}<{{\bf X}_{1}},\ldots ,{{\bf X}_{n}}>$a selfadjoint polynomial in $n$ non-commuting variables ${{\bf X}_{1}},\ldots ,{{\bf X}_{N}}$. We put $P=p({{x}_{1}},\ldots ,{{x}_{n}})$. The following procedure leads to the distribution of $P$.

\begin{enumerate}[step 1]
\item $p$ has a selfadjoint linearization
\[{{L}_{p}}={{b}_{0}}\otimes 1+{{b}_{1}}\otimes {{\bf X}_{1}}+\cdots {{b}_{n}}\otimes {{\bf X}_{n}}\]
with matrices ${{b}_{0}},\ldots ,{{b}_{n}}\in {{M}_{N}}\mathbb{C}{{}_{\text{sa}}}$. We put
\[{{L}_{P}}={{b}_{0}}\otimes 1+{{b}_{1}}\otimes {{x}_{1}}+\cdots {{b}_{n}}\otimes {{x}_{n}}\in {{M}_{N}}(\mathbb{C})\otimes \mathcal{A}.\]

\item The operators ${{b}_{0}}\otimes 1,{{b}_{1}}\otimes {{x}_{1}},...,{{b}_{n}}\otimes {{x}_{n}}$ are freely independent elements in
the operator-valued ${{C}^{*}}$-probability space $({{M}_{N}}(\mathbb{C})\otimes \mathcal{A},\mathbb{E})$, where $\mathbb{E}:={{M}_{N}}(\mathbb{C})\otimes \mathcal{A}\to {{M}_{N}}(\mathbb{C})$ denotes the conditional expectation given by $\mathbb{E}\text{=i}{{\text{d}}_{{{M}_{N}}(\mathbb{C})}}\otimes \phi $. Furthermore, for $j=1,\ldots ,n$ , the ${{M}_{N}}(\mathbb{C})$-valued Cauchy transform ${{G}_{{{b}_{j}}\otimes {{x}_{j}}}}(b)$ is completely determined by the scalar-valued Cauchy transforms ${{G}_{{{x}_{j}}}}$via
\[{{G}_{{{b}_{j}}\otimes {{x}_{j}}}}(b)=\underset{\varepsilon \to 0}{\mathop{\lim }}\,-\frac{1}{\pi }\int_{\mathbb{R}}{(b-t{{b}_{j}}}{{)}^{-1}}\Im ({{G}_{{{x}_{j}}}}(t+i\varepsilon ))dt\] for all $b\in {{\mathbb{H}}^{+}}({{M}_{N}}(\mathbb{C}))$.
\item Due to Step 3, we can calculate the Cauchy transform of
\[{{L}_{p}}-{{b}_{0}}\otimes 1={{b}_{1}}\otimes {{\bf X}_{1}}+\cdots {{b}_{n}}\otimes {{\bf X}_{n}}\]
by using the fixed point iteration for the operator-valued free additive convolution.
The Cauchy transform of ${{L}_{P}}$ is then given by \[{{G}_{{{L}_{P}}}}(b)={{G}_{{{L}_{P}}-{{b}_{0}}\otimes 1}}(b-{{b}_{0}})\quad for\; all\;  b\in {{\mathbb{H}}^{+}}({{M}_{N}}(\mathbb{C}))\].
\item Corollary tells us that the scalar-valued Cauchy transform ${{G}_{P}}(z)$ of $P$ 	is determined by
\[{{G}_{P}}(z)\text{=}\underset{\varepsilon \to \text{0}}{\mathop{\lim }}\,{{\left[ G{}_{{{L}_{P}}}({{\Lambda }_{\varepsilon }}(z)) \right]}_{1,1}}\quad for\; all \;z\in {{\mathbb{C}}^{+}}\]
Finally, we obtain the desired distribution of $P$ by applying the Stieltjes inversion formula.
\end{enumerate}

\newtheorem{example5}{\textbf{Example}}[section]
\begin{example}

We consider the non-commutative polynomial $p\in \mathbb{C}<{{\bf X}_{1}},{{\bf X}_{2}}>$ given by
$p({{\bf X}_{1}},{{\bf X}_{2}})={{\bf X}_{1}}{{\bf X}_{2}}+{{\bf X}_{2}}{{\bf X}_{1}}$.
It is easy to check that
\[{{L}_{p}}=\left[ \begin{matrix}
   0 & {{\bf X}_{1}} & {{\bf X}_{2}}  \\
   {{\bf X}_{1}} & 0 & -1  \\
   {{\bf X}_{2}} & -1 & 0  \\
\end{matrix} \right]\]
is a selfadjoint linearization of $p$.
Now, let ${{X}_{1}},{{X}_{2}}$ be free semicircular or Poisson elements in a non-commutative ${{C}^{*}}$-probability space $(\mathcal{A},\phi )$. Based on the algorithm of Theorem above, we can calculate the distribution of the anticommutator $p({{\bf X}_{1}},{{\bf X}_{2}})={{\bf X}_{1}}{{\bf X}_{2}}+{{\bf X}_{2}}{{\bf X}_{1}}$.

\end{example}

\newtheorem{example6}{\textbf{Example}}[section]
\begin{example}
In the same way, we can deal with the following variation of the anticommutator:
$p({{\bf X}_{1}},{{\bf X}_{2}})={{\bf X}_{1}}{{\bf X}_{2}}+{{\bf X}_{2}}{{\bf X}_{1}}+{{\bf X}_{1}}^{2}$.
It is easy to check that
\[{{L}_{p}}=\left[ \begin{matrix}
   0 & {{\bf X}_{1}} & \frac{\text{1}}{\text{2}}{{\bf X}_{1}}\text{+}{{\bf X}_{2}}  \\
   {{X}_{1}} & 0 & -1  \\
   \frac{\text{1}}{\text{2}}{{\bf X}_{1}}\text{+}{{\bf X}_{2}} & -1 & 0  \\
\end{matrix} \right]\]
is a selfadjoint linearization of $p$.

Then, let ${{\bf X}_{1}},{{\bf X}_{2}}$ be free semicircular or Poisson elements in a non-commutative ${{C}^{*}}$-probability space $(\mathcal{A},\phi )$.Based on the algorithm  above, we can calculate the distribution of the polynomial $p({{\bf X}_{1}},{{\bf X}_{2}})={{\bf X}_{1}}{{\bf X}_{2}}+{{\bf X}_{2}}{{\bf X}_{1}}+{{\bf X}_{1}}^{2}$.
\end{example}

For readers's convenience, we also provide some simulations of the free polynomials of random matrices in the following.

\begin{figure}[!htbp]
  \centering
  \includegraphics[width=7.5cm]{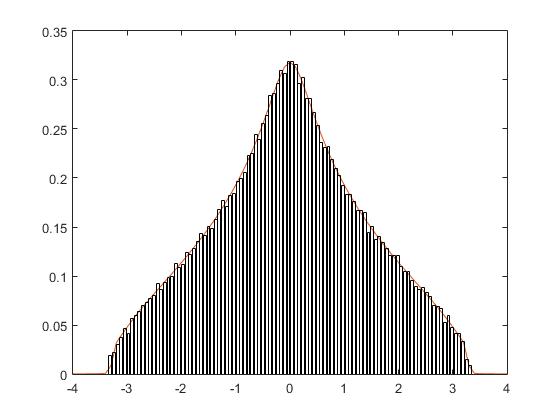}\\
  \caption{Comparison of the distribution of
  $p({{\bf X}_{1}},{{\bf X}_{2}})={{\bf X}_{1}}{{\bf X}_{2}}+{{\bf X}_{2}}{{\bf X}_{1}}$ according to our algorithm, with the histogram of eigenvalues for $p({{\bf X}_{1}}^{(n)},{{\bf X}_{2}}^{(n)})$ according to our algorithm, with the histogram of eigenvalues for $p({{\bf X}_{1}}^{(n)},{{\bf X}_{2}}^{(n)})$, for $n=1000$. ${{\bf X}_{1}},{{\bf X}_{2}}$ are  free semicircular elements and
${{\bf X}_{1}}^{(n)},{{\bf X}_{2}}^{(n)}$ are independent standard Gaussian random matrices.
}
\end{figure}

\begin{figure}[!htbp]
  \centering
  \includegraphics[width=7.5cm]{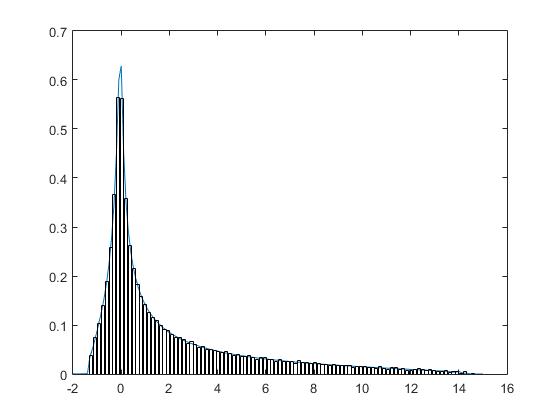}\\
  \caption{Comparison of the distribution of $p({{\bf X}_{1}},{{\bf X}_{2}})={{\bf X}_{1}}{{\bf X}_{2}}+{{\bf X}_{2}}{{\bf X}_{1}}$ according to our algorithm, with the histogram of eigenvalues for $p({{\bf X}_{1}}^{(n)},{{\bf X}_{2}}^{(n)})$, for $n=1000$ . ${{\bf X}_{1}},{{\bf X}_{2}}$ are free poisson elements and
${{\bf X}_{1}}^{(n)},{{\bf X}_{2}}^{(n)}$ are Wishart random matrices.
}
\end{figure}

\begin{figure}[!htbp]
  \centering
  \includegraphics[width=7.5cm]{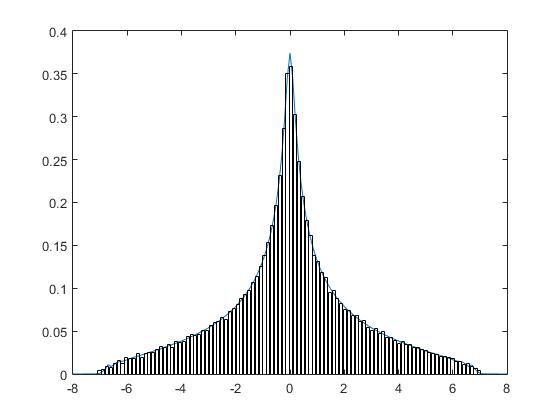}\\
  \caption{Comparison of the distribution of $p({{\bf X}_{1}},{{\bf X}_{2}})={{\bf X}_{1}}{{\bf X}_{2}}+{{\bf X}_{2}}{{\bf X}_{1}}$ according to our algorithm, with the histogram of eigenvalues for $p({{\bf X}_{1}}^{(n)},{{\bf X}_{2}}^{(n)})$, for $n=1000$ . ${{\bf X}_{1}}$ is of free semicircular elements and ${{\bf X}_{2}}$ free Poisson ones. ${{\bf X}_{1}}^{(n)}$ is an independent standard Gaussian random matrix and ${{\bf X}_{2}}^{(n)}$ is a Wishart matrix.
}
\end{figure}
\begin{figure}[!htbp]
  \centering
  \includegraphics[width=7.5cm]{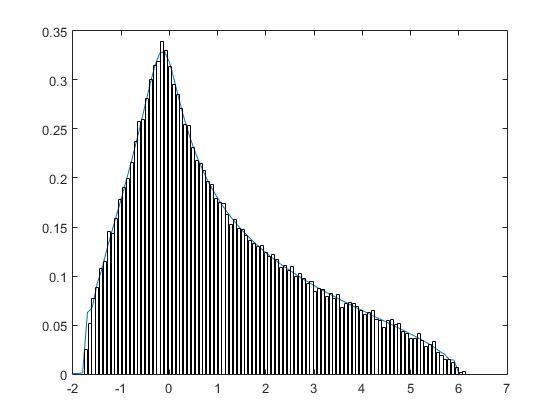}\\
  \caption{Comparison of the distribution of $p({{\bf X}_{1}},{{\bf X}_{2}})={{\bf X}_{1}}{{\bf X}_{2}}+{{\bf X}_{2}}{{\bf X}_{1}}+{{\bf X}_{1}}^{2}$ according to our algorithm, with the histogram of eigenvalues for $p({{\bf X}_{1}}^{(n)},{{\bf X}_{2}}^{(n)})$, for $n=1000$ . ${{\bf X}_{1}},{{\bf X}_{2}}$ are of free poisson elements and
${{\bf X}_{1}}^{(n)},{{\bf X}_{2}}^{(n)}$ are Wishart random matrices.
}
\end{figure}
\begin{figure}[!htbp]
  \centering
  \includegraphics[width=7.5cm]{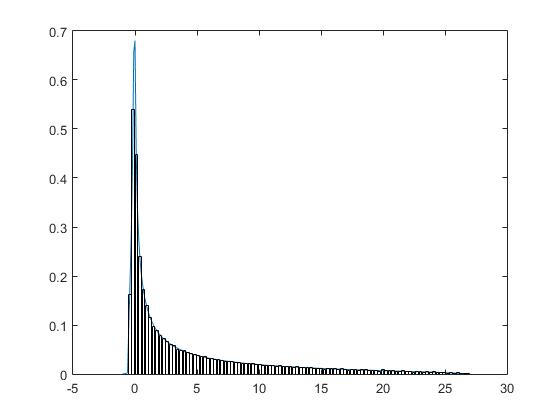}\\
  \caption{Comparison of the distribution of $p({{\bf X}_{1}},{{\bf X}_{2}})={{\bf X}_{1}}{{\bf X}_{2}}+{{\bf X}_{2}}{{\bf X}_{1}}+{{\bf X}_{1}}^{2}$ according to our algorithm, with the histogram of eigenvalues for $p({{\bf X}_{1}}^{(n)},{{\bf X}_{2}}^{(n)})$, for $n=1000$ . ${{\bf X}_{1}},{{\bf X}_{2}}$ are of free poisson elements and
${{\bf X}_{1}}^{(n)},{{\bf X}_{2}}^{(n)}$ are Wishart random matrices.
}
\end{figure}

\begin{figure}[!htbp]
  \centering
  \includegraphics[width=7.5cm]{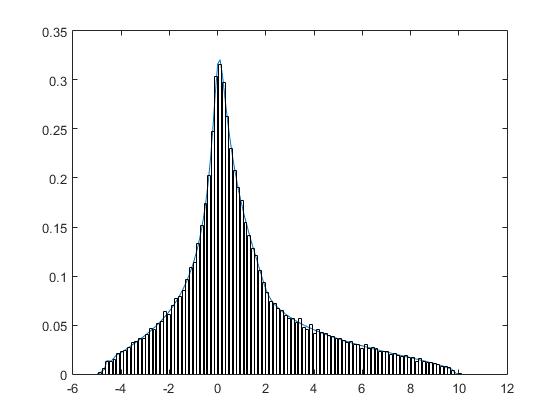}\\
  \caption{Comparison of the distribution of
  $p({{\bf X}_{1}},{{\bf X}_{2}})={{\bf X}_{1}}{{\bf X}_{2}}+{{\bf X}_{2}}{{\bf X}_{1}}+{{\bf X}_{1}}^{2}$according to our algorithm, with the histogram of eigenvalues for $p({{\bf X}_{1}}^{(n)},{{\bf X}_{2}}^{(n)})$, for $n=1000$ . ${{\bf X}_{1}}$ are of free semicircular elements and ${{\bf X}_{2}}$  free Poisson ones. ${{\bf X}_{1}}^{(n)}$ is an independent standard Gaussian random matrix and ${{\bf X}_{2}}^{(n)}$ is a Wishart matrix.
}
\end{figure}

\section{Applications to Smart Grids}
\label{ASG}

In this section, we elaborate some of the more representative problems described in Section \ref{Intro} that capture various features of interest in smart grid and we show how random matrix results have been used to tackle the problems that arise in the large power grid with wide deployment of PMU equipments. Besides, we also conclude some state-of-art data driven methods for comparison.

\subsection{Hypothesis Tests in Smart Grids}
\label{HTSG}

Considering the data model introduced in Section \ref{LDRM}, the problem of testing hypotheses on means of populations and covariance matrices is addressed. We stated by a review of traditional multivariate procedures for these tests. Then we develop adjustments of these procedures to handle with high-dimensional data in smart grid.

As depicted in the Section \ref{LDRM}, a large random matrix flow $\left\{ {{{\bf{Z}}_1},{{\bf{Z}}_2}, \cdots ,{{\bf{Z}}_q}} \right\}$ is adopted to represent the massive streaming PMU data in one sample period. Instead of analyzing the raw individual window-truncated PMU data ${{\bf{Z}}_g}$ \cite{xie2014dimensionality, lim2016svd} or the statistic of ${{\bf{Z}}_g}$ \cite{Xu2015A, ghanavati2016identifying}, a comprehensive analysis of the statistic of $\left\{ {{{\bf{Z}}_1},{{\bf{Z}}_2}, \cdots ,{{\bf{Z}}_q}} \right\}$ is conducted in the following. More specially, denote ${{\bf{\Sigma }}_i}$ as the covariance matrix of $i$th collected PMU measurements, we want to test the hypothesis:
\begin{equation}
\label{HTSG:eq1}
\begin{array}{l}
{H_0}:{{\bf{\Sigma }}_1} = {{\bf{\Sigma }}_2} =  \cdots  = {{\bf{\Sigma }}_q} \\
{H_1}:\exists \ j,k \ \ {\rm{s.t.}} \ \ {{\bf{\Sigma }}_j} \ne {{\bf{\Sigma }}_k}
\end{array}.
\end{equation}

It is worthy noting that the hypothesis \eqref{HTSG:eq1} is a famous testing hypothesis in multivariate statistical analysis  which aims to study samples share or approximately share some same distribution and consider using a set of samples (data streams denoted in equation \eqref{eqC4} in this paper), one from each population, to test the hypothesis that the covariance matrices of these populations are equal.

\subsection{Data Driven Methods for State Evaluation}
\label{DDMSA}

The LR test \cite{bai1996effect} and CLR test \cite{bai2009corrections} as introduced in the Section \ref{HTSG} are most commonly test statistics for the hypothesis in \eqref{HTSG:eq1}. These tests can be understood by replacing the population covariance matrix ${{\bf{\Sigma }}_g}$ by its sample covariance matrix ${\bf{Y}}_g$. While direct substitution of ${{\bf{\Sigma }}_g}$ by ${\bf{Y}}_g$ brings invariance and good testing properties as shown in \cite{bai1996effect} for normally distributed data. The test statistic $V_2$ may not work for high-dimensional data as demonstrated in \cite{ledoit2002some, chen2010two}. Besides, the estimator $V_3$ has unnecessary terms which slow down the convergence considerably when dimension of PMU data is high \cite{chen2010two, chen2012tests}. In such situations, to overcome the drawbacks, trace criterion \cite{chen2010two} is more suitable to the test problem. Specially, instead of estimating the population covariance matrix directly, a well defined distance measure exploiting the difference among data flow $\left\{ {{{\bf{Z}}_1},{{\bf{Z}}_2}, \cdots ,{{\bf{Z}}_q}} \right\}$ is conducted, that is, the trace-based distance measure between ${{\bf{\Sigma }}_s}$ and ${{\bf{\Sigma }}_t}$ is
\begin{equation}
\label{DDMSA:eq3}
{\mathop{\rm tr}\nolimits} \left\{ {{{\left( {{{\bf{\Sigma }}_s} - {{\bf{\Sigma }}_t}} \right)}^2}} \right\} = {\mathop{\rm tr}\nolimits} \left( {{\bf{\Sigma }}_s^2} \right) + {\mathop{\rm tr}\nolimits} \left( {{\bf{\Sigma }}_t^2} \right) - 2{\mathop{\rm tr}\nolimits} \left( {{{\bf{\Sigma }}_s}{{\bf{\Sigma }}_t}} \right),
\end{equation}
where ${\mathop{\rm tr}\nolimits} \left(  \cdot  \right)$ is the trace operator. Instead of estimating ${\mathop{\rm tr}\nolimits} \left( {{\bf{\Sigma }}_s^2} \right)$, ${\mathop{\rm tr}\nolimits} \left( {{\bf{\Sigma }}_t^2} \right)$ and ${\mathop{\rm tr}\nolimits} \left( {{{\bf{\Sigma }}_s}{{\bf{\Sigma }}_t}} \right)$ by sample covariance matrix based estimators, we adopt the merits of the U-statistics \cite{lee1990u}. Specially, for $l = \left\{ {s,t} \right\} \in \Omega  = \left\{ {1 \le s,t \le q,s \ne t} \right\}$,
\begin{small}
\begin{eqnarray}
\label{DDMSA:eq4}
{{{A}}_l} &=& \frac{1}{{{n_g}\left( {{n_g} - 1} \right)}}\sum\limits_{i \ne j} {{{\left( {{\bf{z}}_{li}^{'}{{\bf{z}}_{lj}}} \right)}^2}} \nonumber \\
&-& \frac{2}{{{n_g}\left( {{n_g} - 1} \right)\left( {{n_g} - 2} \right)}}\sum\limits_{i,j,k}^ *  {{\bf{z}}_{li}^{'}{{\bf{z}}_{lj}}{\bf{z}}_{lj}^{'}{{\bf{z}}_{lk}}}  \\
 &+& \frac{1}{{{n_g}\left( {{n_g} - 1} \right)\left( {{n_g} - 2} \right)\left( {{n_g} - 3} \right)}}\sum\limits_{i,j,k,h}^ *  {{\bf{z}}_{li}^{'}{{\bf{z}}_{lj}}{\bf{z}}_{lk}^{'}{{\bf{z}}_{lh}}} \nonumber
\end{eqnarray}
\end{small}is proposed to estimate ${\mathop{\rm tr}\nolimits} \left( {{\bf{\Sigma}}_l^2} \right)$. It is noted that $\sum\nolimits^{*}$ represents summation over mutually distinct indices. For example, $\sum\nolimits_{i,j,k}^ *$ says summation over the set $\left\{ {\left( {i,j,k} \right):i \ne j,j \ne k,k \ne i} \right\}$. Similarly, the estimator for ${\mathop{\rm tr}\nolimits} \left( {{{\bf{\Sigma }}_s}{{\bf{\Sigma }}_t}} \right)$ can be expressed as
\begin{small}
\begin{eqnarray}
\label{DDMSA:eq5}
{C_{st}} &=& \frac{1}{{{n_g^2}}}\sum\limits_i {\sum\limits_j {{{\left( {{\bf{z}}_{si}^{'}{{\bf{z}}_{tj}}} \right)}^2}} } \nonumber \\
&-& \frac{1}{{\left( {{n_g} - 1} \right){n_g^2}}}\sum\limits_{i,h}^* {\sum\limits_j {{\bf{z}}_{si}^{'}{{\bf{z}}_{tj}}{\bf{z}}_{tj}^{'}{{\bf{z}}_{sh}}} } \nonumber \\
 &-& \frac{1}{{\left( {{n_g} - 1} \right){n_g^2}}}\sum\limits_{i,l}^* {\sum\limits_j {{\bf{z}}_{ti}^{'}{{\bf{z}}_{sj}}{\bf{z}}_{sj}^{'}{{\bf{z}}_{th}}} }   \\
 &+& \frac{1}{{\left( {{n_g} - 1} \right)^2{n_g^2}}}\sum\limits_{i,h}^* {\sum\limits_{j,k}^* {{\bf{z}}_{si}^{'}{{\bf{z}}_{tj}}{\bf{z}}_{sk}^{'}{{\bf{z}}_{th}}} } \nonumber .
\end{eqnarray}
\end{small}
The test statistic which measures the distance between ${{\bf{\Sigma }}_s}$ and ${{\bf{\Sigma }}_t}$ is
\begin{equation}
\label{DDMSA:eq6}
{V_{st}} = {A_s} + {A_t} - {C_{st}}.
\end{equation}
Then the proposed test statistic can be expressed as:
\begin{equation}
\label{DDMSA:eq7}
{V_1} = \frac{1}{{{q}\left( {{q} - 1} \right)}}\sum\limits_{\left\{ {s,t} \right\} \in \Omega } {{T_{st}}}.
\end{equation}

As $p,{n_g} \to \infty$, the asymptotic normality \cite{chen2012tests} of the test statistic \eqref{DDMSA:eq6} is presented in the following:
\begin{theorem}
\label{thm1}
Let $\sigma _{st}^2 = \frac{1}{n_g}\left( {{A_s} + {A_t}} \right)$. Assuming the following conditions:
\begin{enumerate}
\item For any $k$ and $l \in \left\{ {s,t} \right\}$, ${\mathop{\rm tr}\nolimits} \left( {{{\bf{\Sigma }}_k}{{\bf{\Sigma }}_l}} \right) \to \infty$ and \[{\mathop{\rm tr}\nolimits} \left\{ {\left( {{{\bf{\Sigma }}_i}{{\bf{\Sigma }}_j}} \right)\left( {{{\bf{\Sigma }}_k}{{\bf{\Sigma }}_l}} \right)} \right\} = O\left\{ {{\mathop{\rm tr}\nolimits} \left( {{{\bf{\Sigma }}_i}{{\bf{\Sigma }}_j}} \right){\mathop{\rm tr}\nolimits} \left( {{{\bf{\Sigma }}_k}{{\bf{\Sigma }}_l}} \right)} \right\}.\]
\item For $i = 1,2, \cdots, n_g$, ${{\bf{z}}^{\left( i \right)}}$ are independent and identically distributed $p$-dimensional vectors with finite $8th$ moment.
\end{enumerate}
Under above conditions,
\[L = \frac{{{V_{st}}}}{{\sigma _{st}}}\mathop  \to \limits^d \mathcal{N}\left( {0,1} \right)\]
\end{theorem}

\begin{corollary}
\label{prop}
For any $q \ge 2$, as $p,{n_g} \to \infty$, the proposed test statistic $V_1$ satisfies
\begin{equation}
\label{eqB8}
{V_1}\mathop  \to \limits^d  \mathcal{N}\left( {\mu ,{\sigma ^2}} \right),
\end{equation}
where $\mu \approx 0, \sigma ^2 = \mathop \sum \nolimits^* \sigma _{st}^2$.
\end{corollary}
Let $R = \frac{{{V_1}}}{{\sigma _{{V_1}}}} $, the false alarm probability (FAP) for the proposed test statistic can be represented as
\begin{eqnarray}
\label{eqB9}
{P_{{FAP}}} &=& P\left( {R > \alpha |{H_0}} \right) \nonumber \\
 &=& \int_R^\infty  {\frac{1}{{\sqrt {2\pi } }}\exp \left( {\frac{{ - {t^2}}}{2}} \right)} dt \nonumber \\
 &=& Q\left( R \right),
\end{eqnarray}
where $Q\left( x \right) = \int_x^\infty  {{1 \mathord{\left/ {\vphantom {1 {\sqrt {2\pi } }}} \right. \kern-\nulldelimiterspace} {\sqrt {2\pi } }}\exp \left( {{{ - {t^2}} \mathord{\left/ {\vphantom {{ - {t^2}} 2}} \right. \kern-\nulldelimiterspace} 2}} \right)} dt$. For a desired FAP $\tau$, the associated threshold should be chosen such that \[\alpha  = {Q^{ - 1}}\left( \tau  \right) .\]
Otherwise, the detection rate (DR) can be denoted as
\begin{equation}
\label{DR}
{P_{{DR}}} = P \left( {R \ge {Q (\alpha) }|{H_1}} \right).
\end{equation}

It is noted that the computation complexity of proposed test statistic in \eqref{eqB8} is $O(\varepsilon n_g^4)$  which limits its practical application. Here, we proposed a effective approach to reducing complexity of the proposed test statistic from $O(\varepsilon n_g^4)$ to $O(\eta n_g^2)$ by principal component calculation and redundant computation elimination. For simplicity, we briefly explained the technical details in our recent work which is available at $https://arxiv.org/abs/1609.03301$.

\begin{figure}[!htp]
{
\includegraphics[width=0.7\textwidth]{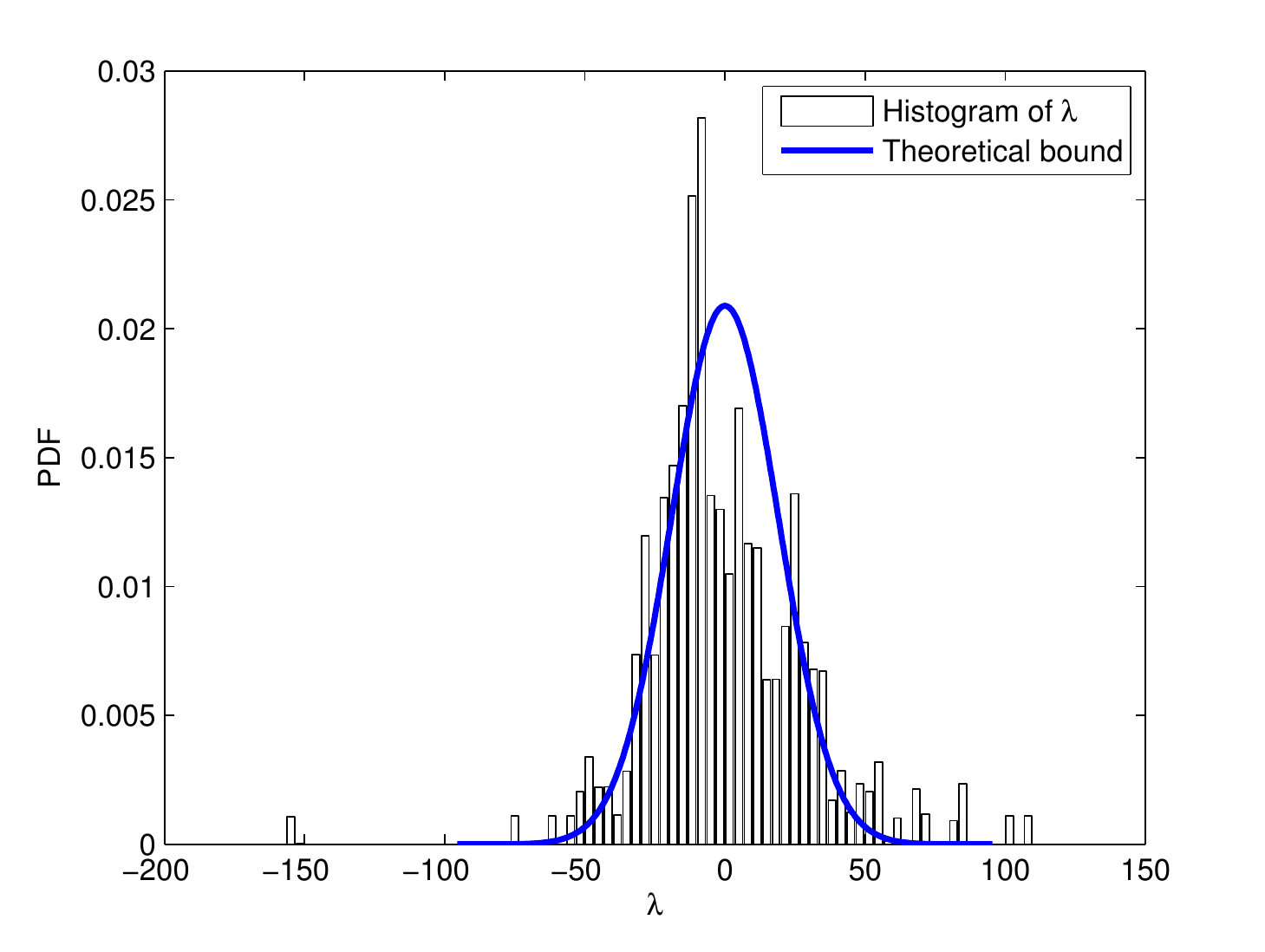}
}
\caption{{\label{real_data_learning}} Parameter learning of the IEEE 118-bus system.}
\end{figure}

\begin{figure}[!htp]
{
\includegraphics[width=0.7\textwidth]{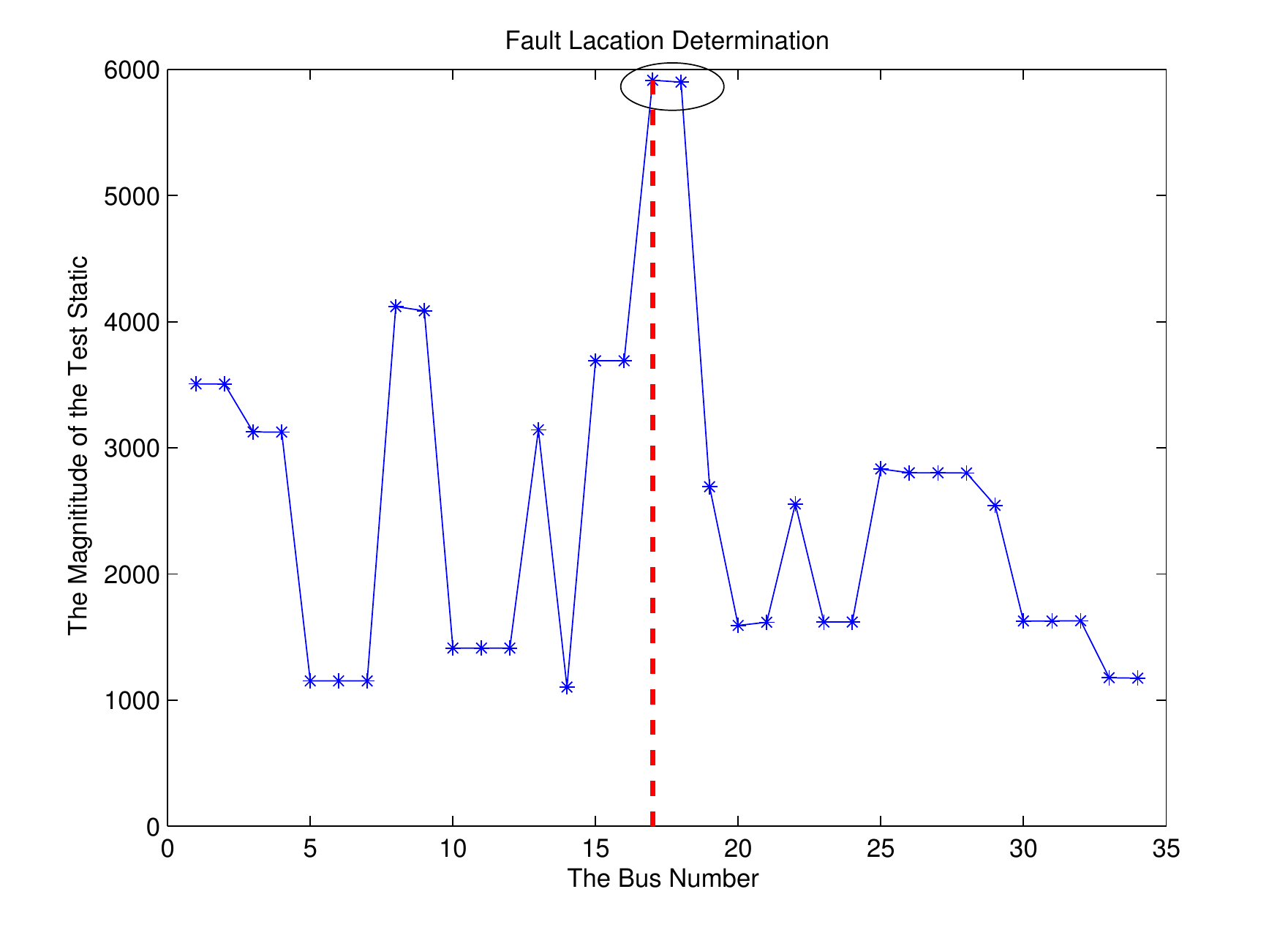}
}
\caption{{\label{real_data_loc}} Data analysis of the realistic 34-PMU power flow around events occurrence.}
\end{figure}

In this section, we evaluate the efficacy of the proposed test statistic for power system stability. For the experiments shown in the following, the real power flow data were of a chain-reaction fault happened in the China power grids in 2013. The PMU number, the sample rate and the total sample time are $p = 34$, $K = 50 Hz$ and $284s$, respectively. The chain-reaction fault happened from $t=65.4s$ to $t=73.3s$. Let $q = 5, n_g = 50$. Fig.\ref{real_data_learning} shows that the mean and variance of $\lambda$ agree well with theoretical ones. Based on the results in Fig.\ref{real_data_learning} and event indicators \eqref{DDMSA:eq7}, the occurrence time and the actual duration of the event can be identified as $t_0 = 65s$ and $t_dur \approx 8s$, respectively. The location of the most sensitive bus can also be identified using the data analysis above. The result shown in Fig.\ref{real_data_loc} illustrates that 17$th$ and 18$th$ PMU are the most sensitive PMUs which are in accordance with the actual accident situation.

\subsection{Situation Awareness based on Linear Eigenvalue Statistics}
\label{SALES}

Situation awareness (SA) is of great significance in power system operation, and a reconsideration of SA is essential for future grids~\cite{bda2016tsg}. These future grids are always huge in size and  complex in topology. Operating under a novel regulation, their management mode is much different from previous one.

All these driving forces demand a new prominence to the term situation awareness (SA). The SA is essential for power grid security; inadequate SA is identified as one of the root causes  for the largest blackout in history---the 14 August 2003 Blackout in the United States and Canada \cite{us2004final}.

In \cite{endsley2011designing}, SA is defined as  the perception of the elements in an environment, the comprehension of their meaning, and the projection of their status in the near future. This chapter is aimed at the use of model-free and data-driven methodology for the comprehension of the power grid.

The massive data compose the profile of the actual grid---present state;
SA aims to translate the present state into perceived state for decision-making \cite{panteli2013assessing}.

\begin{figure}[htbp]
 \centering
 \subfigure[SA for the operational decision-making]{\label{fig:SA}
 \includegraphics[width=0.45\textwidth]{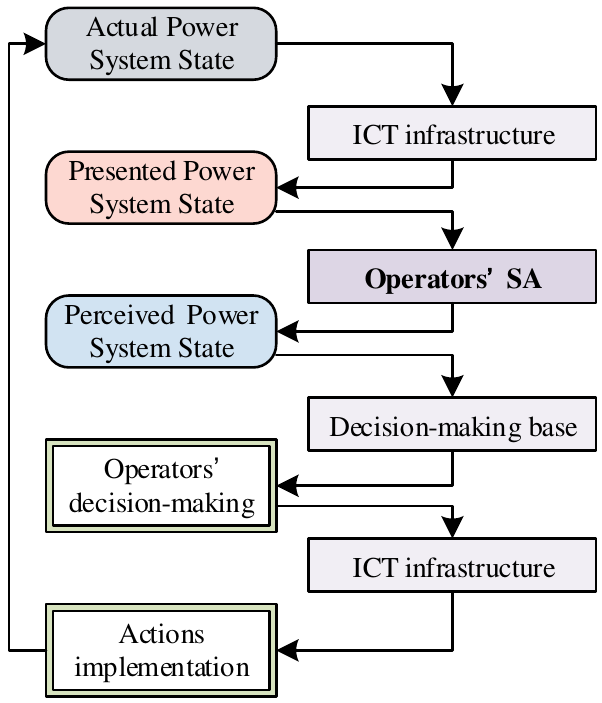}
 }
 \subfigure[SA Methodology based on  RMT]{\label{fig:methodSA}
 \includegraphics[width=0.45\textwidth]{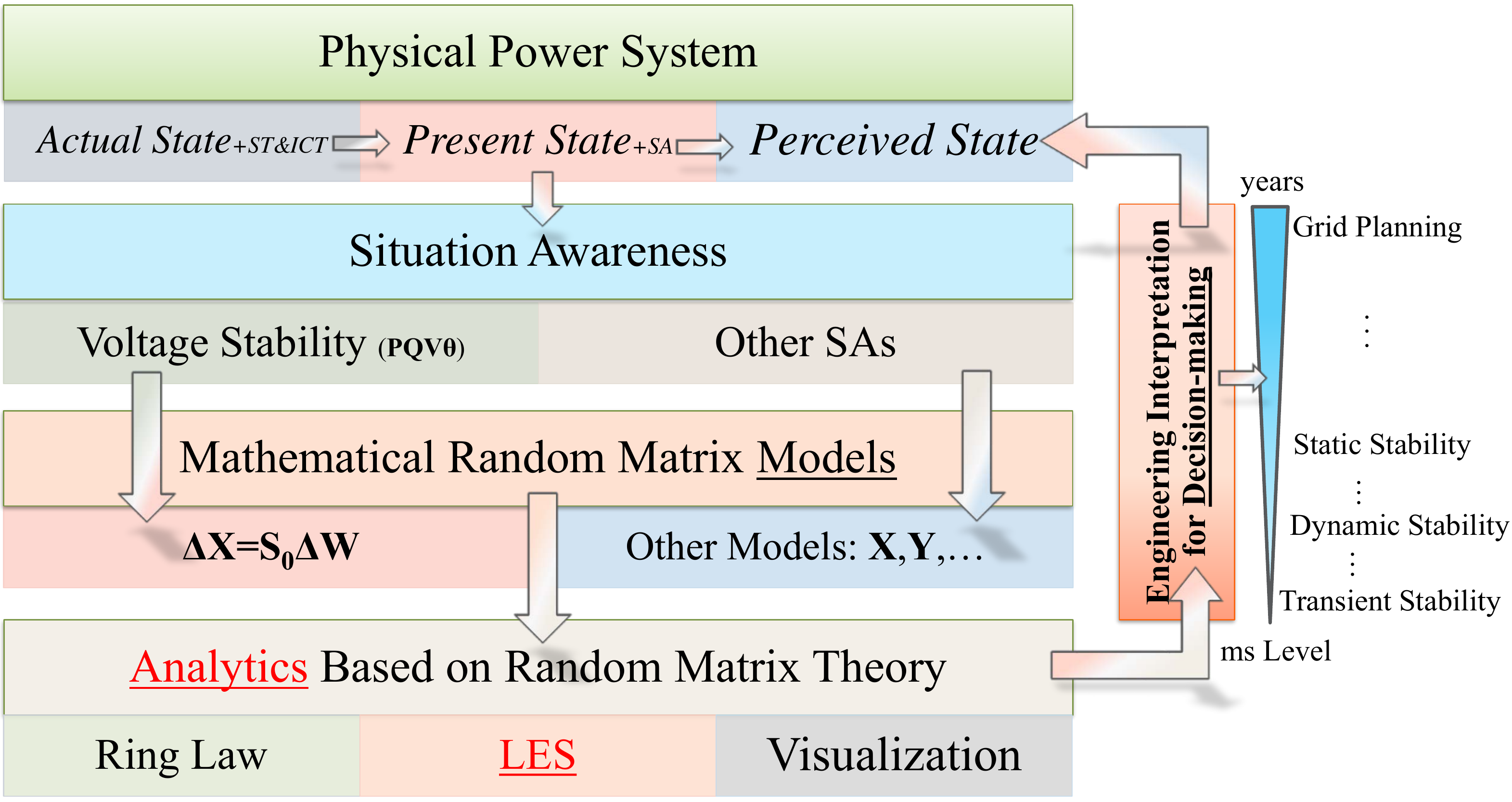}
 }
 \caption{SA and its methodology}
\end{figure}

The proposed methodology consists of  three essential procedures as illustrated in Fig. \ref{fig:methodSA}: 1) big data model---to model the system using experimental data for the RMM; 2) big data analysis---to conduct high-dimensional analyses for the indicator system as the statistical solutions; 3) engineering interpretation---to visualize and interpret the statistical results to human beings for decision-making.

Power grids operate in a balance situation obeying
\begin{equation}
\label{Eq:Balance}
\left\{ \begin{aligned}
  & \Delta {{P}_{i}}={{P}_{is}}-{{P}_{i}}\left( \mathbf{v},\boldsymbol{\theta } \right) \\
 & \Delta {{Q}_{i}}={{Q}_{is}}-{{Q}_{i}}\left( \mathbf{v},\boldsymbol{\theta } \right) \\
\end{aligned} \right.,
\end{equation}
where ${P}_{is}$ and ${Q}_{is}$ are the power injections on node $i$, while ${P}_{i}\left( \mathbf{v},\boldsymbol{\theta } \right)$ and ${Q}_{i}\left( \mathbf{v},\boldsymbol{\theta } \right)$ are the injections of the network satisfying
\begin{equation}
\label{Eq:PiQi}
\left\{ \begin{aligned}
  & {{P}_{i}}={{V}_{i}}\sum\limits_{j=1}^{n}{{{V}_{j}}\left( {{G}_{ij}}\cos {{\theta }_{ij}}+{{B}_{ij}}\sin {{\theta }_{ij}} \right)} \\
 & {{Q}_{i}}={{V}_{i}}\sum\limits_{j=1}^{n}{{{V}_{j}}\left( {{G}_{ij}}\sin {{\theta }_{ij}}-{{B}_{ij}}\cos {{\theta }_{ij}} \right)} \\
\end{aligned} \right..
\end{equation}

For simplicity, combining \eqref{Eq:Balance} and \eqref{Eq:PiQi}, we obtain
\begin{equation}
\label{Eq:WXY}
{{\mathbf{W}}_{0}}=f\left( {{\mathbf{X}}_{0}},{{\mathbf{Y}}_{0}} \right),
\end{equation}
where ${\mathbf{W}}_{0}$ is the vector of power injections on nodes depending on ${P}_{is}$, ${Q}_{is}$.   ${\mathbf{X}}_{0}$ is the system status variables depending on ${V}_{i}$, ${\theta}_{i}$, while  ${\mathbf{Y}}_{0}$ is the network topology parameters depending on ${B}_{ij}$, ${G}_{ij}$.

For a system state with certain fluctuations---thus randomness in datasets, we formulate the system as
\begin{equation}
\label{Eq:WXYDelta}
{{\mathbf{W}}_{0}}\!+\!\Delta \mathbf{W}\!=\!f\left( {{\mathbf{X}}_{0}}\!+\!\Delta \mathbf{X},{{\mathbf{Y}}_{0}}\!+\!\Delta \mathbf{Y} \right).
\end{equation}
With a Taylor expansion,~\eqref{Eq:WXYDelta} is rewitten as
\begin{equation}
\label{Eq:WXYDeltaTaylor}
\begin{aligned}
   {{\mathbf{W}}_{\!0}}\!+\!\Delta\! \mathbf{W}\!=&\!f\!\left( {{\mathbf{X}}_{\!0}},\!{{\mathbf{Y}}_{\!0}} \!\right)\!+\!f{{'}_{\!\mathbf{X}}}\!\left( {{\mathbf{X}}_{\!0}},\!{{\mathbf{Y}}_{\!0}} \!\right)\Delta\! \mathbf{X}\!+\!f{{'}_{\!\mathbf{Y}}}\!\left( {{\mathbf{X}}_{\!0}},\!{{\mathbf{Y}}_{\!0}} \!\right)\Delta\! \mathbf{Y} \\
 & \!+\!\frac{1}{2}f'{{'}_{\!\mathbf{X\!X}}}\!\left( {{\mathbf{X}}_{\!0}},\!{{\mathbf{Y}}_{\!0}} \!\right){{\!\left( \Delta\! \mathbf{X} \!\right)}^{\text{2}}}\!+\frac{1}{2}\!f'{{'}_{\!\mathbf{Y\!Y}}}\!\left( {{\mathbf{X}}_{\!0}},\!{{\mathbf{Y}}_{\!0}} \!\right){{\!\left( \Delta\! \mathbf{Y} \!\right)}^{\text{2}}} \\& +f'{{'}_{\!\mathbf{X\!Y}}}\!\left( {{\mathbf{X}}_{\!0}},\!{{\mathbf{Y}}_{\!0}} \!\right)\Delta\! \mathbf{X}\Delta\! \mathbf{Y}\!\!+\!\cdots   .  \\
\end{aligned}
\end{equation}

Equ. \eqref{Eq:PiQi} shows that ${{\mathbf{W}}_{0}}$ is linear with $\mathbf{Y}_0$; it means that $f'{{'}_{\!\mathbf{Y\!Y}}}\!\left(\! \mathbf{X},\!\mathbf{Y} \!\right)\!=\!0$.
On the other hand, the value of system status variables $\mathbf{X}$ are relatively stable and we can ignore the second-order term ${{\left( \Delta \mathbf{X} \right)}^{2}}$ and higher-order terms. In this way, we turn \eqref{Eq:WXYDeltaTaylor} into
\begin{equation}
\label{Eq:WXYDeltaTaylorSimplify}
\begin{aligned}
  \Delta\! \mathbf{W}\!=&\!f{{'}_{\!\mathbf{X}}}\!\left( {{\mathbf{X}}_{\!0}},\!{{\mathbf{Y}}_{\!0}} \!\right)\Delta\! \mathbf{X}\!+\!f{{'}_{\!\mathbf{Y}}}\!\left( {{\mathbf{X}}_{\!0}},\!{{\mathbf{Y}}_{\!0}} \!\right)\Delta\! \mathbf{Y}   \\&\!+\!f'{{'}_{\!\mathbf{X\!Y}}}\!\left( {{\mathbf{X}}_{\!0}},\!{{\mathbf{Y}}_{\!0}} \!\right)\Delta\! \mathbf{X}\Delta\! \mathbf{Y}\!.
\end{aligned}
\end{equation}

Suppose the network topology is unchanged, i.e., $\!\Delta\!\mathbf{Y}\!=\!0$. From~\eqref{Eq:WXYDeltaTaylorSimplify}, we deduce that
\begin{equation}
\label{Eq:Y_0}
\Delta \mathbf{X}\!=\!{{\left(f{{'}_{\!\mathbf{X}}}\!\left(\! {{\mathbf{X}}_{\!0}},\!{{\mathbf{Y}}_{\!0}}\! \right) \right)}^{-1}}\!\left(\! \Delta\! \mathbf{W} \! \right)\!=\!{{\mathbf{S}}_{0}}\Delta \mathbf{W}.
\end{equation}

On the other hand, suppose the power demands is unchanged, i.e., $\!\Delta\!\mathbf{W}\!=\!0$. From \eqref{Eq:WXYDeltaTaylorSimplify}, we obtain that
\begin{equation}
\label{Eq:W_0}
\Delta \mathbf{X}\!=\!{{\mathbf{S}}_{0}}\Delta \mathbf{W}_y,
\end{equation}
where $\mathbf{W}_y\!=\![\mathbf{I}\!+\!f'{{'}_{\!\mathbf{X\!Y}}}\!\left( {{\mathbf{X}}_{\!0}},\!{{\mathbf{Y}}_{\!0}} \!\right)\!\Delta\!\mathbf{Y}\!\mathbf{S}_0]^{\!-\!1}
[\-\!f{{'}_{\!\mathbf{Y}}}\!\left( {{\mathbf{X}}_{\!0}},\!{{\mathbf{Y}}_{\!0}} \!\right)]     .$

Note that $\mathbf{S}_{0}\!=\!\!{{\left(f{{'}_{\!\mathbf{X}}}\!\left(\! {{\mathbf{X}}_{\!0}},\!{{\mathbf{Y}}_{\!0}}\! \right) \right)}^{-1}}$, i.e., the inversion of the Jacobian matrix $\mathbf{J}_0$, expressed as
\begin{equation}
\label{Eq:Jij}
{\mathbf{J}_{ij0}}={{\left. \left[ \begin{matrix}
   \frac{\partial {{P}_{i}}}{\partial {{U}_{j}}} & \frac{\partial {{P}_{i}}}{\partial {{\theta }_{j}}}  \\
   \frac{\partial {{Q}_{i}}}{\partial {{U}_{j}}} & \frac{\partial {{Q}_{i}}}{\partial {{\theta }_{j}}}  \\
\end{matrix} \right] \right|}_{{{U}_{j}}={{U}_{j0}},{{\theta }_{j}}={{\theta }_{j0}}}}.
\end{equation}

Thus, we describe the power system operation using a random matrix---if there is an unexpected active power change or short circuit,  the corresponding change of system status variables ${\mathbf{X}}_{0}$, i.e. ${V}_{i}$, ${\theta}_{i}$, will obey  \eqref{Eq:Y_0} or \eqref{Eq:W_0} respectively.

For a practical system,  we can always build a relationship in the form of $\mathbf{Y}\!=\!\mathbf{H}\mathbf{X}$ with a similar procedure as \eqref{Eq:WXY} to \eqref{Eq:W_0}; it is linear in high dimensions. For an equilibrium operation system in which the reactive power is almost constant or changes much more slowly than the active one,  the relationship model between  voltage magnitude and active power is just like the Multiple Input Multiple Output (MIMO) model in wireless communication~\cite{qiu2015smart,zhang2015MassiveMIMO}. 
Note that most variables of vector $\mathbf{V}$ are random due to the ubiquitous noises, e.g., small random fluctuations in $\mathbf{P}$. In addition, we can add very small artificial fluctuations to make them random or replace the missing/bad data with random Gaussian variables. Furthermore, with the normalization, we can build the standard random matrix model (RMM) in the form of $\tilde{\mathbb{V}} = \tilde{\mathbf{\Xi}}\mathbf{R}$, where  $\mathbf{R}$ is a standard Gaussian random matrix.

The data-driven approach conducts analysis requiring no prior knowledge of system topologies, unit operation/control mechanism, causal relationship, etc. It is able to handle massive data all at once; the large size of the data, indeed, enhances the robustness of the final decision against the bad data (errors, losses, or asynchronization). Comparing with classical data-driven methodologies (e.g. PCA), the RMT-based counterpart has some unique characteristics:
\begin{itemize}
\item The statistical indicator is generated from all the data in the form of matrix entries. This is not true to principal components---we really do not know the rank of the covariance matrix. Thus, the RMT approach is robust against those challenges in classical data-driven methods,  such as error accumulations and spurious correlations \cite{Xu2015A}.
\item For the statistical indicator, a theoretical or empirical value can be obtained in advance. The statistical indicator such as LES follows Gaussian distribution, and its variance is bounded \cite{shcherbina2011central} and decays very fast in the order of  $O({N^{-2}})$ given a moderate data dimension $N,$ say $N=118.$
\item We can flexibly handle heterogenous data to realize data fusion via matrix operations, such as the blocking \cite{He2015A}, the sum \cite{zhang2015MassiveMIMO}, the product \cite{zhang2015MassiveMIMO}, and the concatenation \cite{Xu2015A} of matrices. Data fusion is guided by the latest mathematical research \cite[Chapter 7]{qiu2015smart}.
\item Only eigenvalues are used for further analyses, while the eigenvectors are omitted. This leads to much smaller required memory space and faster data-processing speed. Although some information is lost in this way,  there is still  rich information contained in the eigenvalues \cite{ipsen2014weak}, especially those outliers \cite{benaych2013outliers, tao2013outliers}.
\item Particularly,  for a certain RMM, various forms of LES can be constructed  by designing test functions without  introducing any physical error (i.e. \Equs{\VLES{F}{}}{\sum\nolimits_{i=1}^{N}{\varphi_F \left( {{\lambda }_{\mathbf{M},i}} \right)}}).
Each LES, similar to a filter, provides a unique view-angle. As a result, the system is systematically understood piece by piece. Finally, with a proper LES, we can trace some specific signal.
\end{itemize}

\begin{figure}[ht]
\centering
\begin{overpic}[scale=0.58
]{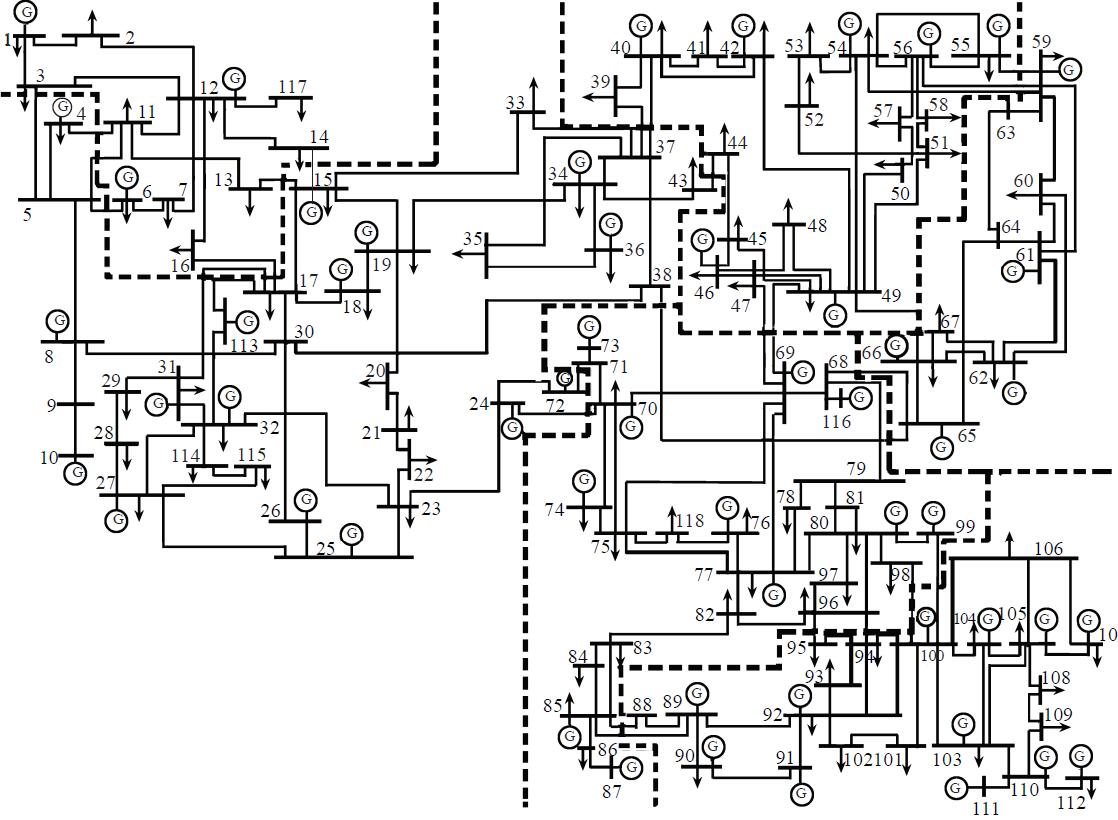}
    \setlength {\fboxsep}{1pt}
    \put(71,60) {\fcolorbox{red}{white}{ \color{blue}{$52$}}} 

      \setlength {\fboxsep}{2pt}

\end{overpic}
\caption{Partitioning network for the IEEE 118-node system.}
\label{fig:IEEE118network}
\end{figure}

We adopt a standard IEEE 118-node system as the grid network (Fig. \ref{fig:IEEE118network}) and the events is shown in Tab. \ref{Tab: Event Series}.

\begin{table}[ht]
\caption {Series of Events}
\label{Tab: Event Series}
\centering

\begin{minipage}[htbp]{0.8\textwidth}
\centering

\begin{tabularx}{\textwidth} { >{\text}l !{\color{black}\vrule width1pt}    >{$}l<{$}|  >{$}l<{$}| >{$}l<{$}|   >{$}l<{$} }  
\toprule[1.5pt]
\textbf {Stage} & \textbf{E}1 & \textbf{E}2 & \textbf{E}3 & \textbf{E}4\\
\midrule[1pt]
Time (s) & $1--500$ & $501--900$ & $901--1300$ & $1301--2500$ \\
\hline
\VPbus{52} (MW) & 0 & \uparrow30  & \uparrow 120 & \nearrow t/4-205\\
\bottomrule[1pt]
\end{tabularx}
\raggedright
\small {$P_{52}$ is the power demand of node 52.
}
\end{minipage}
\end{table}

\begin{figure}[htbp]
 \centering
 \subfigure[Assumed Event, Unavailable.]{\label{fig:Case0Event}
 \includegraphics[width=0.45\textwidth]{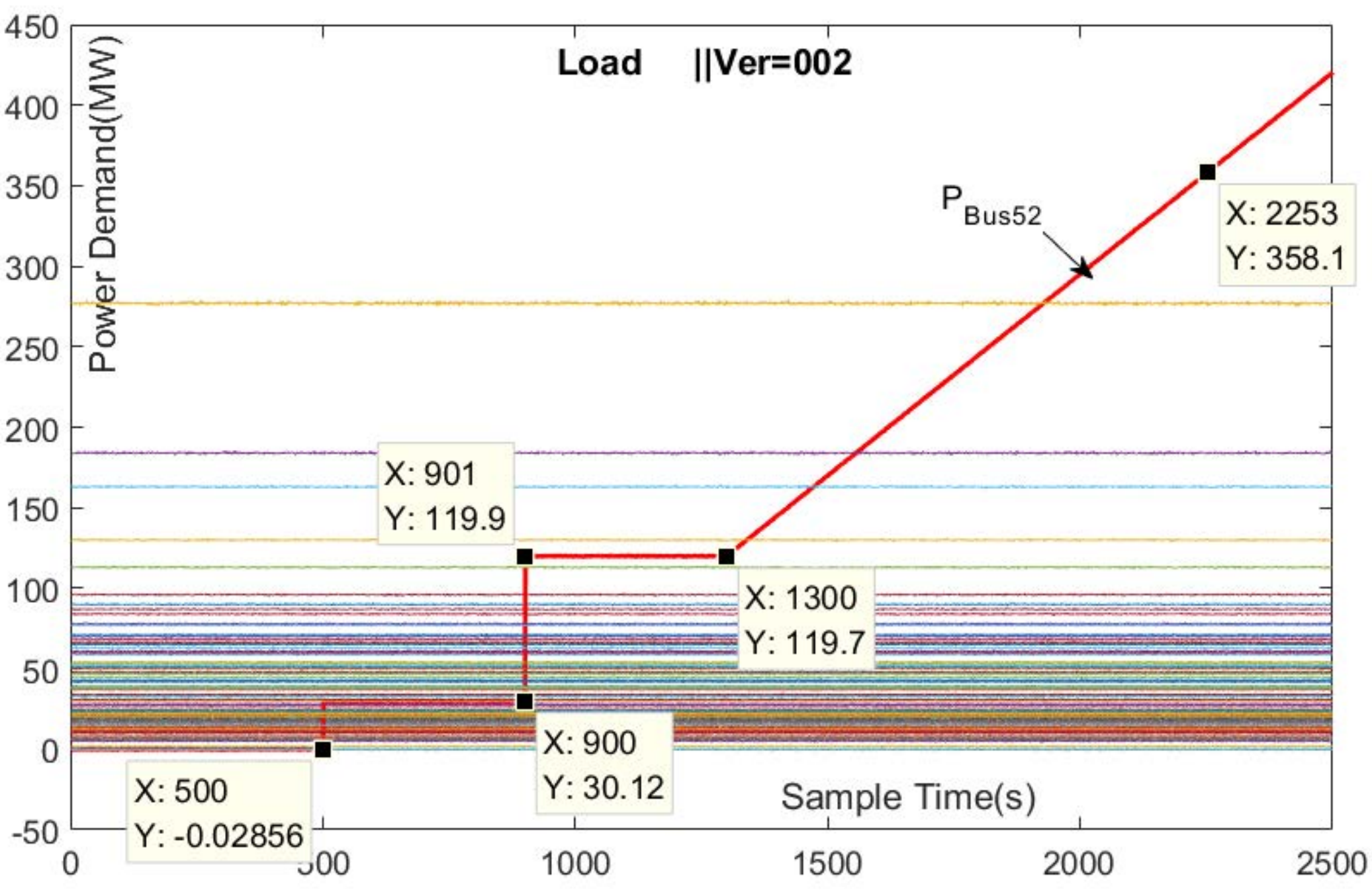}
 }
 \subfigure[Raw Voltage,  $\bm{\Omega}_{\mathbf{V} }$ for Analysis.]{\label{fig:Case0Vol}
 \includegraphics[width=0.45\textwidth]{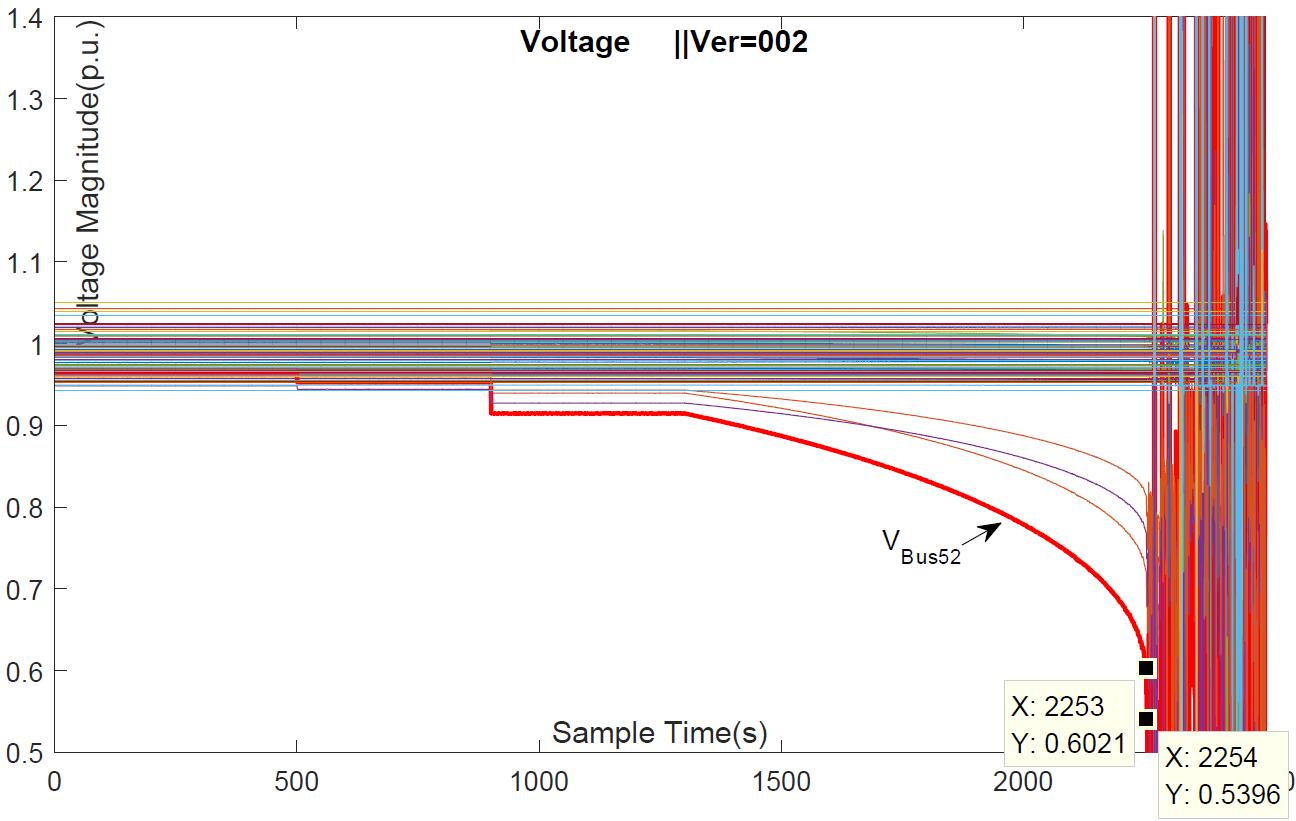}
 }
 \subfigure[Category for Operation Status and Selected Matrix Based on $\bm{\Omega}_{\mathbf{V} }$.]{\label{fig:Case0category}
 \includegraphics[width=0.45\textwidth]{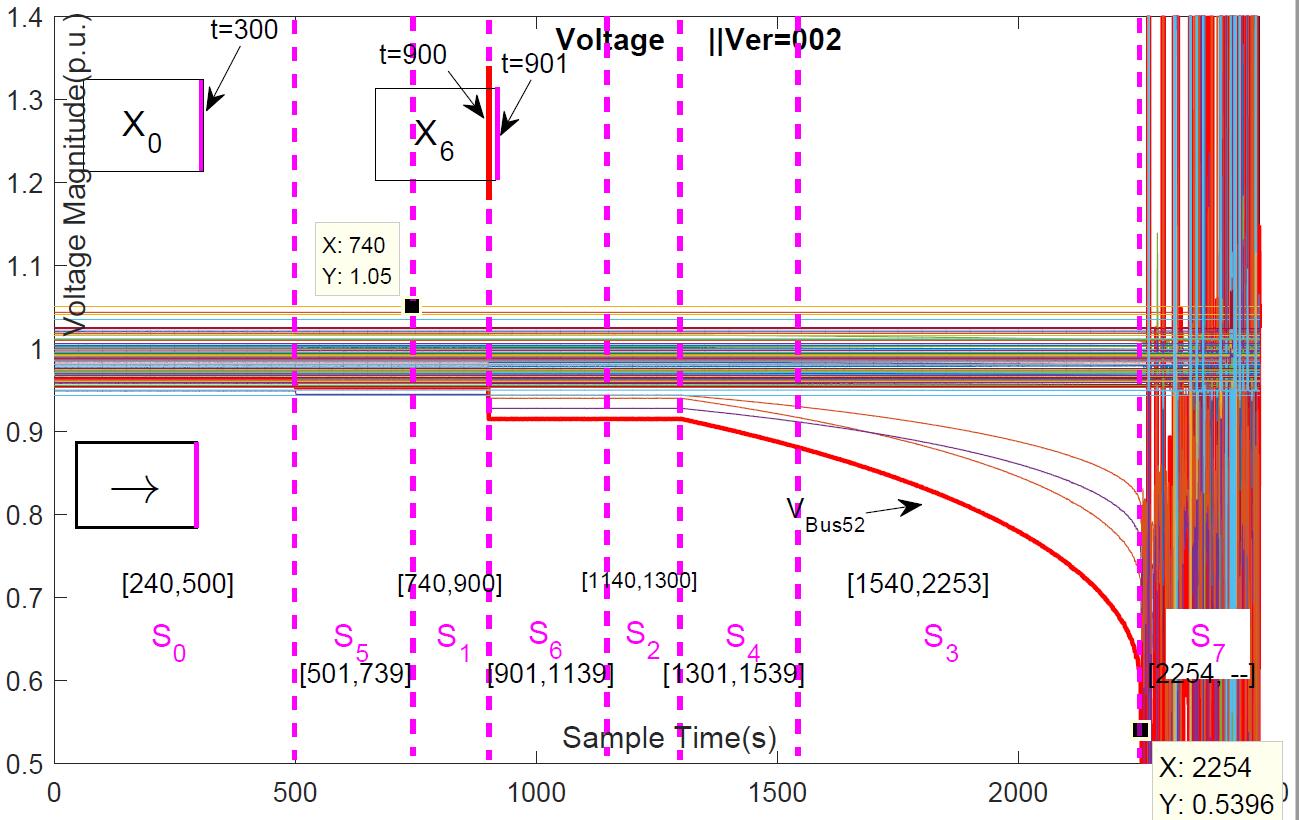}
 }
 \caption{Assumed Event, Data Source, and  Category for Case.}
 \label{fig:Case0A}
 \end{figure}

The power demand of nodes are assigned as
\FuncC{eq:gridfluctuation}{
\Equs {\Index {\Tdata y}{\Text{load\_}nt}}  {
\Index  y{\Text{load\_}nt}\Mul{}{}(1+\Muls{\Vgam{\Text {Mul}}}{r_1})
} +   \Muls{\Vgam{\Text {Acc}}}{r_2},
} 		
where $r_1$ and $r_2$ are the element of standard Gaussian random matrix;  \Vgam{\Text {Acc}}=0.1, \Vgam{\Text {Mul}}=0.001.
Thus, the power demand on each node is obtained as the system injections (Fig. \ref{fig:Case0Event});  the voltage can also be obtained (Fig. \ref{fig:Case0Vol}). Suppose  we sample the voltage data at 1 Hz, the data source is denoted as ${{\bf{\Omega }}_{\bf{V}}}:{\tilde v_{i,j}} \in {\mathbb{R}^{118 \times 2500}}$. The number of dimensions is $n\!=\!118$ and the sampling time span is $t\!=\!2500$.

Suppose that the power demand data (Fig. \ref{fig:Case0Event}) are \textit{unknown} or unqualified for SA due to the low sampling frequency or the bad quality.
For further analysis, we just start with data source $\bm{\Omega}_{\mathbf{V} }$ (Fig. \ref{fig:Case0Vol}) and assign the analysis matrix as ${\bf{X}} \in {\mathbb{R}^{118 \times 240}}$ (4 minutes' time span).
First, we conduct category for the system operation status; the results are shown as Fig. \ref{fig:Case0category}.
In general, according to the raw data source and the analysis matrix size, we divide our system into 8 stages. Note that it is a statistical division---$\textbf{S}4, \textbf{S}5$, and $\textbf{S}6$ are transition stages, and their time span is right equal to the length of the analysis matrix minus one, i.e, $T\!-\!1\!=\!239$.
These stages are described as follows:
\begin{itemize}
\item For $\textbf{S}0, \textbf{S}1, \textbf{S}2$, the white noises play a dominant part. \VPbus{52} is rising in turn.
\item For $\textbf{S}3$, \VPbus{52} maintains stable growth.
\item $\textbf{S}4$, transition stage. Ramping signal exists.
\item $\textbf{S}5, \textbf{S}6$, transition stages. Step signal exists.
\item For $\textbf{S}7$, voltage collapse.
\end{itemize}

We also select two typical data cross-sections  for stage $\textbf{S}0$ and $\textbf{S}6$:  ${X_0} \in {\mathbb{R}^{118 \times 240}}$ during period $t\!=\![61\!:\!300]$ at the sampling time $t_\text{end}\!=\!300$, and 2) ${X_6} \in {\mathbb{R}^{118 \times 240}}$ during period $t\!=\![662\!:\!901]$ at the sampling time $t_\text{end}\!=\!901$.

\begin{figure}[htbp]
 \centering
 \subfigure[Ring Law for $\Vector X_0$]{\label{fig:Case0X0ring}
 \includegraphics[width=0.45\textwidth, height= 0.45\textwidth]{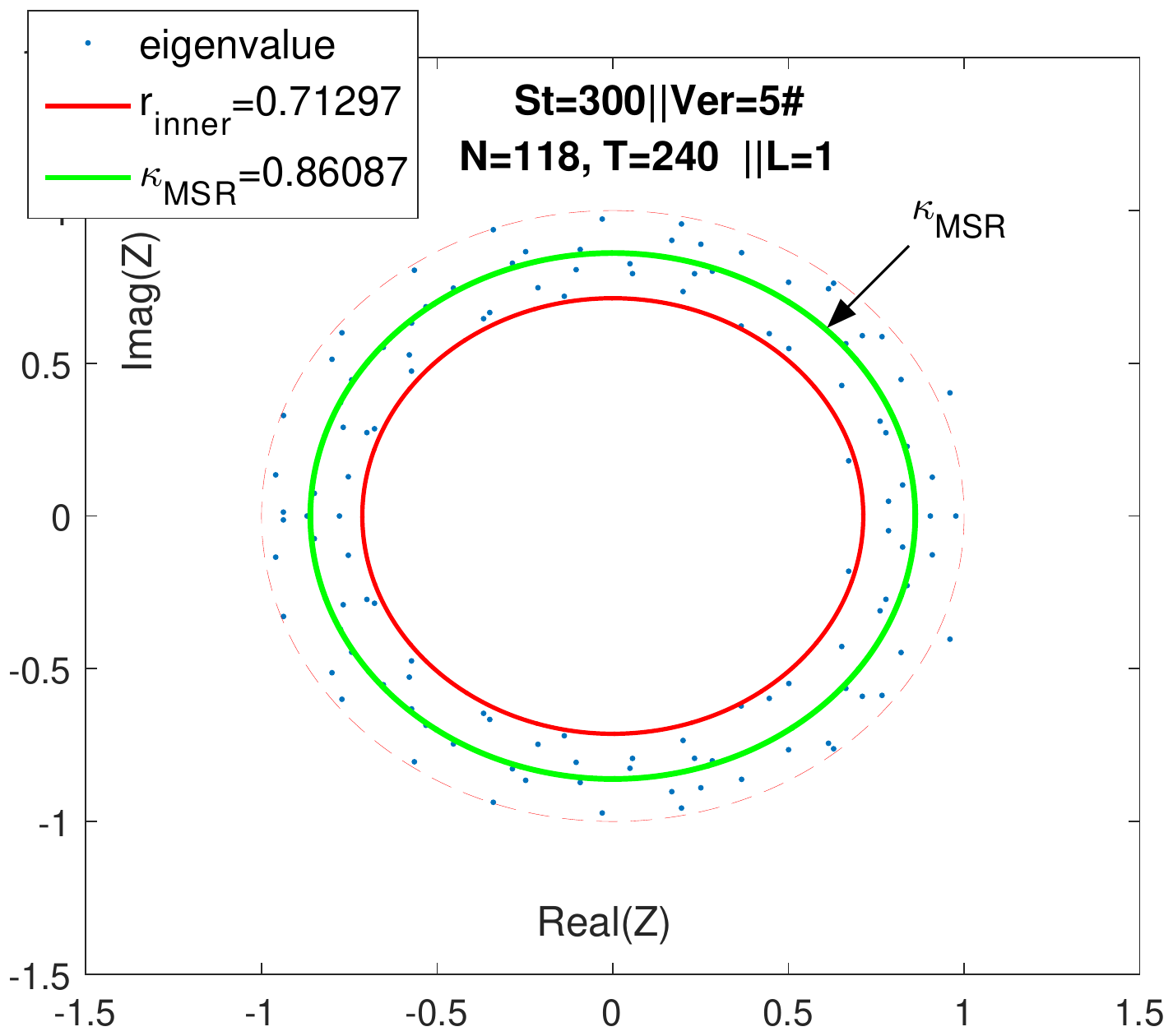}
 }
 \subfigure[M-P Law  for $\Vector X_0$]{\label{fig:Case0X0mp}
 \includegraphics[width=0.45\textwidth]{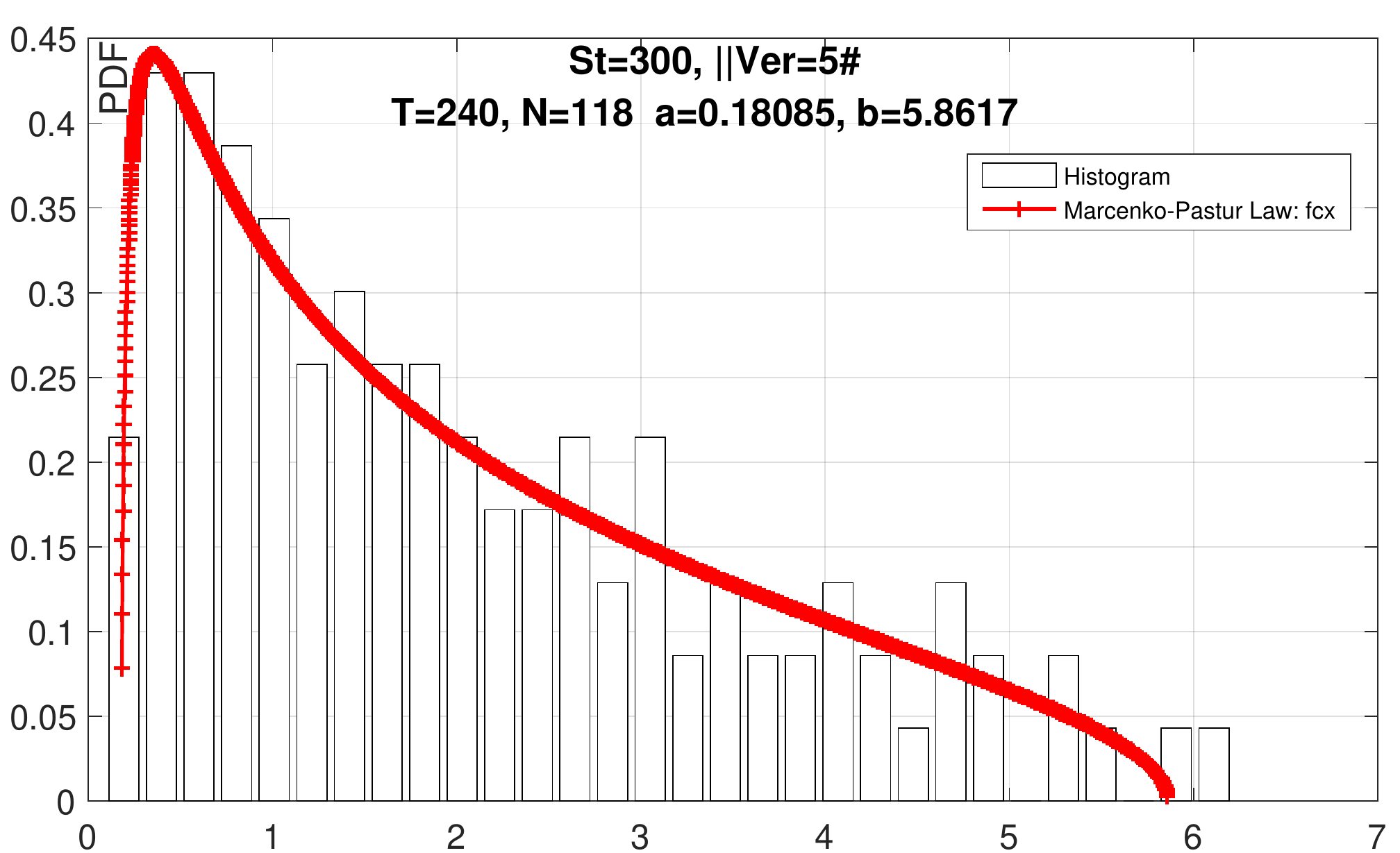}
 }
 \subfigure[Ring Law  for $\Vector X_6$]{\label{fig:Case0X6ring}
 \includegraphics[width=0.4\textwidth]{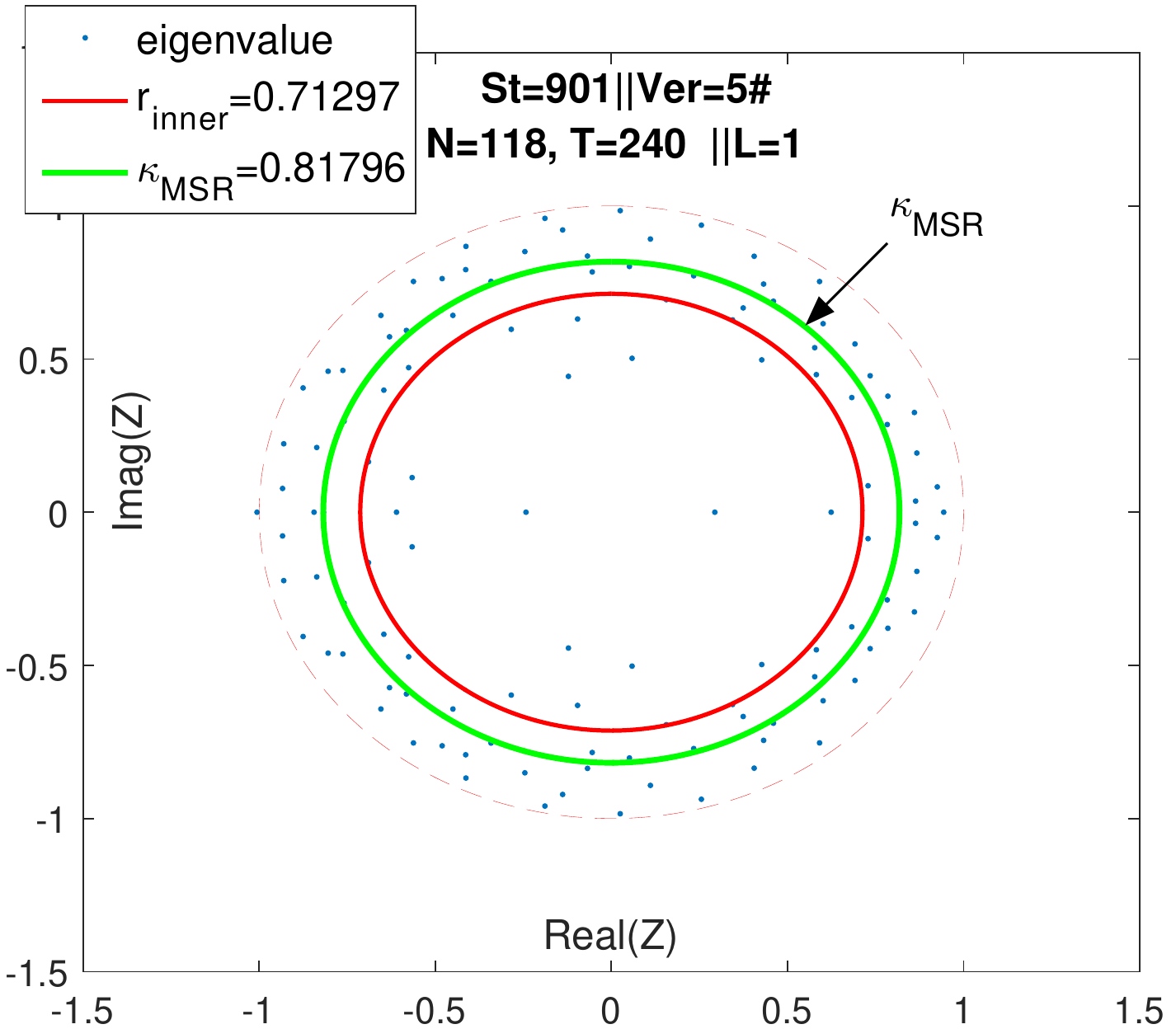}
 }
 \subfigure[M-P Law  for $\Vector X_6$]{\label{fig:Case0X6mp}
 \includegraphics[width=0.5\textwidth]{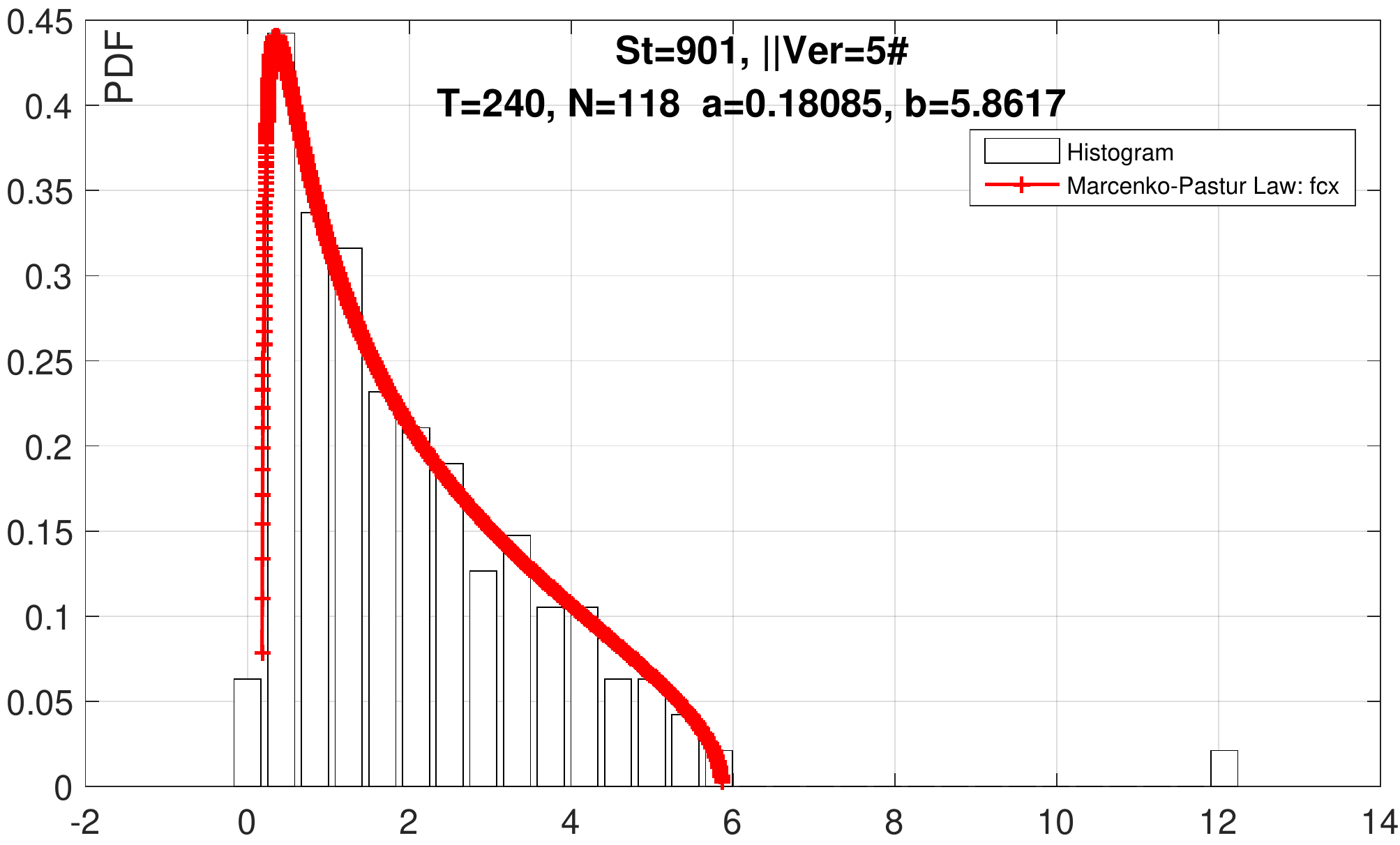}
 }
 \caption{Anomaly Detection Result.}
 \label{fig:Case00A}
 \end{figure}

Besides, as discussed in \ref{LDRM}, we build up the RMM $\tilde{\mathbb{V}}$ from the raw voltage data. Then,  $\tau_{\text{MSR}}$ is employed as a statistical indicator to conduct anomaly detection. For the selected  data cross-section $\Vector X_0$ and $\Vector X_6$, their M-P Law and Ring Law Analysis are shown as Fig \ref{fig:Case0X0ring}, \ref{fig:Case0X0mp}, \ref{fig:Case0X6ring} and  \ref{fig:Case0X6mp}. 

\begin{figure}[htbp]
\centering
\includegraphics[width=0.65\textwidth]{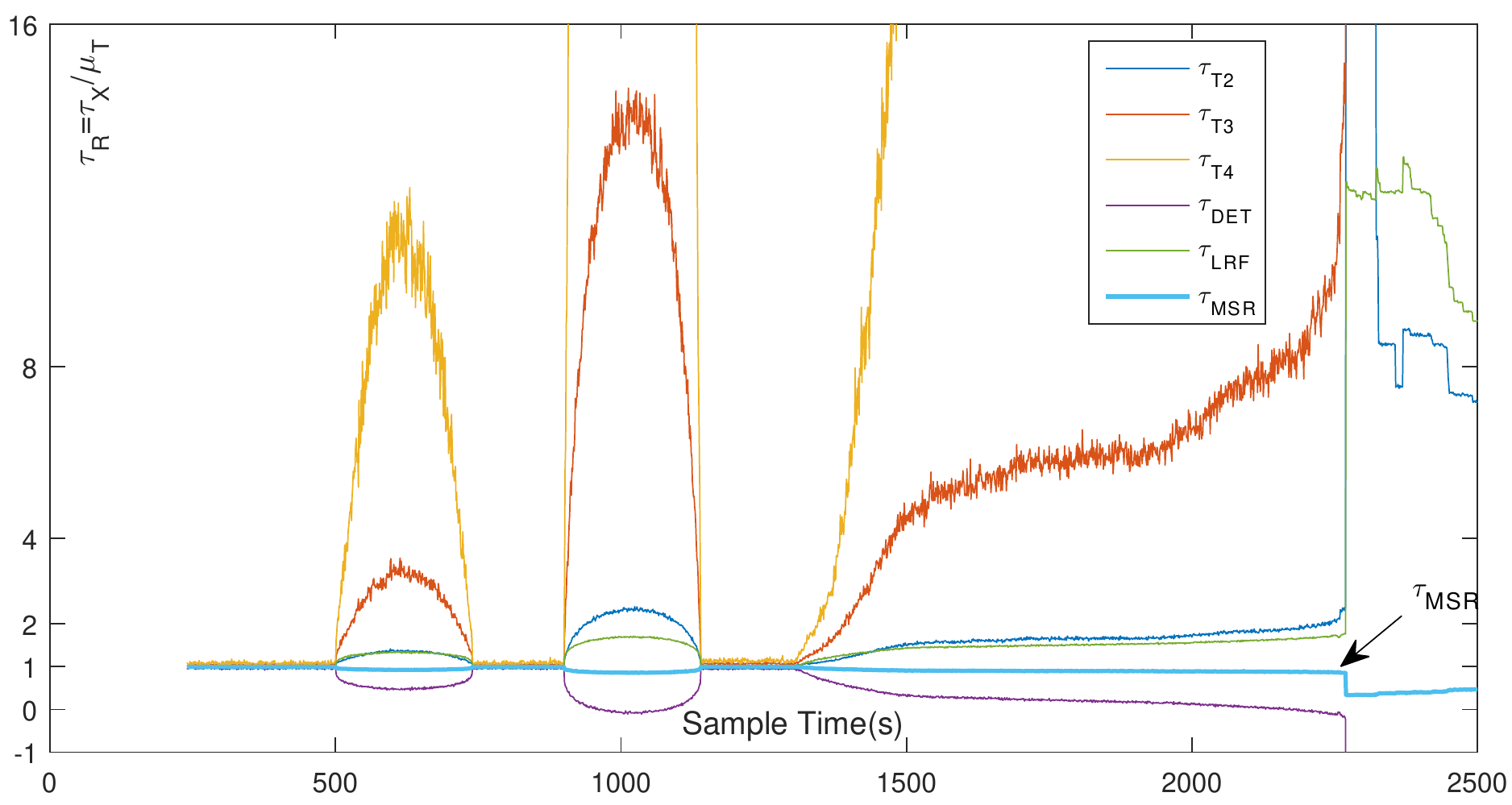}
\caption{Illustration of Various LES Indicators.}
\label{fig:Case0LESs}
\end{figure}

Fig \ref{fig:Case00A} shows that  when there is no signal in the system, the experimental RMM well matches Ring Law and M-P Law, and the experimental value of LES is approximately equal to the theoretical value.  This validates the theoretical justification for modeling rapid fluctuation at each node with additive white Gaussian noise, as shown in Section \ref{LDRM}. On the other hand, Ring Law and M-P Law are violated at the very beginning ($t_\text{end}\!=\!901$) of the step signal.
Besides, the proposed high-dimensional indicator  $\tau_{\text{MSR}}$, is extremely sensitive to the anomaly.  At $t_\text{end}\!=\!901$, the $\tau_{\text{MSR}}$ starts the dramatic change as shown in the \Cur{\tau_{\text{MSR}}}{t} curve, while the raw voltage magnitudes remain still in the normal range as shown in Fig. \ref{fig:Case0category}. 
Moreover, we design numerous kinds of LES $\tau$ and define $\mu_0\!=\!\tau/\STE{\tau}.$ The results are shown in Fig. \ref{fig:Case0LESs} and prove that  different indicators have different characteristics and effectiveness; this suggests another topic to explore in the future.

\begin{figure}[htbp]
 \centering
 \subfigure[ESD of $\Vector Y_0$ (Normal)]{
 \includegraphics[width=0.45\textwidth]{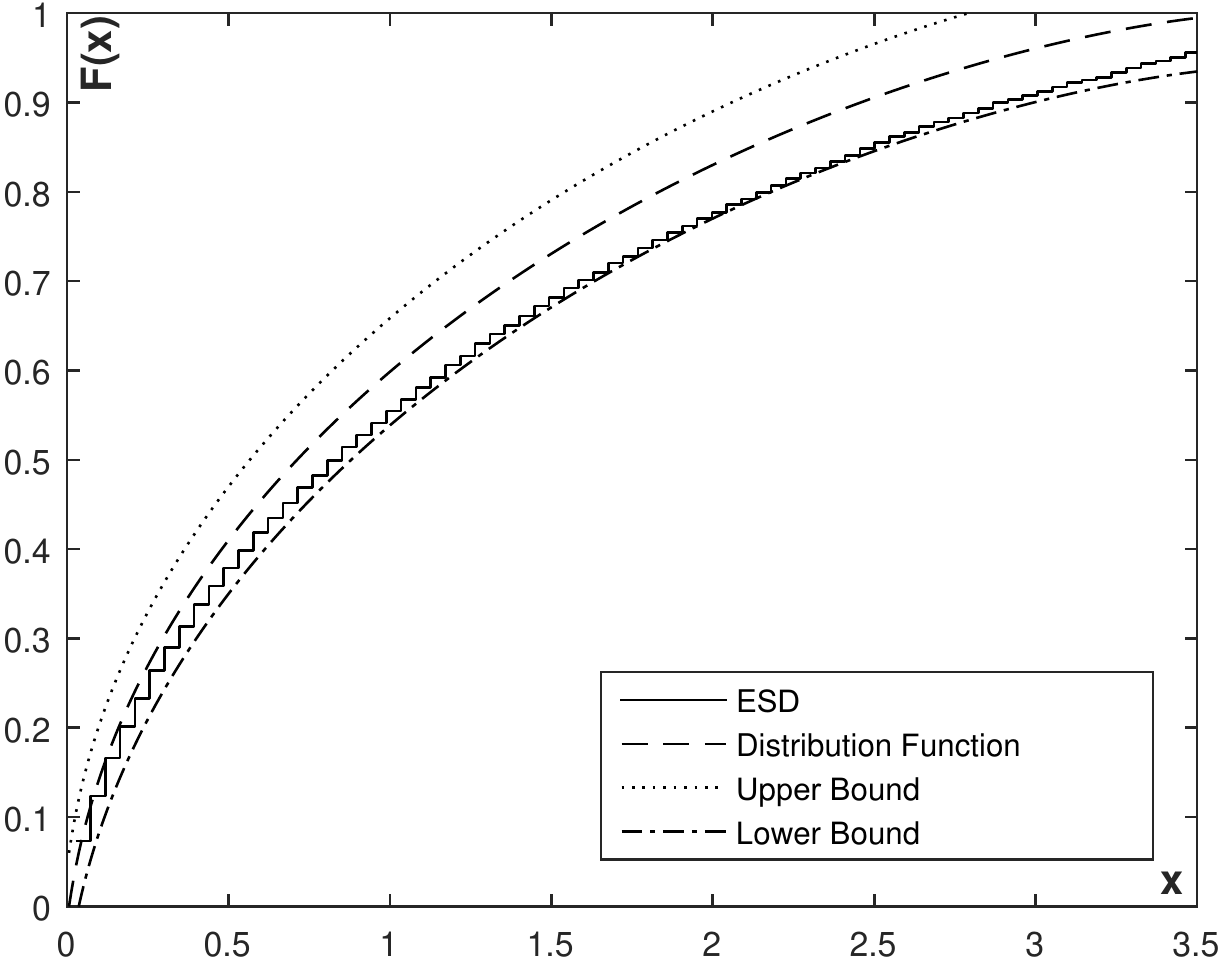}
 }
 \subfigure[ESD of $\Vector Y_6$ (Abnormal)]{
 \includegraphics[width=0.45\textwidth]{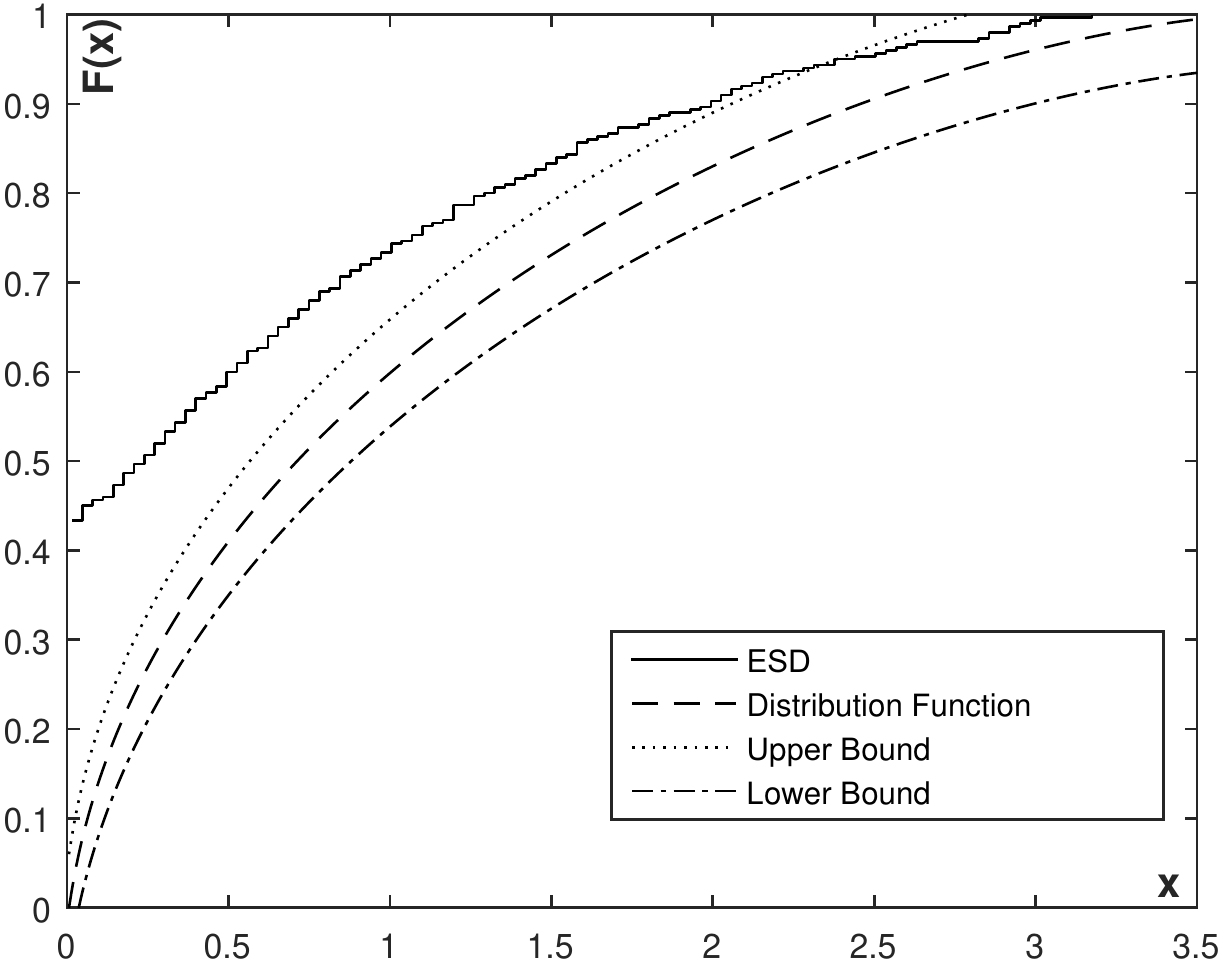}
 }
 \caption{Anomaly Detection Using LUE matrices}
 \label{MP_Law}
\end{figure}

\begin{figure}[htbp]
 \centering
 \subfigure[Density of $\Vector Z_0$ (Normal)]{
 \includegraphics[width=0.45\textwidth]{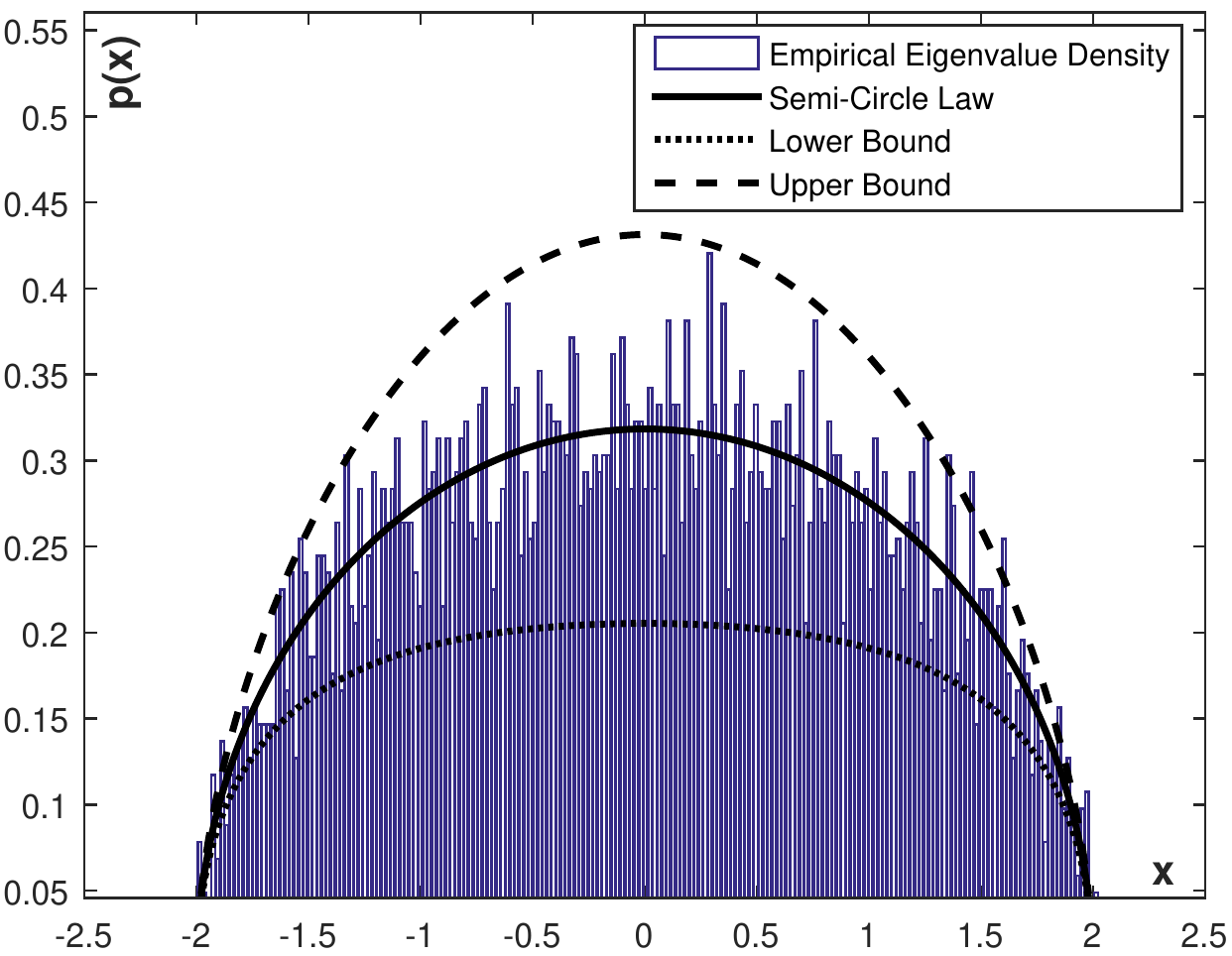}
 }
 \subfigure[Density of $\Vector Z_6$ (Abnormal)]{
 \includegraphics[width=0.45\textwidth]{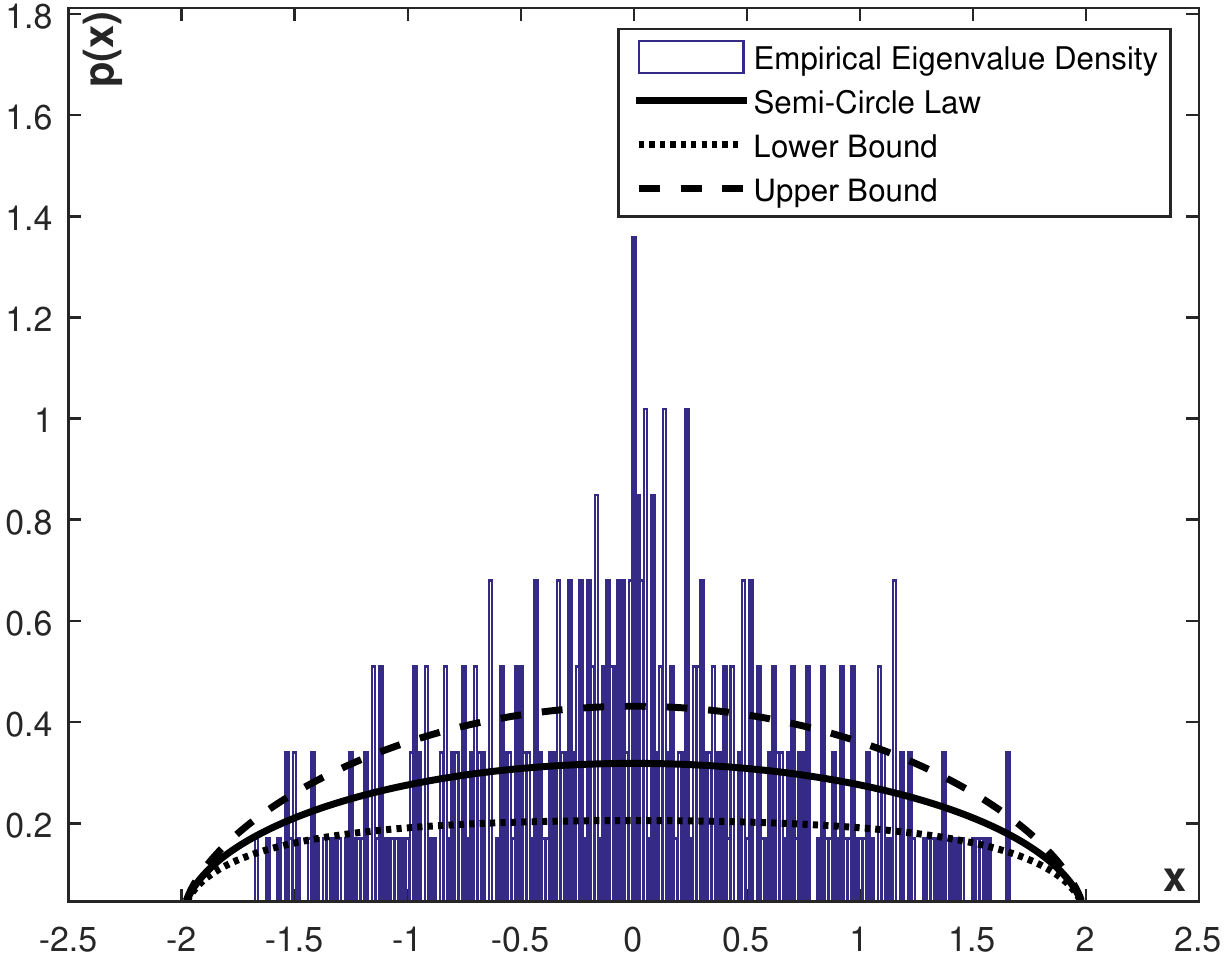}
 }
 \subfigure[ESD of $\Vector Z_0$ (Normal)]{
 \includegraphics[width=0.45\textwidth]{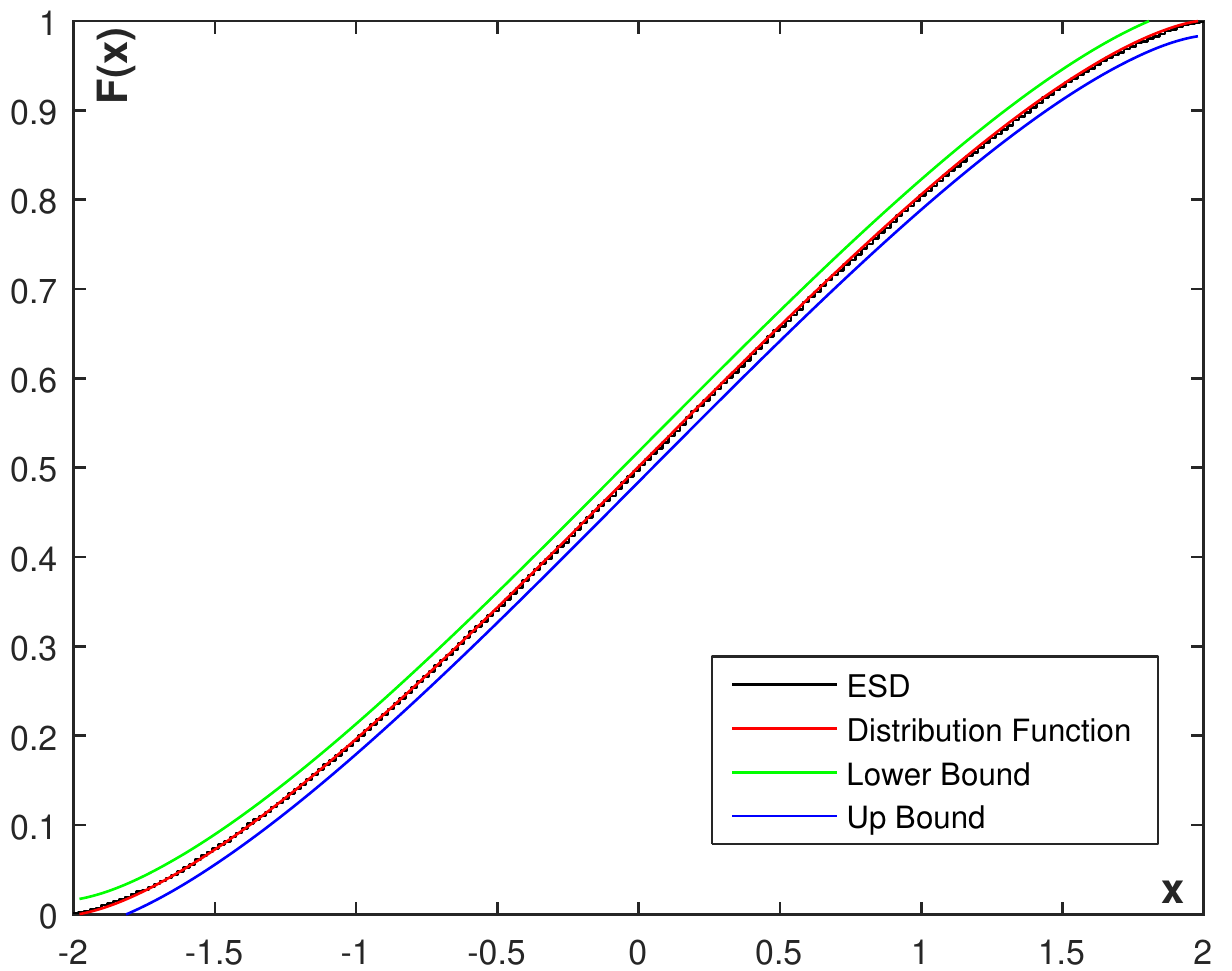}
 }
 \subfigure[ESD of $\Vector Z_6$ (Abnormal)]{
 \includegraphics[width=0.45\textwidth]{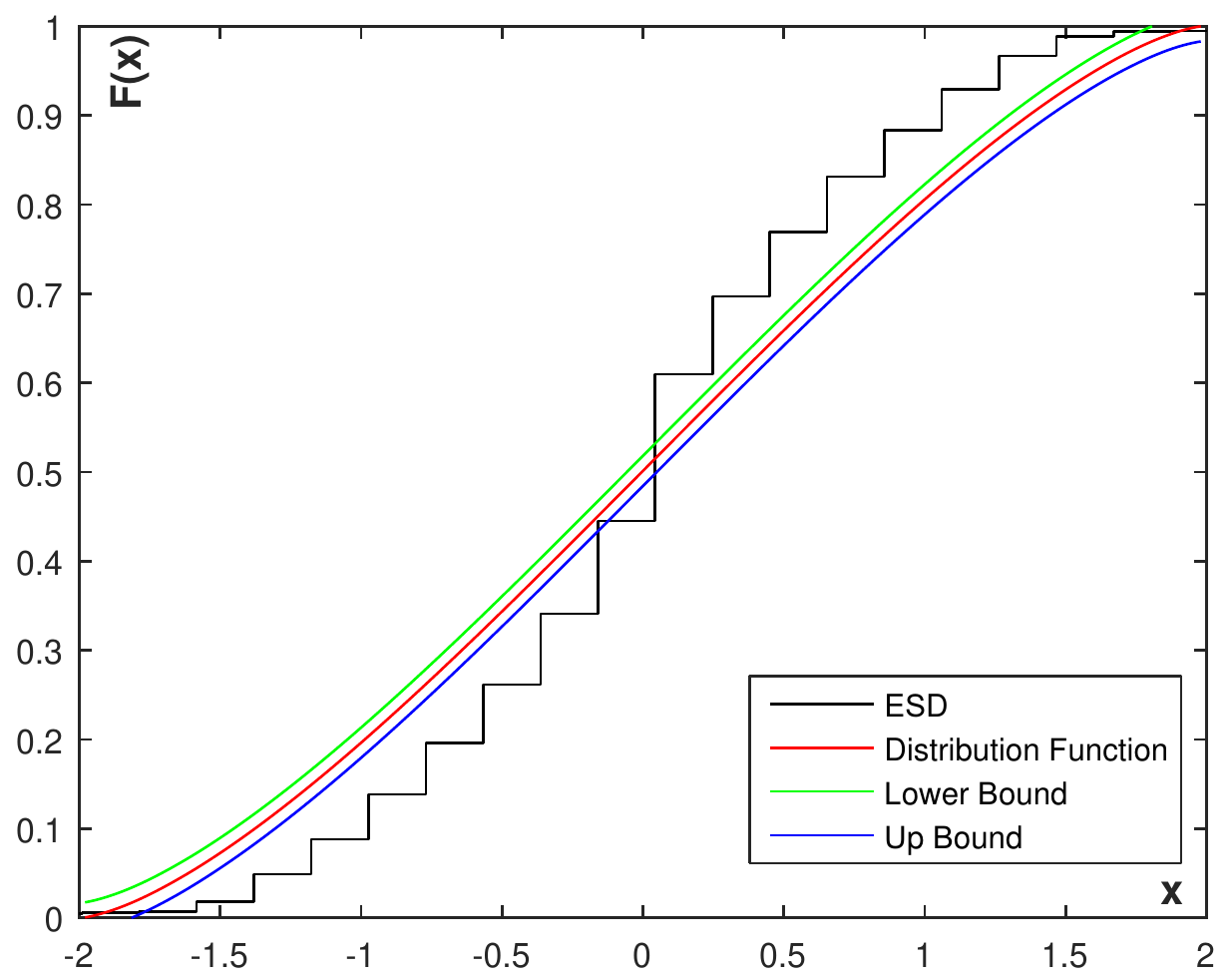}
 }
 \caption{Anomaly Detection Using GUE matrices}
 \label{Semi_Circle_Law}
\end{figure}

Furthermore, we investigate the SA based on the high dimensional spectrum test. The sampling time is set as $t_\text{end}\!=\!300$ and $t_\text{end}\!=\!901$.
Following Lemma \ref{thm2:Convergence for Spectra} and Lemma \ref{thm5:Convergence for Spectra},

$\Vector Y_0, {Y_6} \in {\mathbb{R}^{118 \times 240}}$ (span $t\!=\![61\!:\!300]$ and $t\!=\![662\!:\!901]$), and
$\Vector Z_0, {Z_6} \in {\mathbb{R}^{118 \times 118}}$ (span $t\!=\![183\!:\!300]$ and $t\!=\![784\!:\!901]$)
 are selected. The results are shown in Fig. \ref{MP_Law} and Fig. \ref{Semi_Circle_Law}.
These results validate that empirical spectral density test is competent to conduct anomaly detection---when the power grid is under a normal condition, the empirical spectral density ${f_{\bf{A}}}\left( x \right)$ and the ESD function ${F_{\bf{A}}}\left( x \right)$ are almost strictly bounded between the upper bound and the lower bound of their asymptotic limits. On the other hand, these results also  validate that GUE and LUE are proper mathematical tools to model the power grid operation.

\begin{figure}[htbp]
 \centering
 \subfigure[$V-P$ Curve]{\label{fig:Case0NosecurveVP}
 \includegraphics[width=0.45\textwidth]{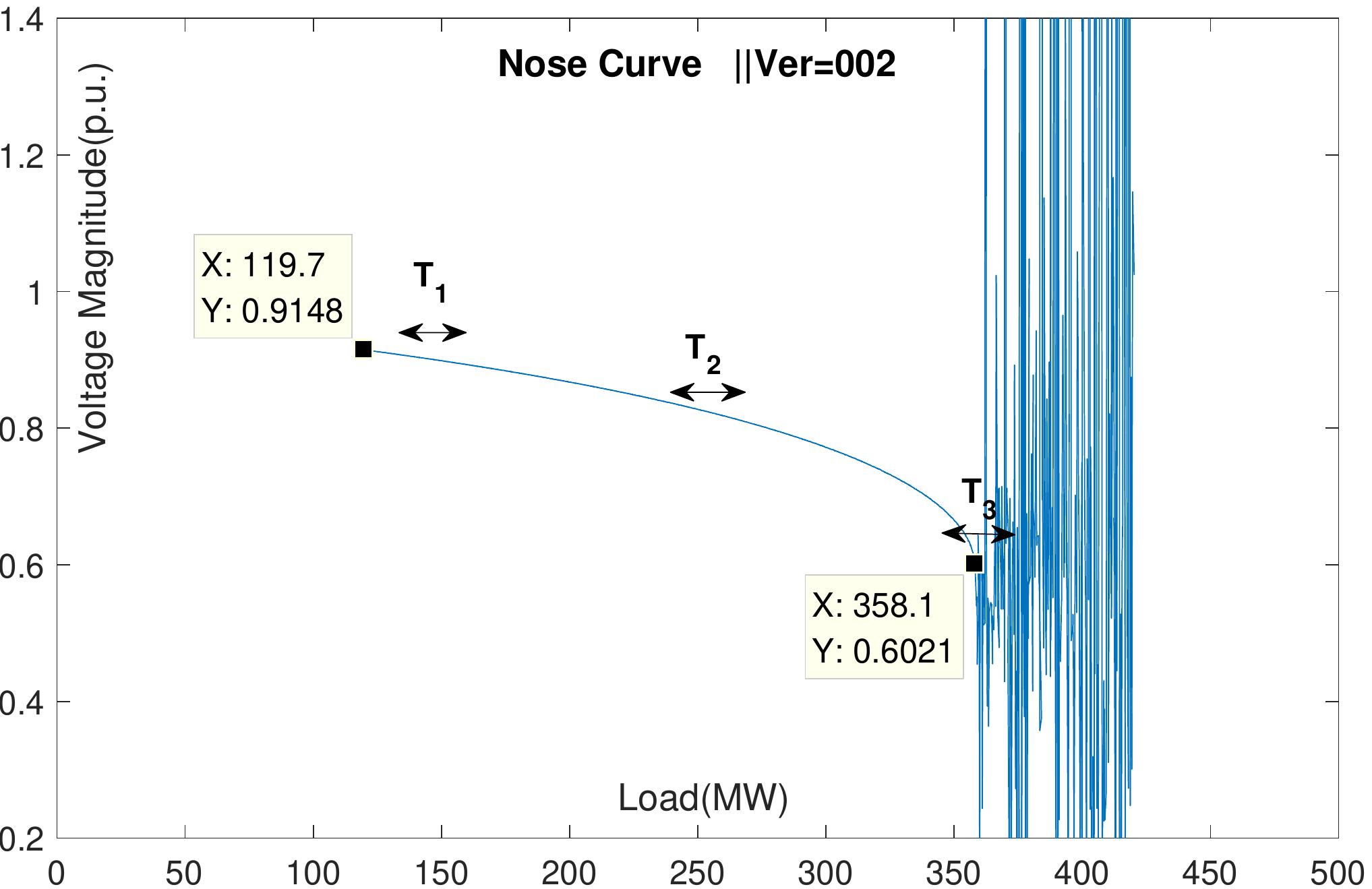}
 }
 \subfigure[$\lambda-P$ Curve]{\label{fig:Case0NosecurveEigP}
 \includegraphics[width=0.45\textwidth]{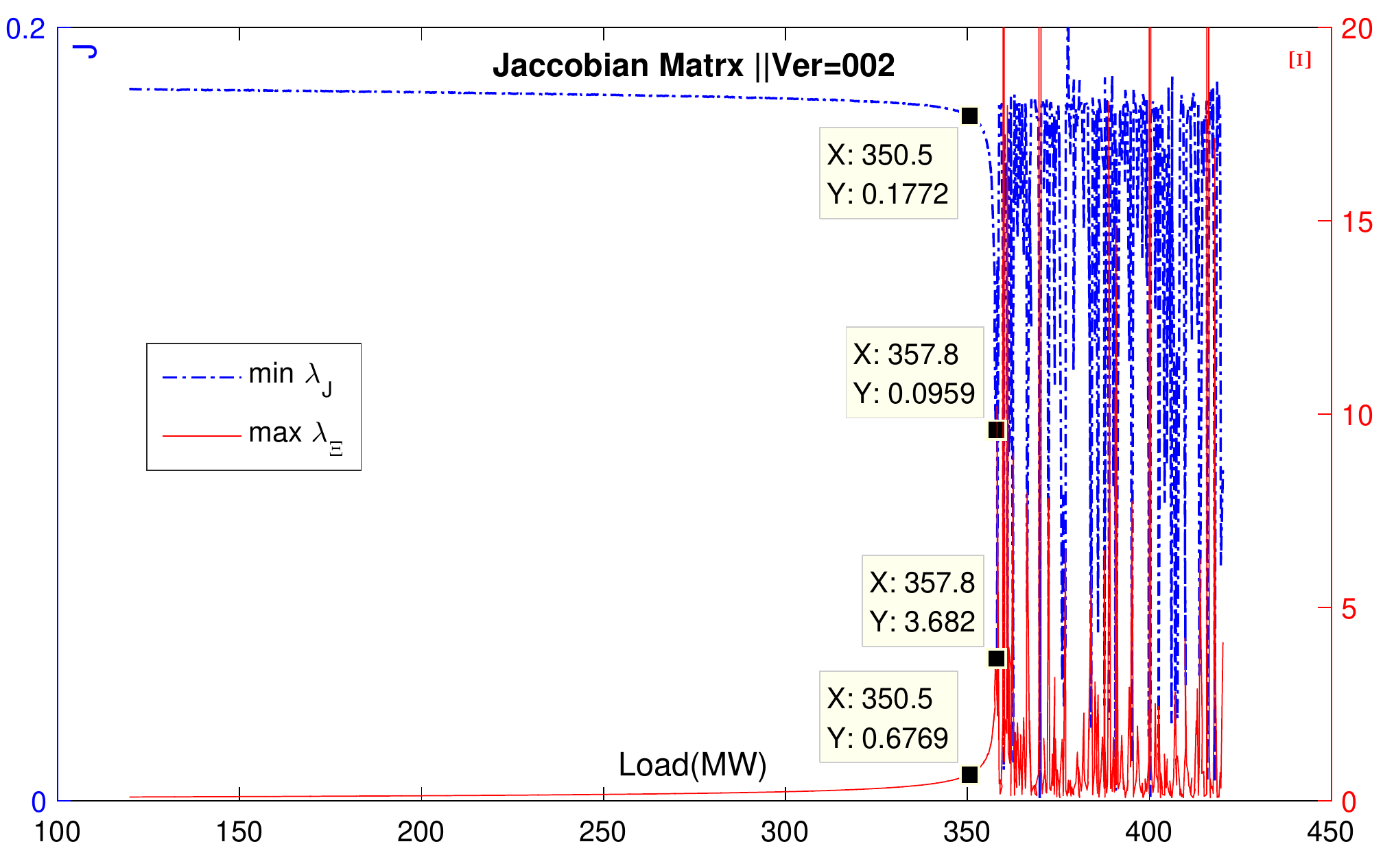}
 }
 \caption{The $V-P$ curve and  $\lambda-P$ curve.}
 \label{fig:Case0C}
\end{figure}

\begin{figure}[htbp]
 \centering
 \subfigure[Ring Law for $\Vector T_1$]{\label{fig:Case0T1ring}
 \includegraphics[width=0.4\textwidth]{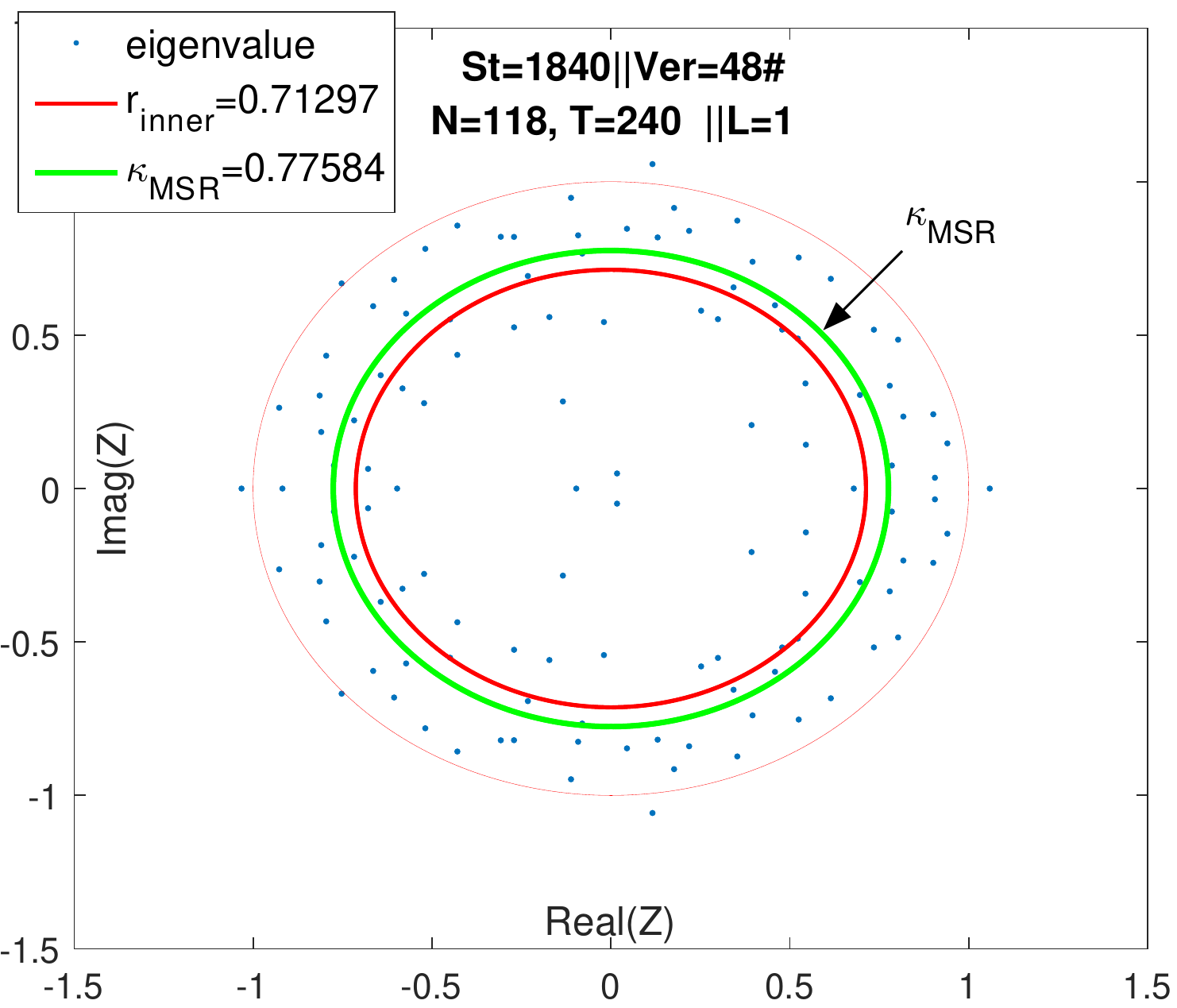}
 }
 \subfigure[M-P Law for $\Vector T_1$]{\label{fig:Case0T1mp}
 \includegraphics[width=0.5\textwidth]{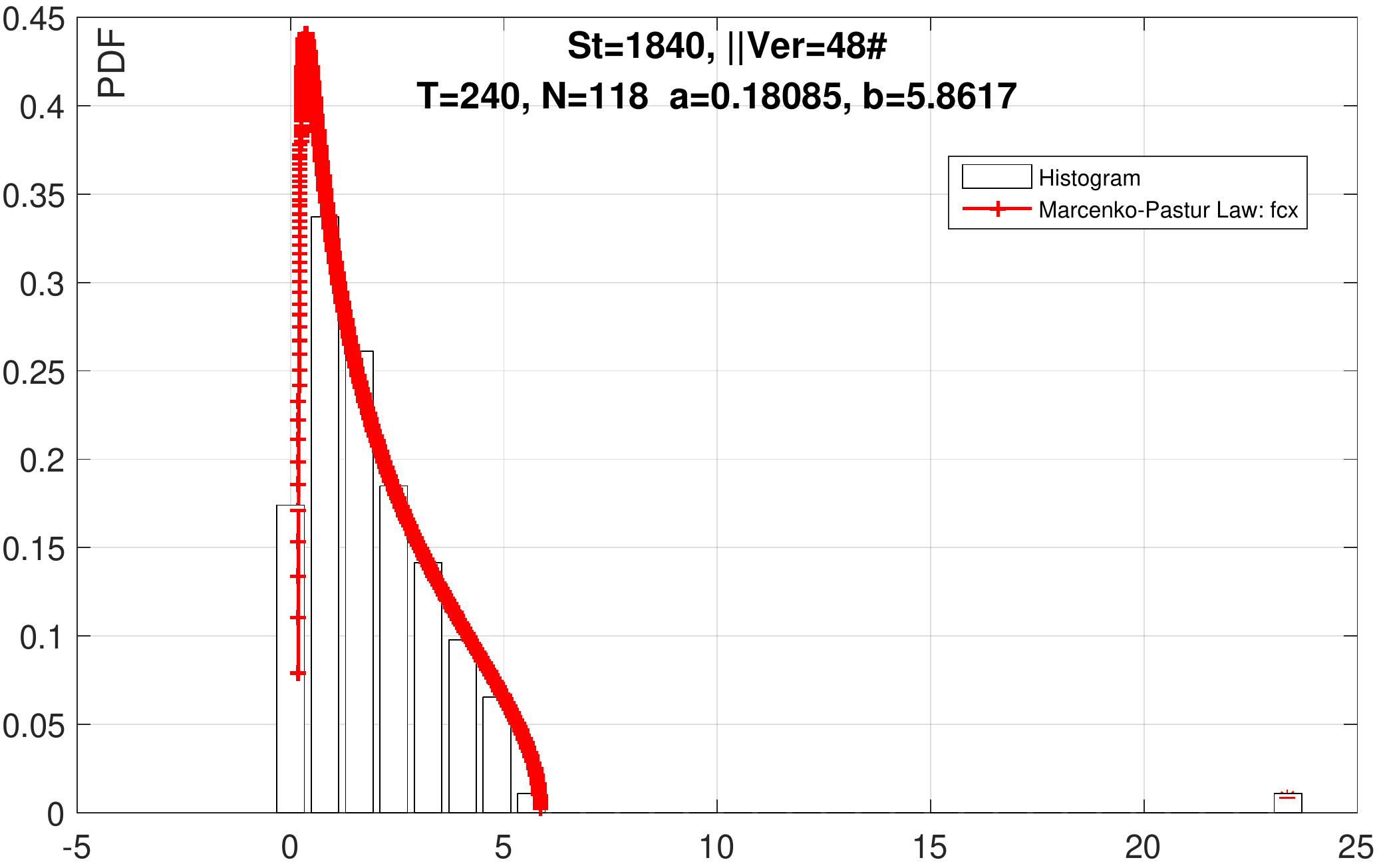}
 }
 \subfigure[Ring Law for $\Vector T_2$]{\label{fig:Case0T2ring}
 \includegraphics[width=0.4\textwidth]{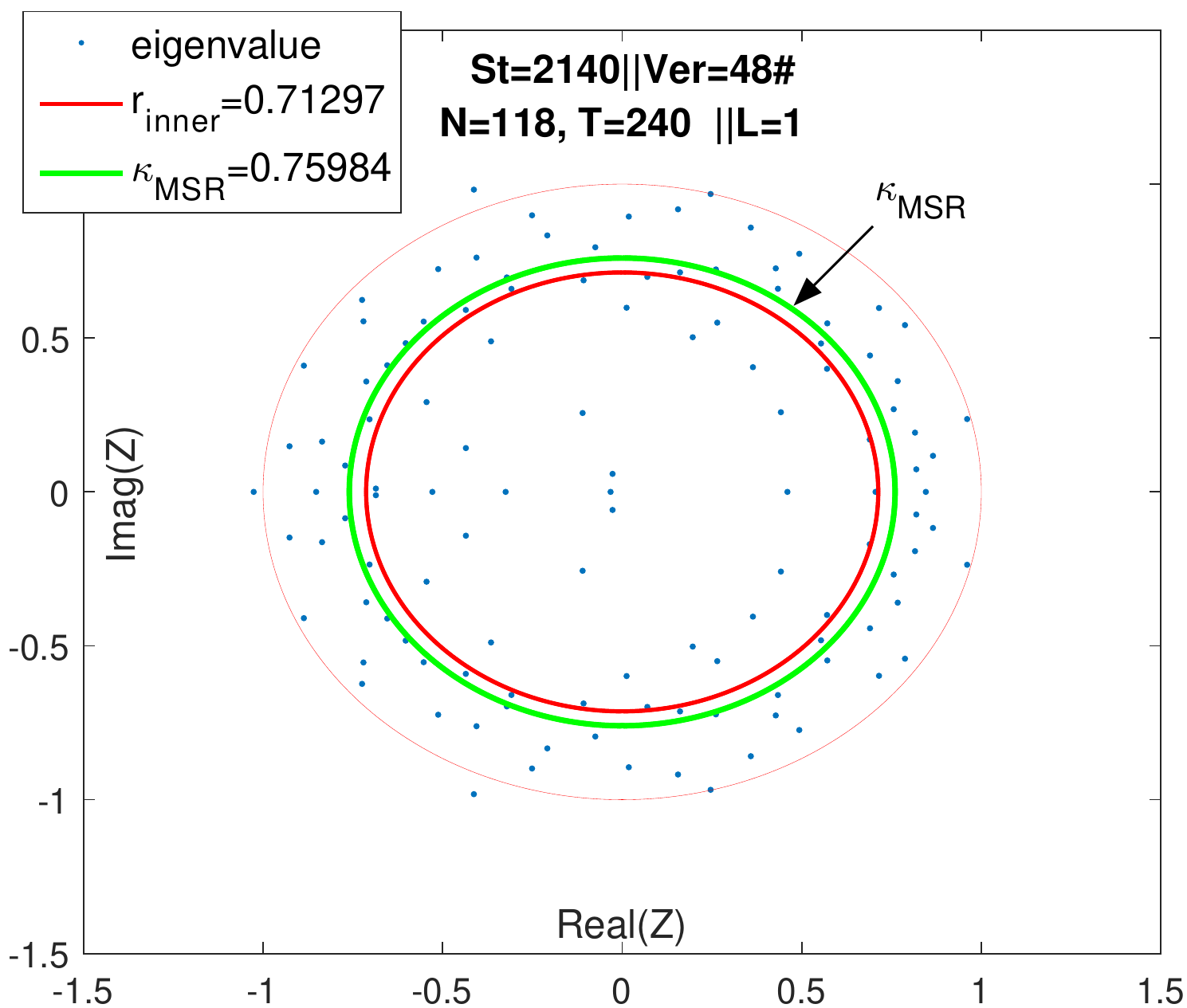}
 }
 \subfigure[M-P Law for $\Vector T_2$]{\label{fig:Case0T2mp}
 \includegraphics[width=0.5\textwidth]{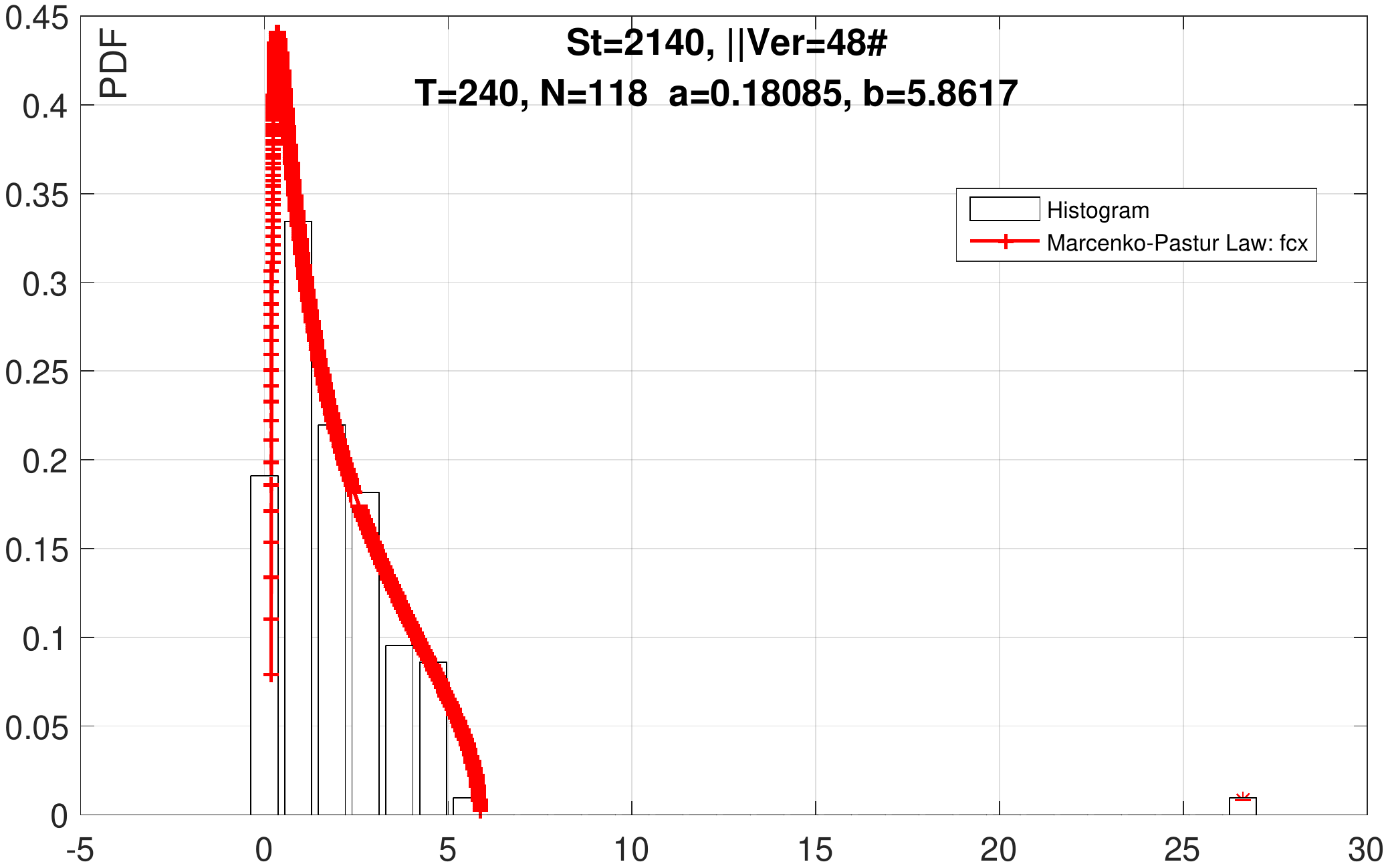}
 }
 \subfigure[Ring Law for $\Vector T_3$]{\label{fig:Case0T3ring}
 \includegraphics[width=0.4\textwidth]{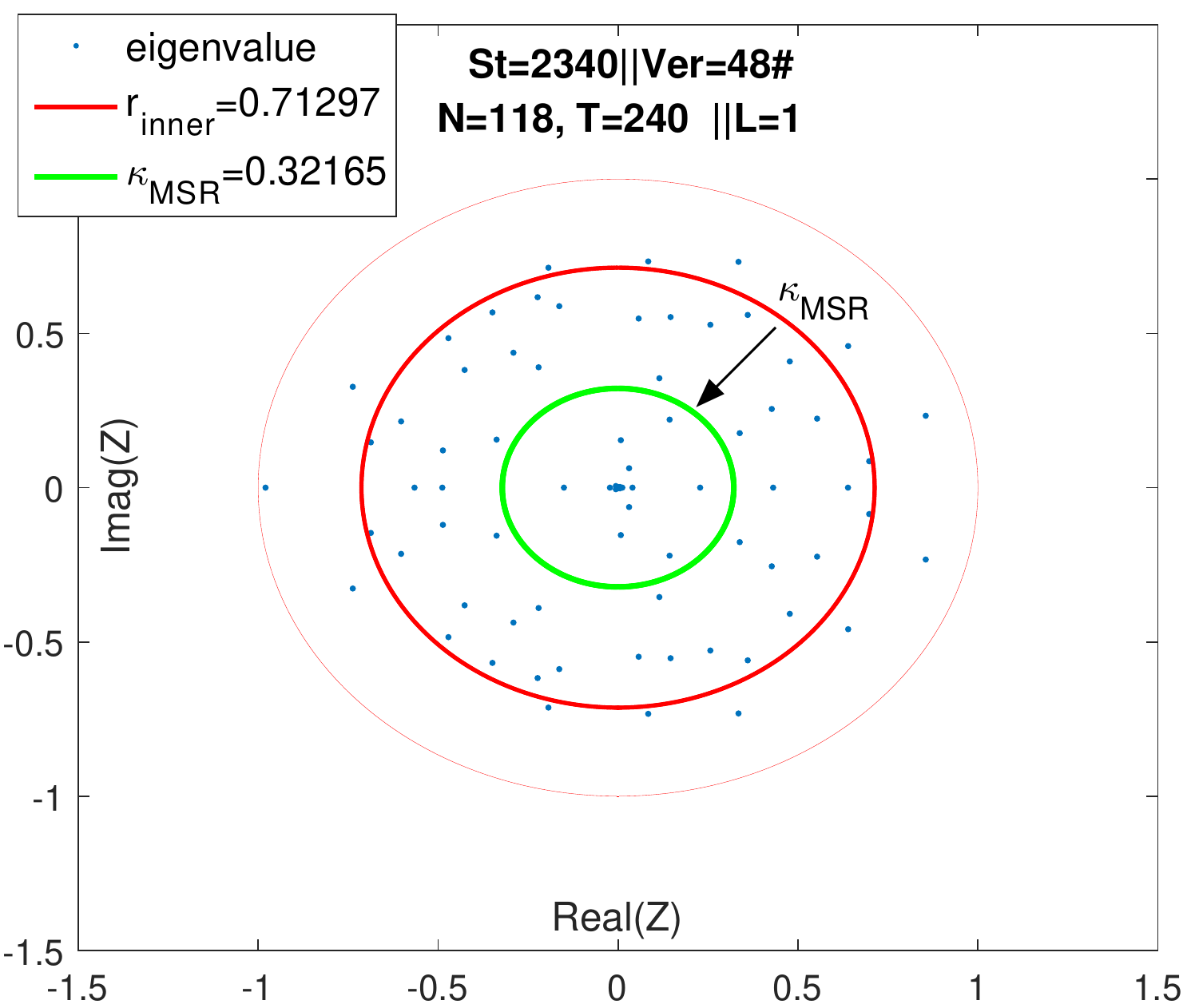}
 }
 \subfigure[M-P Law  for $\Vector T_3$]{\label{fig:Case0T3mp}
 \includegraphics[width=0.5\textwidth]{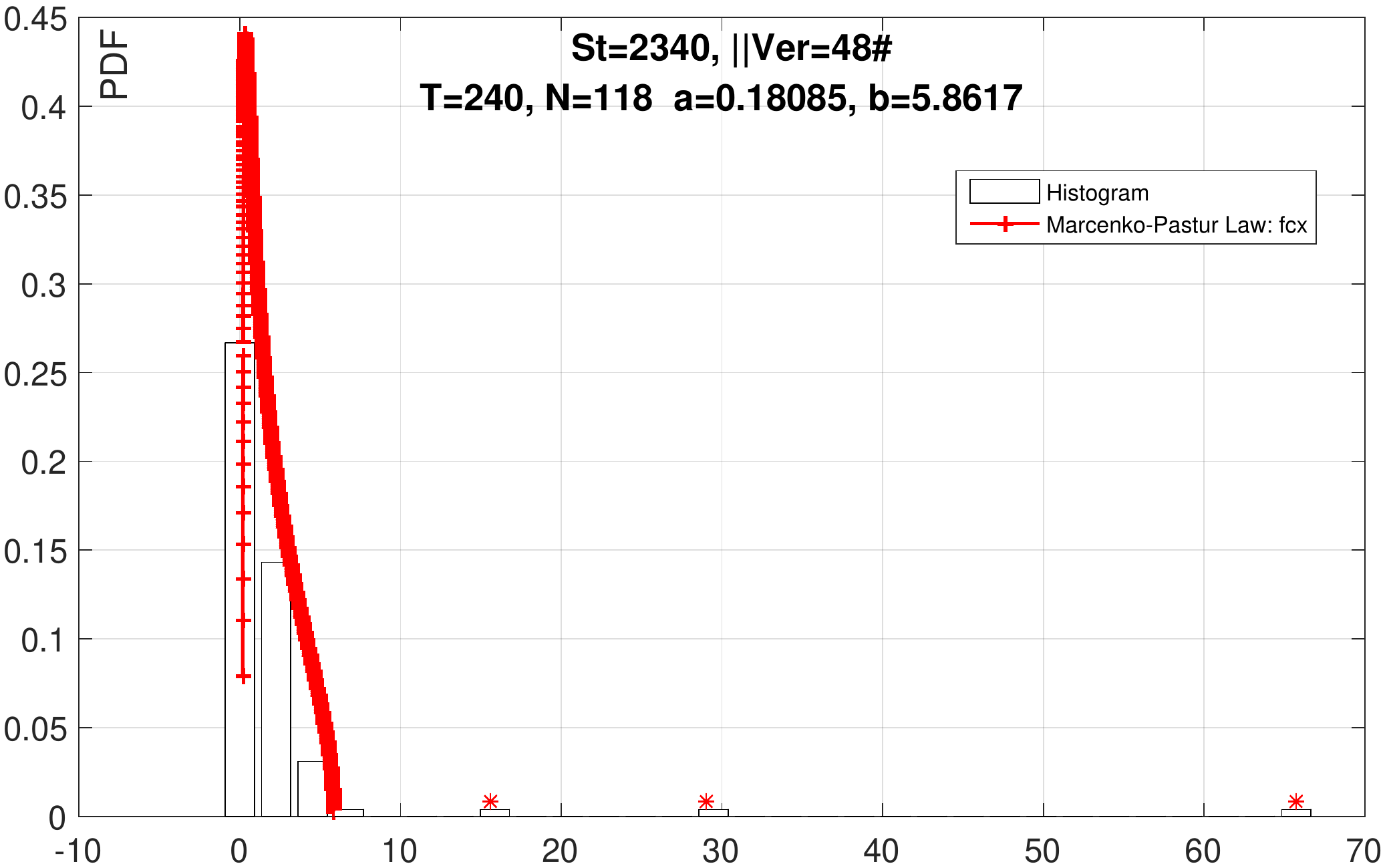}
 }
 \caption{RMT-based Results for Voltage Stability Evaluation.}
 \label{fig:Case0F}
\end{figure}

The $V-P$ curve (also called nose curve) and the smallest eigenvalue of the Jacobian matrix \cite{lim2016svd} are two clues for steady stability evaluation.   In this case, we focus on   the \textbf{E}4 part during which \VPbus{52} keep increasing to break down the steady stability. The related $V-P$ curve and  $\lambda-P$ curve, respectively, are given in Fig. \ref{fig:Case0NosecurveVP} and Fig. \ref{fig:Case0NosecurveEigP}. Only using the data source $\bm{\Omega}_{\mathbf{V} }$, we choose some data cross-section, $\mathbf T_1\!:\![1601\!:\!1840]; \quad \mathbf T_2\!:\![1901\!:\!2140]; \quad \mathbf T_3\!:\![2101\!:\!2340],$  as shown in  Fig. \ref{fig:Case0NosecurveVP}. The RMT-based results are shown as Fig. \ref{fig:Case0F}. The outliers become more evident as the stability degree decreases. The statistics of the outliers are similar to the smallest eigenvalue of Jacobian Matrix, Lyapunov Exponent or the entropy in some sense.

For further analysis, we take the signal and stage division into account. Generally speaking, sorted by the stability degree, the stages are ordered as  $\textbf{S}0>\textbf{S}1>\textbf{S}2>\textbf{S}3 \gg \text{max}(\textbf{S}4, \textbf{S}5)\gg \textbf{S}6 \gg \textbf{S}7$. According to Fig. \ref{fig:Case0LESs}, we make the Table \ref{Tab: Case0LESs}. The high-dimensional indicators $\overline{\tau{}_\mathbf{X}}_\text{R}$ and
$V_{\text{R}}$ have the same trend as the stability degree order. These statistics have the potential for data-driven stability evaluation.

\begin{table}[htbp]
\caption{Indicator of Various LESs at Each Stage.}
\label{Tab: Case0LESs}
\centering

\begin{minipage}[!h]{0.8\textwidth}
\centering

\footnotesize
\begin{tabularx}{\textwidth} { >{\scshape}l !{\color{black}\vrule width1pt}  >{$}l<{$}  >{$}l<{$}  >{$}l<{$}   >{$}l<{$}  >{$}l<{$}  >{$}l<{$}   >{$}l<{$}  >{$}l<{$}}   
\toprule[1.5pt]
\hline
 & {\text{MSR}} &   {\text{T}_2} & {\text{T}_3} & {\text{T}_4} & {\text{DET}}  & {\text{LRF}}\\
\hline
\hline
\multicolumn{7}{l} {$\textbf{E}_0$: Theoretical Value}\\
\hline
\STE{\tau}&0.8645&1338.3&10069 &8.35\text{E}4&48.322&73.678\\
$\Fdata {D}_{\text{T}}({\tau})$&-     &665.26 &93468 &1.30\text{E}7&1.3532&1.4210\\
\toprule[1pt]

\multicolumn{7}{l} {$\textbf{S}0$ [{0240:0500}, {261}]: Small fluctuations around 0 MW}\\
\hline
$\overline{\tau{}_\mathbf{X}}_\text{R}$&0.995&1.010&1.040&1.080&0.959&1.014\\
$V$  &6\text{E}\!-\!6&78.38&3.03\text{E}4&7.14\text{E}6&0.4169&0.3908\\
$V_{\text{R}}$  &1&1&1&1&1&1\\
\toprule[1pt]

\multicolumn{7}{l} {$\textbf{S}5$ [{0501:0739}, {239}]: A step signal (0 MW $\uparrow$ 30 MW) is included}\\
\hline
$\overline{\tau{}_\mathbf{X}}_\text{R}$&0.9331&1.280&2.565&7.661&0.5453&1.284\\
$V_{\text{R}}$ &1.49\text{E}1 &1.64\text{E}2&1.16\text{E}3&8.63\text{E}3&3.43\text{E}1&3.97\text{E}1\\
\toprule[1pt]

\multicolumn{7}{l} {$\textbf{S}1$ [{0740:0900}, {161}]: Small fluctuations around 30 MW}\\
\hline
$\overline{\tau{}_\mathbf{X}}_\text{R}$&0.9943&1.010&1.039&1.084&0.9568&1.015\\
$V_{\text{R}}$&0.8608  &0.9121&0.9476&1.234&0.8972&1.101\\
\toprule[1pt]

\multicolumn{7}{l} {$\textbf{S}6$ [{0901:1139}, {239}]: A step signal (30 MW $\uparrow$ 120 MW) is included}\\
\hline
$\overline{\tau{}_\mathbf{X}}_\text{R}$&0.8742&2.054&1.06\text{E}1&7.22\text{E}1&7\text{E}\!-\!2&1.597\\
$V_{\text{R}}$ &5.49\text{E}1 &2.06\text{E}3&3.87\text{E}4&8.54\text{E}5&1.52\text{E}2&1.62\text{E}2\\
\toprule[1pt]
\multicolumn{7}{l} {$\textbf{S}2$ [{1140:1300}, {161}]: Small fluctuations around 120 MW}\\
\hline
$\overline{\tau{}_\mathbf{X}}_\text{R}$&0.9930&1.019&1.067&1.135&0.9488&1.021\\
$V_{\text{R}}$ &0.7823 &1.053&1.189&1.135&0.7310&0.9255\\
\toprule[1pt]
\multicolumn{7}{l} {$\textbf{S}4$ [{1301:1539}, {239}]: A ramp signal (119.7 MW $\nearrow$) is included}\\
\hline
$\overline{\tau{}_\mathbf{X}}_\text{R}$&0.9337&1.295&2.787&9.615&0.5316&1.294\\
$V_{\text{R}}$ &8.50\text{E}1 &7.41\text{E}2&5.63\text{E}3&5.17\text{E}4&2.14\text{E}2&2.30\text{E}2\\
\toprule[1pt]
\multicolumn{7}{l} {$\textbf{S}3$ [{1540:2253}, {714}]: Steady increase ($\nearrow$ 358.1 MW)}\\
\hline
$\overline{\tau{}_\mathbf{X}}_\text{R}$&0.8906&1.717&6.530&3.48\text{E}1&0.1483&1.545\\
$V_{\text{R}}$ &1.35\text{E}1 &3.28\text{E}2&5.33\text{E}3&1.10\text{E}5&6.11\text{E}1&6.85\text{E}1\\
\toprule[1pt]
\multicolumn{7}{l} {$\textbf{S}7$ [{2254:2500}, {247}]: Static voltage collapse (361.9 MW $\nearrow$)}\\
\hline
$\overline{\tau{}_\mathbf{X}}_\text{R}$&0.4259&1.02\text{E}1&2.11\text{E}2&4.65\text{E}3&-1.4\text{E}1&1.08\text{E}1\\
$V_{\text{R}}$ &1.94\text{E}3 &5.81\text{E}5&1.20\text{E}8&3.2\text{E}10&9.02\text{E}4&9.62\text{E}4\\
\toprule[1pt]

\hline
\toprule[1pt]
\end{tabularx}
\raggedright
 {\small{}
 *$\overline{\tau{}_\mathbf{X}}_\text{R}=\overline{\tau{}_\mathbf{X}}/\mathbb{E}({\tau})$;
 $V_{\text{R}}({\tau{}_\mathbf{X}})=V{({\tau{} _\mathbf{X}})}/V{({\tau{} _{\mathbf{X}_0}})}$.
 \normalsize{}
}
\end{minipage}
\end{table}

\begin{figure}[htbp]
\centering
\includegraphics[width=0.6\textwidth]{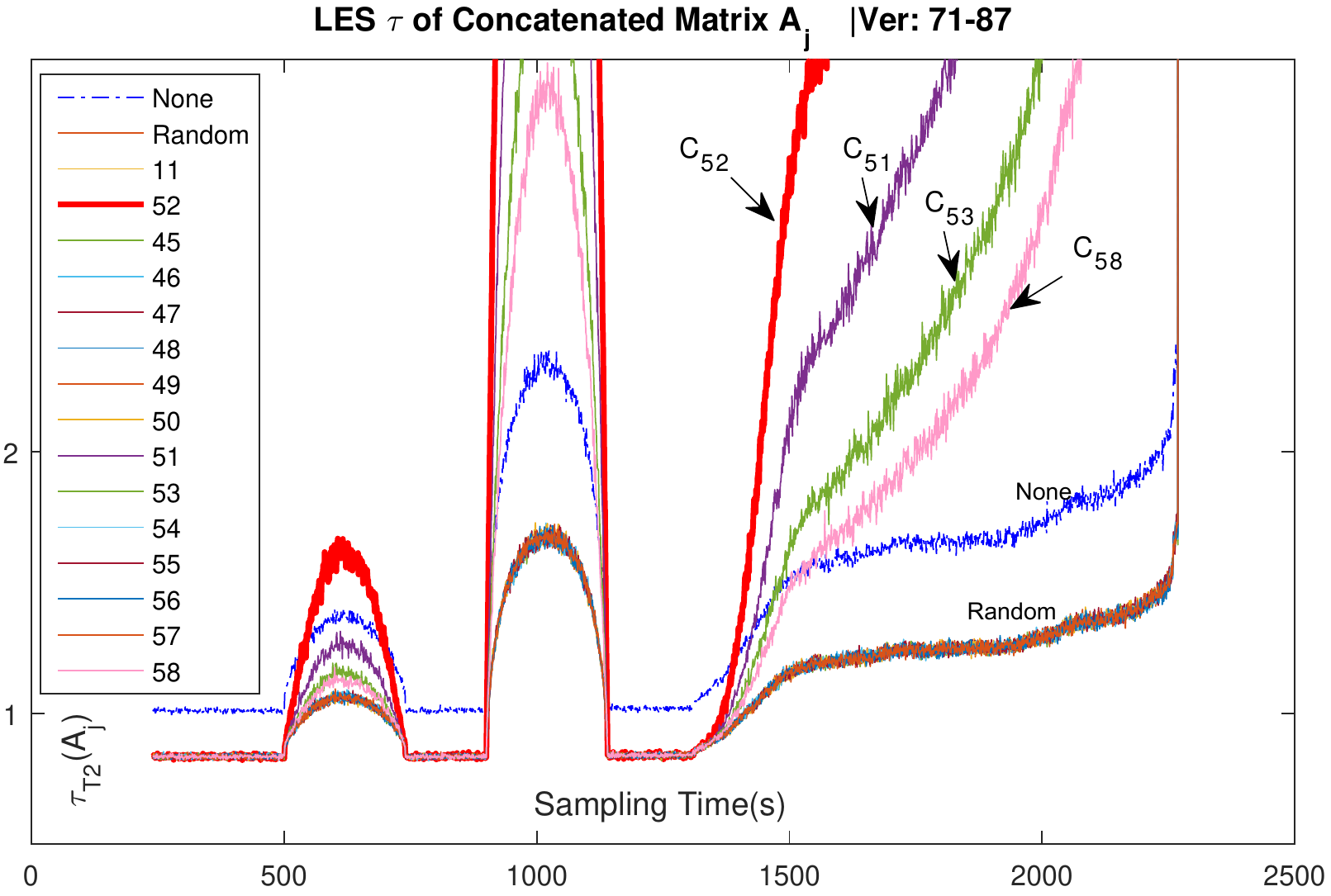}
\caption{Sensitivity Analysis based on Concatenated Matrix.}
\label{fig:Case0Concatenation}
\end{figure}

The key for correlation analysis is the  concatenated matrix $\mathbf{A}_i$, which consist of two part---the basic matrix $\mathbf{B}$ and a certain factor matrix $\mathbf{C}_i$, i.e., $\mathbf{A}_i\!=\![\mathbf{B}; \mathbf{C}_i]$. For more details, see our previous work \cite{Xu2015A}.
The LES of each $\mathbf{A}_i$ is computed in parallel, and Fig. \ref{fig:Case0Concatenation} shows the results.

In Fig. \ref{fig:Case0Concatenation}, the blue dot line (marked with None) shows the LES of basic matrix  $\mathbf{B}$, and the orange line (marked with Random) shows the LES of the concatenated matrix $[\mathbf{B}; \mathbf{R}]$ ($\mathbf{R}$ is the standard Gaussian Random Matrix). Fig. \ref{fig:Case0Concatenation} demonstrates that: 1) node 52 is the causing factor of the anomaly; 2) sensitive nodes are 51, 53, and 58; and 3) nodes 11, 45, 46, etc, are not affected by the anomaly. Based on this algorithm, we can continue to conduct behavior analysis, e.g.,  detection and estimation of residential PV installations \cite{Zhang2016A}. Behavior analysis is a big topic. Limited to the space, we will not expand it here.

\subsection{Early Event Detection using Free Probability}
\label{EEDFP}

$\bf{Problem \ Modeling:}$
Following \cite{He2016Designing}, we build the statistic model for power grid.
Considering   $T$ random vectors observed at time instants $i=1,...,T,$ we form a random matrix as follows
 \begin{equation}
\label{eq100}
	\left[ {\Delta {{{\mathbf{V}}_1}} , \cdots ,\Delta  {{{\mathbf{V}}_T}}} \right] = \left[ { {{\mathbf{\Xi}}_1\Delta{\mathbf{P}}_1}, \cdots , {{{\mathbf{\Xi}}_T\Delta \mathbf{P}}_T}} \right].
\end{equation}

In an equilibrium operating system, the voltage magnitude vector injections ${\bf V}$ with entries $V_{i},i=1,\cdots,N$ and the phase angle vector injections $\boldsymbol{\theta}$ with entries $\theta_{i},i=1,\cdots,N$ experience slight changes. Without dramatic topology changes, rich statistical empirical evidence indicates that the Jacobian matrix $\mathbf{J}$  keeps nearly constant, so does $\mathbf{\Xi}$. Also, we can estimate the changes of ${\bf V},$ $\boldsymbol{\theta},$ and $\mathbf{\Xi}$ only with the classical approach.
Thus  we rewrite \eqref{eq100} as:

\begin{equation}
\label{Eq:pm1}
\mathbb{V} = {\bm{\Xi}_N}{\mathbb{P}}_{N \times T}
\end{equation}
where $\mathbb{V} = \left[ {\Delta {{{\mathbf{V}}_1}} , \cdots ,\Delta  {{{\mathbf{V}}_T}}} \right]$, $ {\bm {\Xi}}={\mathbf{\Xi}}_1=\cdots={\mathbf{\Xi}}_T,$ and $\mathbb{P} = \left[ {\Delta {{{\mathbf{P}}_1}} , \cdots ,\Delta  {{{\mathbf{P}}_T}}} \right].$
Here $\mathbb{V}$ and $\mathbb{P}$ are random matrices. In particular, $\mathbb{P}$ is a random matrix with Gaussian entries.

$\bf{Model \ Designs:}$ Multivariate  linear or nonlinear polynomials perform a significant role in problem modeling, so we build our  models on the basis of random matrix polynomials. Here, we study two typical random matrix polynomial models.

The first case is the multivariate linear polynomial: \[{P_1}({S_0},{S_1}) = {S_0} + {S_1}.\]

The second one is the selfadjoint multivariate nonlinear polynomial:  \[{P_2}({S_0},{S_1}) = {S_0}{S_1} + {S_1}{S_0}.\]

Here, both ${S_0}$ and ${S_1}$ are the sample covariance matrices. The asymptotic eigenvalue distributions of $P_1$ and $P_2$  can be obtained via basic principles of the free probability theory as introduced above. The asymptotic eigenvalue distributions of $P_i$ are regarded as the theoretical bounds.

$\bf{Hypothesis \ Testing \ and \ Anomaly \ Detection:}$

We formulate our problem of anomaly detection in terms of the same hypothesis testing as \cite{He2016Designing}: no outlier exists ${\cal H}_0$, and outlier exists ${\cal H}_1$.

\begin{equation}
\left|
\begin{array}{*{20}{c}}
{{{\cal H}_0}:\widetilde{\mathbb{V}}  = \widetilde{\Xi}}{R}_{N\times T}\\
{{{\cal H}_1}:\widetilde{\mathbb{V}}  \neq \widetilde{\Xi}}{R}_{N\times T}
\end{array}
\right.
\end{equation}
where ${R}$ is the standard Gaussian random matrix.

Generate ${S_0}$, ${S_1}$ from the sample data through the  preprocess in \ref{The Processing of the Grid Data} . Compare the theoretical bound with the spectral distribution of raw data polynomials.
If outlier exists, ${\cal H}_0$ will be rejected, i.e. signals exist in the system.

$\bf{The \ Processing \ of \ the \ Grid \ Data:}$
\label{The Processing of the Grid Data}
The data sampled from power grid is always non-Gaussian, so we adopt a  normalization procedure in \cite{He2015A} to conduct data preprocessing. Meanwhile, we employ Monte Carlo method to compute the spectral distribution of raw data polynomial according to the asymptotic property theory. See details in Algorithm.\ref{alg1}.

\begin{algorithm}[htbp]

\caption{}
\label{alg1}
\begin{algorithmic}[1]
\REQUIRE ~~\\
The sample data matrices: ${V_0}$ and ${V_1}$ ;\\
The number of repetition times : $M$;\\
The size of ${V_0}$ and ${V_1}$: $N,T$;\\
SNR: $\eta$
        \FOR{$i \leq M$}

            \STATE add  small white noises to sample data matrices；\\
            $ \widetilde{V_0}= V_0 + \eta$ randn($N,N$);\\
            $ \widetilde{V_1}= V_0 + \eta$ randn($N,N$);
            \STATE standardize $\widetilde{V_0}$ and $\widetilde{V_1}$, i.e. mean=0, variance=1;
            \STATE generate the covariance matrices: ${S_0}=\widetilde{V_0}\widetilde{V_0}^{'}/N$,\\
             ${S_0}=\widetilde{V_1}\widetilde{V_0}^{'}/N$;
            \STATE compute the eigenvalues of $P({S_0},{S_1})$;
        \ENDFOR
\STATE Computer the frequency of different eigenvalues and draw the spectral distribution histogram;
\ENSURE ~~\\
The spectral distribution  histogram.
\end{algorithmic}
\end{algorithm}

$\bf{Simulation \ Results:}$

\begin{figure}[htbp]
\centering
\begin{overpic}[scale=0.45]{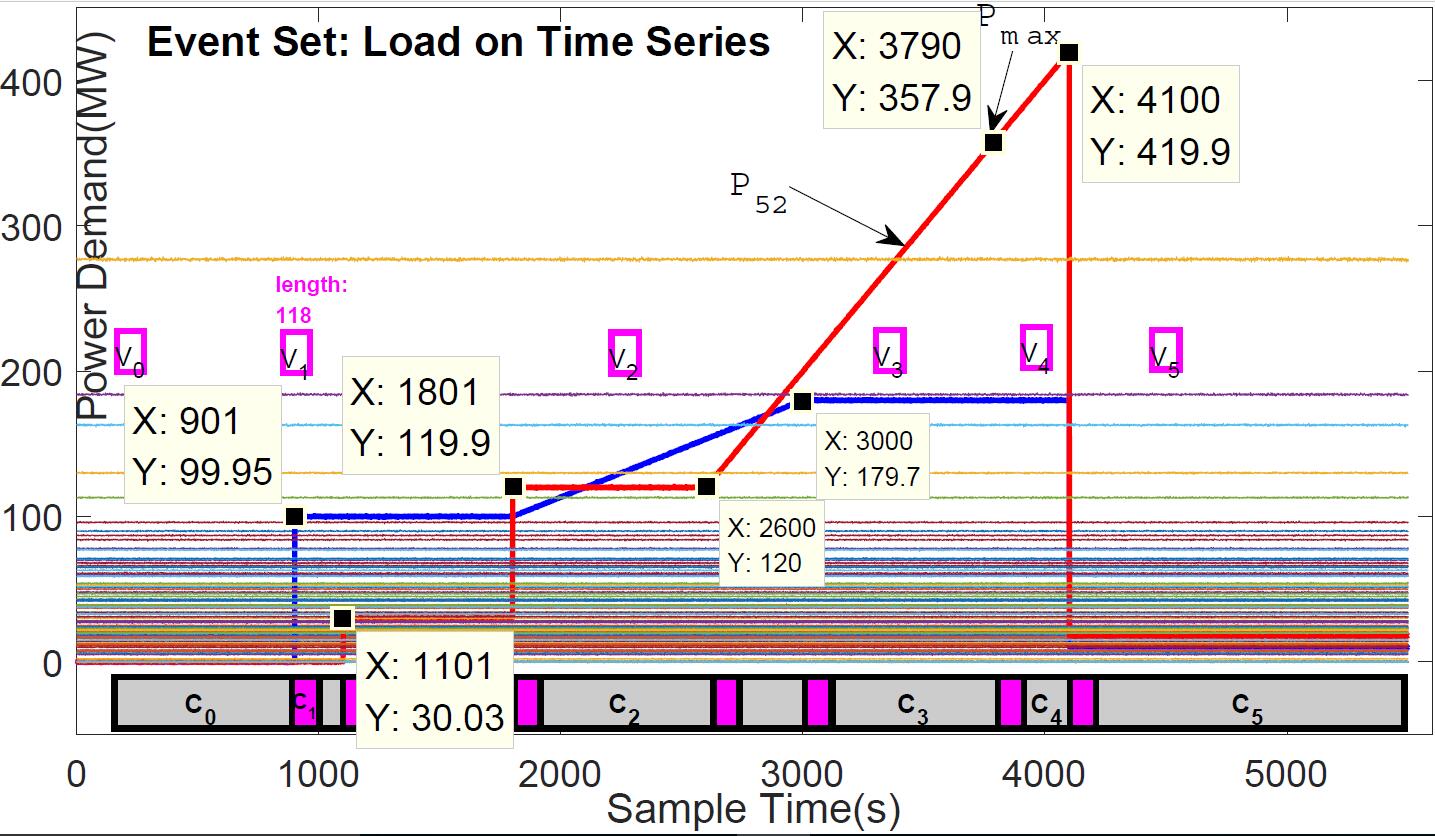}

\end{overpic}
\caption{The event assumptions on time series.}
\label{fig:loadevent}
\end{figure}

\begin{figure}[htbp]
 \centering
 \subfigure[White noises $V_5$ \& $V_0$]{
 \includegraphics[width=0.45\textwidth]{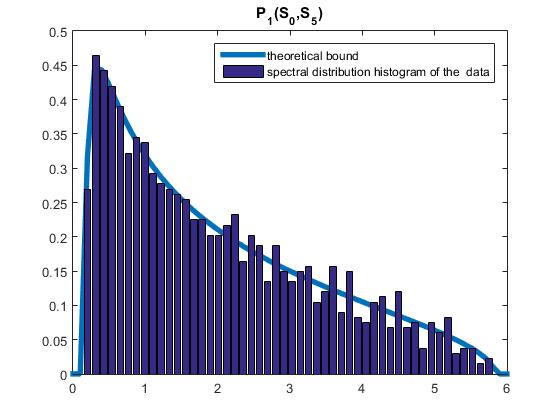}
 }
 \subfigure[Step signal $V_1$ \& $V_0$]{
 \includegraphics[width=0.45\textwidth]{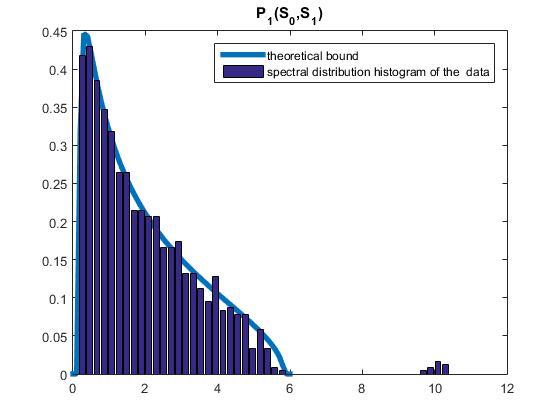}
 }
 \subfigure[Stable growth A $V_2$ \& $V_0$]{
 \includegraphics[width=0.45\textwidth]{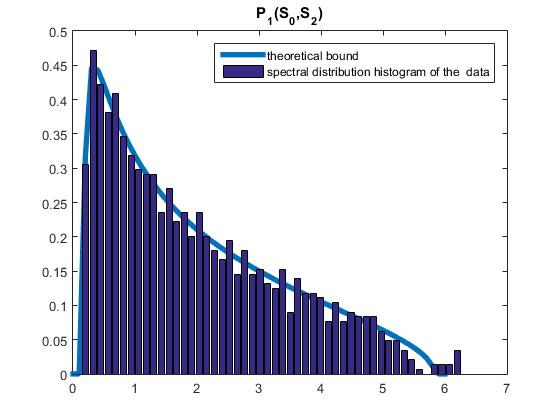}
 }\\
 \subfigure[Stable growth B $V_3$ \& $V_0$]{
 \includegraphics[width=0.45\textwidth]{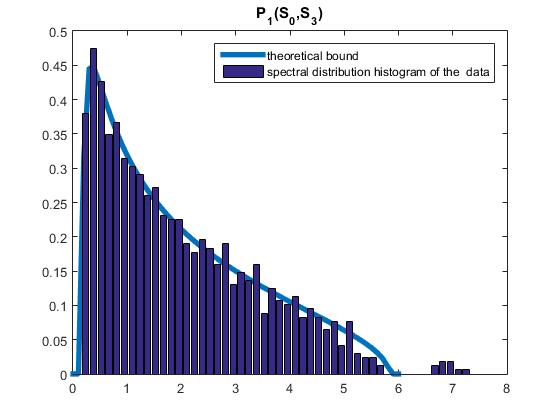}
 }
 \subfigure[Voltage collapse $V_4$ \& $V_0$]{
 \includegraphics[width=0.45\textwidth]{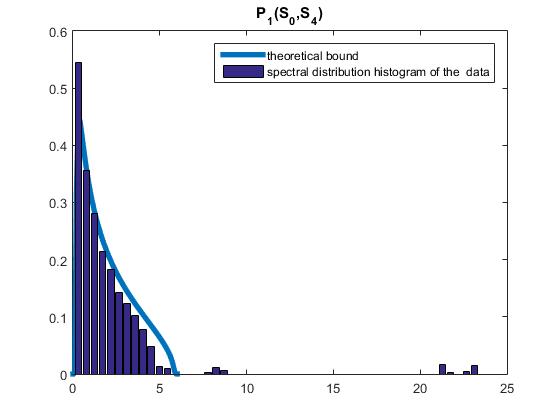}
 }
 \caption{Data fusion using multivariate linear polynomial $P_1$}
 \label{fig12}
 \end{figure}

\begin{figure}[htbp]
 \centering
 \subfigure[White noises $V_5$ \& $V_0$]{
 \includegraphics[width=0.45\textwidth]{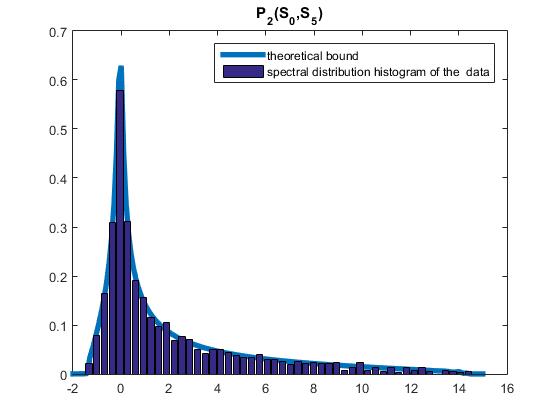}
 }

 \subfigure[Step signal $V_1$ \& $V_0$]{
 \includegraphics[width=0.45\textwidth]{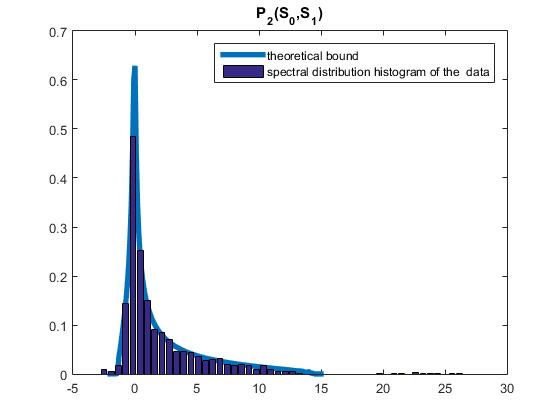}
 }
 \subfigure[Stable growth A $V_2$ \& $V_0$]{
 \includegraphics[width=0.45\textwidth]{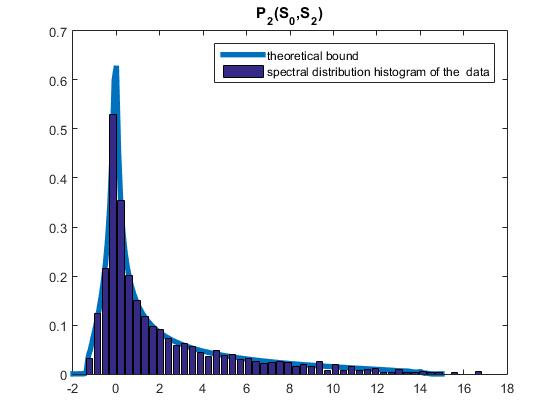}
 }\\
 \subfigure[Stable growth B $V_3$ \& $V_0$]{
 \includegraphics[width=0.45\textwidth]{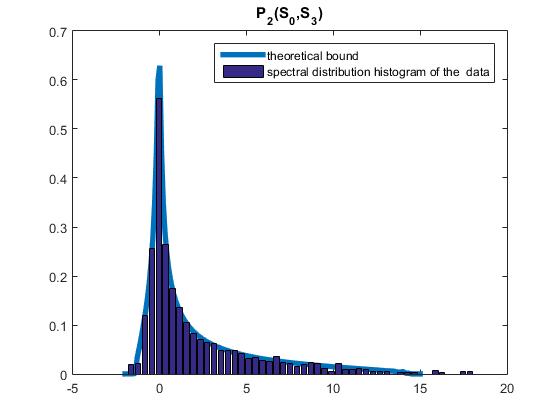}
 }
 \subfigure[Voltage collapse $V_4$ \& $V_0$]{
 \includegraphics[width=0.45\textwidth]{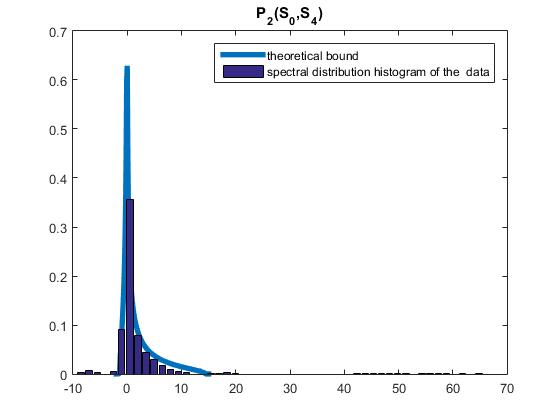}
 }
 \caption{Data fusion using multivariate nonlinear polynomial $P_2$.}
 \label{fig13}
 \end{figure}

Our data fusion method is tested with simulated data in the standard IEEE 118-bus system. Detailed information of the system is referred to the case118.m in Matpower package and Matpower 4.1 User's Manual \cite{Zimmerman2011MATPOWER}. For all cases, let the sample dimension $N=118$.
In our simulations, we set the sample length equal to $N$, i.e. $T=118$, $c=T/N=1$ and select six sample voltage matrices presented in Tab. \ref{tab1}, as shown in Fig. \ref{fig:loadevent}.
The results of our simulations are presented in Fig.~\ref{fig12} and Fig.~\ref{fig13}. The outliers existed when the system was abnormal  and its sizes become large when the anomaly become serious.

\begin{table}[htbp]

\centering

\caption{System status and sampling data}

\begin{tabular}{p{4cm}|p{4cm}|p{5cm}}
\hline
  Cross Section (s)& Sampling (s) &Descripiton\\
\hline
  $\textbf{C}_0:118-900$ & $V_0:100\sim217$&Reference, no signal\\
  $\textbf{C}_1:901-1017$ & $V_1:850 \sim 967$&Existence of a step signal \\
  $\textbf{C}_2:1918-2600$ & $V_2:2200 \sim 2317$&Steady load growth for Bus 22\\
  $\textbf{C}_3:3118-3790$ & $V_3:3300 \sim 3417$&Steady load growth for Bus 52\\
  $\textbf{C}_4:3908-4100$ & $V_4:3900 \sim 4017$&Chaos due to voltage collapse\\
  $\textbf{C}_5:4118-5500$ & $V_5:4400 \sim 4517$&No signal\\
\hline
\end{tabular}
\label{tab1}
\raggedright
\\
{*We choose the temporal end edge of the sampling matrix as the marked time for the cross section. E.g., for $V_0:100\sim217$, the temporal label is 217 which belong to $\textbf{C}_0:118-800$. Thus, this method is able to be applied to conduct real-time analysis.}
 \end{table}

\section{Conclusion and Future Directions}
\label{ConC}

Motivated by the immediate demands of tackling the tricky problems raised from large scale smart grids, this chapter introduced RMT-based schemes for spatio-temporal big data analysis.  Firstly, we represent the spatio-temporal PMU data as a sequence of large random matrices. This is a crucial part for power state evaluation as it turning the big PMU data into tiny data for the practical use. Rather than employing the raw PMU data, a comprehensive analysis of PMU data flow, namely, RMT-based techniques, is then proposed to indicate the state evaluation state. The core techniques include streaming PMU data modelling, asymptotic properties analysis and data fusion methods (based on free probability). Besides, the case studies based on synthetic data and real data are also included with the aim to bridge the technology gap between RMT and spatio-temporal data analysis in smart grids.

The current works based on RMT, provides a fundamental exploration of data analysis for spatio-temporal PMU data. Much more attentions are to be paid along this research direction, such as classification of power events and load forecasting. It is also noted that this work provide data-driven methods which are new substitutes for power system state estimation. The combination of power system scenario analysis, spectrum sensing mechanisms, networking protocols and big data techniques \cite{qiu2014cognitive, Khan2015Cognitive, Khan2017Requirements, qiu2015smart} is encouraged to be investigated for better understanding of the power system state.

\bibliographystyle{unsrt}
\bibliography{leochu}

\end{document}